\documentclass[author-year, prd, amsmath, amssymb, longbibliography, reprint,superscriptaddress]{revtex4-1}

\usepackage{graphicx}
\usepackage[caption=false]{subfig}
\usepackage{dcolumn}
\usepackage{bm}
\usepackage{natbib}
\usepackage{xfrac}
\usepackage[toc,page]{appendix}
\usepackage[alsoload=hep]{siunitx}
\usepackage{tikz}
\usepackage{tikzscale}
\usepackage{pgfplots}
\usepackage{pgfplotstable}
\usepackage{adjustbox}
\usepackage{tikz-3dplot}
\usepackage[mode=buildnew]{standalone}
\usepackage{etoolbox}
\usepackage{mdframed}
\usepackage[shortlabels]{enumitem}
\usepackage{color}
\usepackage{array}
\usepackage{float}
\usepackage{bigints}
\usepackage{subfloat}
\usepackage{subfig}
\usepackage{tabularx}
\usepackage{booktabs}
\usepackage{upgreek}
\usepackage{xparse}
\usepackage{xstring}
\usepackage{ifthen}
\usepackage{tensor}
\usepackage{mathrsfs}
\usepackage{ragged2e}
\usepackage{xfrac}
\usepackage{wrapfig}

\AtBeginDocument{
\heavyrulewidth=.08em
\lightrulewidth=.05em
\cmidrulewidth=.03em
\belowrulesep=.65ex
\belowbottomsep=0pt
\aboverulesep=.4ex
\abovetopsep=0pt
\cmidrulesep=\doublerulesep
\cmidrulekern=.5em
\defaultaddspace=.5em
}

\newcommand{\gw}{\textsc{gw}}

\newcommand{\vect}[1]{\bm{\mathrm{#1}}}
\newcommand{\bigo}[1]{\mathcal{O}\!\left(#1\right)}

\newcolumntype{P}[1]{>{\centering\arraybackslash}p{#1}}

\newcommand{\rmi}{\mathrm{i} \mkern1mu} 
\newcommand{\dd}{\textrm{d}}
\newcommand{\Si}{\textrm{Si}}
\newcommand{\Ci}{\textrm{Ci}}

\newcommand{\np}[1]{^{\!\!#1}}
\newcommand{\cn}{\hspace*{-0.75pt}\mathsc{N}\hspace*{-0.75pt}}

\newcommand{\Imath}{\ensuremath{\hat{\imath}}} 
\newcommand{\Jmath}{\ensuremath{\hat{\jmath}}} 

\usepackage{soul}

\newcommand{\mathsc}[1]{\mbox{\tiny{\(#1\)}}}

\renewcommand{\thetable}{\arabic{table}}
   
\makeatletter 
	\renewcommand{\fnum@figure}{Fig.~\thefigure}
\makeatother

\makeatletter 
	\renewcommand{\fnum@table}{Table~\thetable}
\makeatother

\newmdenv[leftline=false,rightline=false]{topbotfig}

\allowdisplaybreaks[4]

\newcolumntype{R}[1]{>{\raggedleft\arraybackslash}p{#1}}
\newcolumntype{C}[1]{>{\centering\arraybackslash}p{#1}}
\newcolumntype{L}[1]{>{\raggedright\arraybackslash}p{#1}}

\begin{document}

\def\apjl{ApJ}
\def\prd{Phys.~Rev.~D}
\def\araa{Annual~Review~of~Astron.~and~Astrophys.}


\title[Astrometric Effects of Gravitational Wave Backgrounds with non-Einsteinian Polarizations]{Astrometric Effects of Gravitational Wave Backgrounds\\with non-Einsteinian Polarizations}

\author{Deyan P. Mihaylov}
\email{d.mihaylov@ast.cam.ac.uk}
\affiliation{Institute of Astronomy, University of Cambridge, Madingley Road, Cambridge, CB3 0HA, UK}

\author{Christopher J. Moore}
\email{cjm96@cam.ac.uk}
\affiliation{Centro de Astrof{\'{i}}sica e Gravita{\c{c}}{\~{a}}o -- CENTRA, Departamento de F{\'{i}}sica, Instituto Superior T{\'{e}}cnico -- IST, Universidade de Lisboa -- UL, Av.\ Rovisco Pais 1, 1049-001 Lisboa, Portugal}
\affiliation{DAMTP, Centre for Mathematical Sciences, University of Cambridge, Wilberforce Road, Cambridge, CB3 0WA, UK}

\author{Jonathan~R.~Gair}
\email{j.gair@ed.ac.uk}
\affiliation{School of Mathematics, University of Edinburgh,~Peter Guthrie Tait Road, Edinburgh, EH9 3FD, UK}

\author{Anthony Lasenby}
\email{a.n.lasenby@mrao.cam.ac.uk}
\affiliation{Astrophysics Group, Cavendish Laboratory, J J Thomson Avenue, Cambridge CB3 0HE, UK}
\affiliation{Kavli Institute for Cosmology, Madingley Road, Cambridge CB3 0HA, UK}

\author{Gerard Gilmore}
\email{gil@ast.cam.ac.uk}
\affiliation{Institute of Astronomy, University of Cambridge, Madingley Road, Cambridge, CB3 0HA, UK}

\date{\today}

\begin{abstract}
The \emph{Gaia} mission offers a new opportunity to search for the low frequency gravitational wave background using astrometric measurements. In this paper, the astrometric effect of gravitational waves is reviewed, with a particular focus on the effect of non-Einsteinian gravitational wave polarizations. A stochastic gravitational wave background generates a correlated vector field of astrometric deflections on the sky. A convenient decomposition for the correlation matrix is introduced, enabling it to be calculated for all possible gravitational wave polarizations and compared to the redshift correlations from the pulsar-timing literature; in the case of a \textsc{gr} background of transverse traceless gravitational waves, this also allows us to identify an astrometric analog of the famous Hellings--Downs curve. Finally, the cross-correlation between the redshift and astrometric signal is also calculated; this may form the basis for future joint pulsar-timing and astrometry searches for arbitrarily polarized gravitational wave backgrounds.
\end{abstract}

\maketitle

\section{Introduction}
\noindent
The first direct detection of gravitational waves (\gw s) from a binary black hole was achieved in September 2015 by the \textsc{lvc} collaboration using the ground-based Advanced \textsc{ligo} detectors \citep{2016PhRvL.116f1102A}. This event, as well as subsequent binary black hole detections \citep{2016PhRvL.116x1103A, 2017PhRvL.118v1101A, 2017PhRvL.119n1101A, 2017arXiv171105578T}, and the first multi-messenger observations of a binary neutron star merger \citep{2017PhRvL.119p1101A}, have paved the way for the development of a host of new research topics in astronomy, astrophysics and cosmology \citep{Blair:2016idv, Hughes:2001ch, TheLIGOScientific:2016htt, Abbott:2017xzu}. In particular, gravitational waves afford the possibility to test the theory of general relativity (\textsc{gr}) in regimes which are inaccessible through light-based observations (e.g. strong gravity), and to constrain the deviations from general relativity far better than has previously been possible \citep{2016PhRvL.116v1101A, 2016PhRvX...6d1015A, 2017ApJ...848L..13A}.

There are, however, \textsc{gw} tests of \textsc{gr} which are difficult or impossible to perform with current instruments. For example, the geometry of the two \textsc{ligo} detectors alone hinders accurate determination of the \textsc{gw} polarization; this has been improved by the addition of Virgo which helped constrain the polarization content of GW170814 \citep{2017PhRvL.119n1101A}. However, there may still exist combinations of \textsc{gw} polarizations (the transverse and longitudinal scalar modes) which laser interferometers cannot distinguish \citep{2017arXiv170401899H,Will:2014kxa}. Pulsar timing arrays (\textsc{pta}s) provide one method for distinguishing between all 6 possible \textsc{gw} polarizations \citep{0004-637X-685-2-1304}, and here another such technique is considered.

The Gaia mission, launched in 2013 by \textsc{esa}, is carrying out a thorough mapping of more than a billion objects in the Milky Way \citep{2016A&A...595A...1G}. As part of this survey, star positions, velocities, and accelerations (among other observables) will be mapped with an unprecedented precision; this new map of the Milky Way objects will allow for an entirely updated understanding of the dynamics and composition of our Galaxy and of the Local Group \citep{1998SPIE.3350..541G}.

The idea of using high-precision astrometric measurements as an alternative avenue to detecting gravitational wave signals is not new: \citep{1990NCimB.105.1141B} were the first to examine the physical principles at play, and later \citep{Gwinn:1996gv} applied them to quasar proper motions. More recently \citep{Book:2010pf} published an in-depth analysis of the concept applied to astrometric measurements, \citep{2012A&A...547A..59M} evaluated the viability of finding this effect in detector data, and \citep{Klioner:2017asb} discussed it in direct relation to the Gaia mission. The approach has been investigated numerically in the context of Gaia, and future Data Releases promise better sensitivity to single \textsc{gw} sources than existing techniques in a certain frequency range \citep{PhysRevLett.119.261102}. This article explores the possibility of using such high-precision astrometric measurements to ascertain the polarization of a stochastic \textsc{gw} background, and hence constrain modified theories of gravity.

Within \textsc{gr}, gravitational waves travel at the speed of light and carry a superposition of two transverse, traceless polarization modes; usually denoted \(+\) and \(\times\) \citep{2005NJPh....7..204F}. Modified theories of gravity can have up to four additional polarization modes; a purely transverse scalar mode, two vectorial modes with mixed transverse and longitudinal properties, and a purely longitudinal scalar mode \citep{Will:2014kxa}. In general, these waves can travel at speeds other than the speed of light; however this is strongly constrained by recent joint electromagnetic and \textsc{gw} observations \citep{2017ApJ...848L..13A}. Detections of \textsc{gw}s using astrometric methods would allow for additional constraints to be placed on the polarisation and speed of \textsc{gw}s. This paper considers all six possible \textsc{gw} polarizations, but the discussion is restricted to waves travelling at the speed of light.

While still not explicitly detected, another possible target for gravitational wave detectors is a stochastic gravitational wave background (\textsc{sgwb}) created by a large number of random, uncorrelated, and individually unresolvable sources. According to the their origin, backgrounds can be classed as astrophysical or cosmological. Astrophysical sources range from supernovae and mergers of compact objects (\textsc{bh}-\textsc{bh}, \textsc{bh}-\textsc{ns}, and \textsc{ns}-\textsc{ns} binaries) to supermassive black hole binaries. Backgrounds of this origin are generally believed to have a power law spectrum and exist at frequencies above \(\sim 10^{-12}\,\si{\textrm{Hz}}\) \citep{Perrodin:2017bxr}. Cosmological backgrounds, on the other hand, include early-Universe events like reheating or inflation, or even more exotic alternatives like primordial black hole mergers or \textsc{qcd} phase transitions. These generally have more unusual spectra, and are more powerful at lower frequencies than astrophysical backgrounds. Astrophysical backgrounds, specifically \textsc{smbh} binaries corresponding to galaxy mergers, are by far the most promising in the context of astrometric and \textsc{pta}-based methods for detection. The most constraining upper limits on the stochastic \textsc{gw} background at frequencies of several nanohertz come from pulsar timing. The North American Nanohertz Observatory for Gravitational Waves (\textsc{nano}Grav) \cite{2016ApJ...821...13A} and the European Pulsar Timing Array (\textsc{epta}) \cite{2015MNRAS.453.2576L} placed 95\% Bayesian upper limits on the amplitude of a stochastic gravitational-wave background of \SI{1.5e-4}{} and \SI{3.0E-15}{} respectively at a frequency of 1 year\(^{-1}\). The Parkes Pulsar Timing Array (\textsc{ppta}) \cite{2013Sci...342..334S} placed a 95\% frequentist upper limit on the amplitude of \SI{2.4E-15}{} at a frequency of 2.8 nanohertz whilst the International Pulsar Timing Array (\textsc{ipta}) \cite{2016MNRAS.458.1267V} placed a \(2\sigma\) upper limit of \SI{1.7E-15}{} at a frequency of 1 year\(^{-1}\).

This article analyses the astrometric response to a gravitational wave and the detector cross-correlation functions (or overlap reduction functions) for an astrophysical background in the context of all possible gravitational wave polarization states. In Section~\ref{sec:astrometricResponse} the relativistic principles that are responsible for the periodic shift in a star's apparent position on the sky due to a gravitational wave are presented. Similar analyses have been published previously (e.g. \citep{Book:2010pf}), but it is repeated here with a focus on the astrometric response to non-\textsc{gr} polarization states. Section~\ref{sec:gravitationalWavepolarizations} summarizes the mathematical formulation of \textsc{gw} polarization tensors. The characteristic astrometric patterns produced by each polarization are then plotted and analyzed in terms of the geometry of the astrometric response function. The expected value of the correlation between the response function of two independent detectors is defined and derived for each of the polarization states in Section~\ref{sec:backgroundMapping}. Their significance and possible applications are also presented, alongside suitable depictions of their geometry.

The astrometric correlations due to a stochastic background of \textsc{gw}s with non-Einsteinian polarizations has not been explored previously. However, the redshift correlations due to such backgrounds have been explored previously, in the context of pulsar timing arrays. Notably, \citep{0004-637X-685-2-1304} sets out a thorough overview of the topic, and \citep{Gair:2015hra} calculates all relevant correlations for scalar and vector backgrounds. Throughout this paper, the astrometric correlations derived will be compared with the existing results for the redshift correlations.

\section{Astrometric response to a gravitational wave}
\label{sec:astrometricResponse}
\begin{figure}[b]
\includestandalone[scale=1]{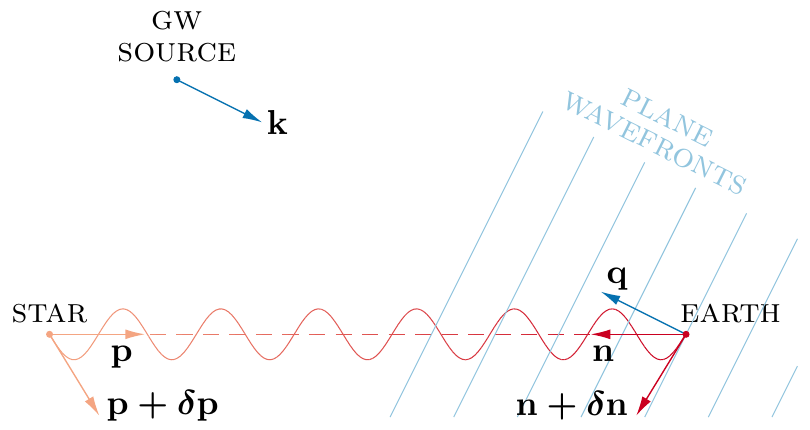}
\caption{An illustration of the geometric setup under consideration. The star lies on the past light cone of the Earth, and the two are linked by an unperturbed null geodesic with tangent vector \(p^{\mu}\) and the apparent position of the star is given by \(n_{\Imath}\). When the metric is perturbed by a \textsc{gw} with wavevector \(k^{\mu} = - q^{\mu}\), the tangent to the null geodesic is perturbed, \(p^{\mu}+\delta p^{\mu}\), along with the apparent position of the star, \(n_{\Imath} + \delta n_{\Imath}\). The apparent position of the \textsc{gw} source is~\(q_{\Imath}\).}
\label{fig:sourceObject}
\end{figure}

\noindent
Astrometric measurements of any distant objects could, in principle, be used to detect \gw s (the term ``star'' will be used to refer to any such distant object, although the same reasoning applies to any light source). The telescope used for the astrometric measurements will not be at rest (Gaia orbits at the Sun-Earth L2 point, and moves on a Lissajous-type orbit) and it will be necessary to correct for the detector motion; the term ``Earth'' will be used to refer to the location of an idealized stationary observer, and for simplicity it will be assumed that the necessary corrections in the data have already been made. Fig.~\ref{fig:sourceObject} depicts the relationship between the ``star", the gravitational wave source, and the ``Earth".

In this section the periodic astrometric deflection of a distant ``star'' as viewed from the ``Earth'' due to a weak, plane-fronted \textsc{gw} signal is derived. The background space-time is assumed to be flat (the mostly positive metric \(\eta_{\mu\nu}\!=\!\textrm{diag} (-, +, +, +)\) is used here) and only leading-order terms in the metric perturbation are retained. We designate coordinates in which the Earth is at the spatial origin, \(x^{\mu}_{\mathsc{E}}(t)=(t, 0)\), and the Star is at fixed spatial position, \(x^{\mu}_{\mathsc{S}}(t)=(t, x_{\mathsc{S}}^{i})\), with \(x_{\mathsc{S}}^{i} \equiv \text{\emph{const}}\). While the results in this section are not new (in particular our derivation is similar to that in \cite{Book:2010pf}) it will be useful to re-derive them here, paying special attention to the case of non-Einsteinian polarization states. The reader might be interested to consult \cite{Schutz} for a simplified discussion of this effect and a comparison of the interferometric, pulsar timing, and astrometric methods for \textsc{gw} detection.

In the absence of a metric perturbation the star and the Earth are joined by the following one-parameter family of null geodesics in the flat space-time, labeled by \(t_{0}\),
\begin{align}
{x_{t_{0}}^{\mu}(\lambda)=(\Omega\lambda + t_{0}, -\Omega\lambda n^{i})}.
\end{align}
Here \(\Omega\) is the unperturbed frequency of the photons coming from the star, \(-n^{i}\) is a unit three-vector in the direction of its propagation, the affine parameter \(\lambda\) varies from \({\lambda_{\mathsc{S}}\!\equiv -\left|\vect{x}_{\mathsc{S}}\right|/\Omega}\) to \(0\) between the star and the observer, and the tangent vector (4-momentum) \({p^{\mu}\equiv\textrm{d}x^{\mu}/\textrm{d}\lambda=\Omega(1, -n^{i})}\) is null.

The observer (source) is equipped with a time independent orthonormal tetrad \(\epsilon^{\mu}_{\hat{a}}\) (\(\sigma^{\mu}_{\hat{a}}\)), satisfying \({\eta_{\mu\nu}\epsilon^{\mu}_{\hat{a}}\epsilon^{\nu}_{\hat{b}}\!=\!\eta_{\mu\nu}\sigma^{\mu}_{\hat{a}} \sigma^{\nu}_{\hat{b}}\!= \!\eta_{\hat{a}\hat{b}}}\), where the time-like basis vector is taken to be the 4-velocity, \({\epsilon^{\mu}_{\hat{0}}\!=\!\sigma^{\mu}_{\hat{0}}\!=\!(1,0,0,0)}\). The tetrad is parallel-transported along the worldline of the observer (source), who measures the following tetrad components (denoted with a hat) of the 4-momentum,
\begin{align}
p_{\hat{a}} = \eta_{\mu\nu} p^{\mu} \epsilon^{\nu}_{\hat{a}} = -\Omega\left(1, n_{\Imath}\right),\; \textrm{where } n_{\Imath} \equiv \epsilon^{\mu}_{\Imath}n_{\mu}\,, \label{eq:tetradcomponents_fourmomentum}
\end{align}
(and similarly for the source). The observed frequency is given by \(\Omega_{\textrm{obs}}\!\equiv\! -p_{\hat{0}}\) whilst the astrometric position is given by \({n_{\Imath}\equiv p_{\Imath}/p_{\hat{0}}}\). 

Now consider perturbing this setup with a monochromatic, plane-fronted \textsc{gw} with angular frequency \(\omega\) radiated from a distant source in the direction of the vector \(q^{i}\), extending over a region including the source and the observer (Fig.~\ref{fig:sourceObject}). The metric perturbation is given by
\begin{align}\label{eq:metricperturbation}
h_{\mu\nu}\left(t, x^{i}\right)=\Re\,\big\{H_{\mu\nu} \exp\left(\rmi k_{\rho} x^{\rho}\right)\!\big\}\,,
\end{align}
where the \(H_{\mu\nu}\) are small complex constants satisfying \({H_{\mu\nu}=H_{\nu\mu}}\), and the \textsc{gw} wavevector is \({k^{\mu}=(\omega, -\omega q^{i})}\), where \(q^{i}\) is a 3-vector pointing from the Earth to the \textsc{gw} source. Additionally, diffeomorphism invariance allows a spatial gauge choice to be made for the metric perturbation, \({H_{0\mu}=0}\). In \textsc{gr} it is possible to further restrict to the well known ``transverse-traceless'' gauge; this is not done here as our focus will be on alternative polarizations. The perturbed space-time is \({\tensor{g}{_{\mu\nu}} = \tensor{\eta}{_{\mu\nu}} + \tensor{h}{_{\mu\nu}}}\), and its inverse \({\tensor{g}{^{\mu\nu}} = \tensor{\eta}{^{\mu\nu}} - \tensor{h}{^{\mu\nu}}}\). The connection coefficients for this metric are
\begin{align}\label{eq:ChristoffelSymbols0}
\begin{split}
\Gamma^{\mu}_{\nu\rho} &= \frac{1}{2}\,\tensor{g}{^{\mu\sigma}}\!\left(\tensor{\partial}{_{\nu}} \tensor{g}{_{\sigma\rho}} + \tensor{\partial}{_{\!\rho}} \tensor{g}{_{\nu\sigma}} - \tensor{\partial}{_{\sigma}} \tensor{g}{_{\mu\nu}}\right) = \\
&= \frac{1}{2}\left(\tensor{\partial}{_{\nu}} \tensor{h}{^\mu_\rho} + \tensor{\partial}{_{\!\rho}} \tensor{h}{_{\nu}^{\mu}} - \tensor{\partial}{^{\mu}} \tensor{h}{_{\nu\rho}}\right)\!+ \bigo{h^{2}},
\end{split}
\end{align}
and the non-zero components are given explicitly by
\begin{align}
\begin{split} \label{eq:ChristoffelSymbols}
\Gamma^{0}_{ij} = \frac{1}{2}\,\partial_{0}h_{ij}\, , \quad \Gamma^{i}_{0j} = \Gamma^{i}_{j0} = \frac{1}{2}\,\partial_{0} h_{ij} \,, \\
\textrm{and} \quad \Gamma^{i}_{jk} = \frac{1}{2}\left(\partial_{k}h_{ij} + \partial_{j}h_{ik} - \partial_{i}h_{jk}\right).
\end{split}
\end{align}

The worldlines of the Earth (\({x^{\mu}_{\mathsc{E}}(t)=(t, 0)}\)) and star (\({x^{\mu}_{\mathsc{S}}(t)=(t,x_{\mathsc{S}}^{i})}\)) are geodesics in both the unperturbed (\(\eta_{\mu\nu}\)) and perturbed (\({\eta_{\mu\nu}\!+\!h_{\mu\nu}}\)) metrics; these worldlines are unaffected by the \textsc{gw}. However, the null geodesics \(x_{t_{0}}^{\mu}(\lambda)\) and the tetrads \(\epsilon^{\mu}_{\hat{a}}\) and \(\sigma^{\mu}_{\hat{a}}\) are affected, and so are the tetrad components \(p_{\hat{a}}\) defined in eq.~(\ref{eq:tetradcomponents_fourmomentum}).

Firstly, consider the tetrad along the Earth's worldline; this is perturbed \(\epsilon^{\mu}_{\hat{a}} \! \mapsto \! \epsilon^{\mu}_{\hat{a}} \! + \! \delta \epsilon^{\mu}_{\hat{a}}(t)\) according to the parallel transport equations (the 4-velocity is \(\epsilon^{\mu}_{\hat{0}}\)),
\begin{align}\label{eq:partransbasistetrad}
\frac{\dd}{\dd t}\,\delta \epsilon^{\mu}_{\hat{a}}(t)  = - \Gamma^\mu_{\nu\rho}\,\epsilon^{\nu}_{\hat{a}}\,\epsilon^{\rho}_{\hat{0}} + \bigo{h^{2}} \, ,
\end{align}
where the symbols \(\Gamma^{\mu}_{\nu\rho}\) are to be understood as being evaluated along the worldline of the Earth, \(\!x_{\mathsc{E}}^{\mu}(t)\). When integrating eq.~(\ref{eq:partransbasistetrad}) the constants of integration may be discarded as it is the time-dependent change in astrometric position which we ultimately seek to measure. Substituting for the connection coefficients from eq.~(\ref{eq:ChristoffelSymbols}) and integrating eq.~(\ref{eq:partransbasistetrad}) along the worldline of the observer shows that the only nonzero components of \(\delta \epsilon_{\hat{a}}^{\mu}(t)\) are
\begin{align}
\delta \epsilon_{\Imath}^{j}(t) = - \frac{1}{2}\,\tensor{H}{^{j}_{\Imath}} \exp\!\left(-\rmi \omega t_{0}\right) \,.
\end{align}
Note that the time-like component of the tetrad is unaffected. Similar equations hold for the deviation \(\delta\sigma_{\Imath}^{j}(t)\).

Secondly, consider the null geodesics and their tangents; these are perturbed \(x^{\mu}\!\mapsto \!x^{\mu}\!+\!\delta x^{\mu}(\lambda)\) and \(p^{\mu}\!\mapsto \!p^{\mu}\!+\!\delta p^{\mu}(\lambda)\) according to the geodesic equation,
\begin{align}\label{eq:PerturbedGeodesic}
\frac{\dd^{2}}{\dd\lambda^{2}}\,\delta x_{t_{0}}^{\mu}(\lambda)  =  \frac{\textrm{d}}{\textrm{d}\lambda}\,\delta p_{t_{0}}^{\mu}(\lambda)  =  - \Gamma^{\mu}_{\nu\rho}\,p^{\nu} p^{\rho} + \bigo{h^{2}} \, .
\end{align}
In the above equation the symbols \(\Gamma_{\nu \rho}^{\mu}\) are to be understood as being evaluated along the unperturbed trajectory \(x_{t_{0}}^{\mu}(\lambda)\) (corrections to this trajectory enter at order of \(h^{2}\)). Substituting for the connection coefficients from eq.~(\ref{eq:ChristoffelSymbols}) and for the metric perturbation from eq.~(\ref{eq:metricperturbation}) gives
\begin{subequations}
\begin{align}
& \frac{\dd}{\dd\lambda}\,\delta p_{t_{0}}^{0}(\lambda) = \frac{\rmi \omega \Omega^{2}}{2}\,H_{ij}n^{i}n^{j} f(t_{0}, \lambda) \, , \label{eq:deltap0}\\
\begin{split}
&\frac{\dd}{\dd\lambda}\,\delta p_{t_{0}}^{i}(\lambda) = \frac{\rmi \omega\Omega^{2}}{2}\,\Big(\!-2\tensor{H}{^{i}_{j}} n^{j} \Big. \\
& \quad\quad\quad \Big. + \left(q_{j} \tensor{H}{^{i}_{k}} + q_{k} \tensor{H}{^{i}_{j}} - q^{i} \tensor{H}{_{\!jk}}\!\right)n^{j}n^{k}\Big) f(t_{0}, \lambda),
\end{split} \label{eq:deltapi}
\end{align}
\end{subequations}
where \(f(t_{0}, \lambda)\!= \exp\!\left(\textrm{--}\rmi\omega\Omega\lambda\!\left(1 - q^{i}n_{i}\right)\!- \rmi\omega t_{0}\right)\). Provided $q^{i}n_{i}\neq 1$, integrating eq.~(\ref{eq:deltap0}) with respect to \(\lambda\) with the initial condition that the emission frequency in the rest frame of the star is unaffected by the \textsc{gw} (\({\Omega_{\textrm{emit}}\!\equiv \eta_{\mu\nu}\sigma^{\mu}_{\hat{0}}p^{\nu}=\Omega}\), which reduces to \(\delta p_{t_{0}}^{0}(\lambda_{\mathsc{S}})\!=\!0\) because \(\delta\sigma^{\mu}_{\hat{0}}=0\)) gives
\vspace*{-5pt}
\begin{align}\label{eq:deltapintegrated}
\delta p^{0}(\lambda)\!=\!\frac{-\Omega}{2(1 - q^{k}n_{k})}\,H_{ij}n^{i}n^{j} \big(f(t_{0}, \lambda)\!-\!f(t_{0}, \lambda_{\textrm{s}})\big) \,.
\vspace*{-5pt}
\end{align}
Integrating eq.~(\ref{eq:deltapi}) twice with respect to \(\lambda\) gives
\begin{align}\label{eq:deltaxintegrated}
\begin{split}
& \delta x^{i}(\lambda) = \frac{-\rmi}{2\omega(1 - q^{l}\,n_{l})^{2}}\Big(\!-2\tensor{H}{^{i}_{j}} n^{j} + \left(q_{j}\tensor{H}{^{i}_{k}} \right.\Big. \\ 
& \quad\quad \Big.\left. +\,q_{k}\tensor{H}{^{i}_{j}} - q^{i}H_{jk}\right)n^{j}n^{k}\Big)f(t_{0}, \lambda)+\lambda A^{i} + B^{i} \, ,
\end{split}
\end{align}
where \(A^{i}\) and \(B^{i}\) are constants of integration. 
The boundary conditions needed to determine \(A^{i}\) and \(B^{i}\) are:
\begin{enumerate}[(a)]
\item The family of geodesics \(x_{t_{0}}^{\mu}(\lambda) + \delta x_{t_{0}}^{\mu}(\lambda)\) remain null; at leading order this condition becomes
\begin{align}
h_{\mu\nu}(\Omega\lambda+t_{0},-\Omega \lambda n^{i})p^{\mu}p^{\nu}+2\eta_{\mu\nu}p^{\mu}\delta p_{t_{0}}^{\nu}(\lambda)\!=\!0.
\end{align}
\vspace*{-20pt}
\item The geodesics intersect the Earth's worldline, the freedom to reparametrize the geodesics \(\lambda\!\mapsto\!\lambda\!+\!\lambda_{0}\) may be used to ensure that the intersection occurs at \(\lambda\!=\!0\), so this condition becomes \(\delta x_{t_{0}}^{i}(0)\!=\!0\).
\item The geodesics intersect the star's worldline at some value \(\lambda_{\mathsc{S}} + \delta\lambda_{\mathsc{S}}\) (where \(\delta\lambda_{\mathsc{S}}\!=\!\mathcal{O}(h)\)), so this condition becomes \({x_{t_{0}}^{i}(\lambda_{\mathsc{S}}+\delta\lambda_{\mathsc{S}})\!+\!\delta x_{t_{0}}^{i}(\lambda_{\mathsc{S}}+\delta\lambda_{\mathsc{S}})\!=\!x^{i}_{\mathsc{S}}}\).
\end{enumerate}
Boundary condition (b) fixes the constants \(B^{i}\) leaving
\begin{align}
& \delta x^{i}(\lambda) = \frac{-\rmi}{2\omega (1 - q^{\ell}\,n_{\ell})^{2}}\Big(\!- 2\tensor{H}{^{i}_{j}}n^{j} + \left(q_{j}\tensor{H}{^{i}_{k}} + q_{k}\tensor{H}{^{i}_{j}} \right.\Big. \nonumber\\
& \quad\quad \Big.\left. -\,q^{i}H_{jk}\right) n^{j} n^{k}\Big) \big(f(t_{0}, \lambda) - f(t_{0}, 0)\big)\!+ \lambda A^{i} \,.
\end{align}
Next, we decompose \(A^{i}\!=\!A^{i}_{\perp}\!+\!A^{i}_{\parallel}\) in parallel and perpendicular directions to \(n^{i}\). It is sufficient to enforce condition (a) at \(\lambda\!=\!\lambda_{\mathsc{S}}\) (where it is most straightforward to do so because \(\delta p_{t_{0}}^{0}(\lambda_{\mathsc{S}})\!=\!0\)) as the norm of a tangent is automatically  preserved along a geodesic; this gives
\begin{align}
A^{i}_{\perp} = \frac{-\Omega n^{i}}{2(1 - q^{\ell}\,n_{\ell})}\,H_{jk} n^{j} n^{k} f(t_{0}, \lambda_{\mathsc{S}}+q_{k} \tensor{H}{^{i}_{j}}\!)\,.
\end{align}
Finally, condition (c) is applied to fix \(A^{i}_{\parallel}\) and \(\delta\lambda_{\mathsc{S}}\), although only the former is needed here,
\begin{align}
A^{i}_{\parallel} = &\frac{\rmi}{2\lambda_{\mathsc{S}}\omega\left(1 - q^{\ell}\,n_{\ell}\right)^{2}}\,\Big(2 \tensor{H}{^{i}_{j}} n^{j} - \left(q_{j} \tensor{H}{^{i}_{k}} + q_{k} \tensor{H}{^{i}_{j}} \right. \nonumber\\
& \quad\quad\quad\quad\quad\quad \left.  - q^{i} H_{ij}\right) n^{j} n^{k} - 2 n^{i} H_{jk} n^{j} n^{k} \\
& \quad\quad + n^{i} q^{m} n_{m} H_{jk}n^{j}n^{k} \Big) \big(f(t_{0}, 0)\!-\!f(t_{0}, \lambda_{\mathsc{S}})\big)\,. \nonumber
\end{align}

As discussed following eq.~(\ref{eq:tetradcomponents_fourmomentum}), the observed frequency is given by the temporal tetrad component of the photon \(\Omega_{\textrm{obs}}\equiv-p_{\hat{0}}\),
\begin{align}
\Omega_{\textrm{obs}}  = & \,\big(\eta_{\mu\nu}+h_{\mu\nu}(t_{0}, 0)\big)\big(p^{\mu}+\delta p_{t_{0}}^{\mu}(0)\big)\big(\epsilon_{\hat{0}}^{\nu}+\delta\epsilon_{\hat{0}}^{\nu}(t_{0})\big) \nonumber\\
= & \, \Omega \left(\!1-\frac{H_{ij}n^{i}n^{j}}{2(1 - q^{k} n_{k})}\,\big(f(t_{0}, 0)\!-\!f(t_{0}, \lambda_{\mathsc{S}})\big)\!\right). \label{eq:omegaobs}
\end{align}

Therefore, the redshift, defined as \(1+z\equiv \Omega/\Omega_{\textrm{obs}}\), where \(\Omega\) is the emitted frequency, is obtained from eq.~(\ref{eq:omegaobs}) as
\begin{align}\label{eq:formulaRedshift}
z = \frac{n^{i}n^{j}}{2(1 - q^{k} n_{k})}\,\big(h_{ij}(\textrm{E}) - h_{ij}(\textrm{S})\big) \,.
\end{align}
The redshift depends anti-symmetrically on the metric perturbations at the ``emission'' and ``absorption'' events at the star and the Earth, respectively (\({h_{ij}(\textrm{E}) \equiv H_{ij}f(t_{0}, 0)}\) and \({h_{ij}(\textrm{S})\equiv H_{ij}f(t_{0}, \lambda_{\textrm{s}})}\)). This symmetry arises from the endpoints of the integral along the null geodesic linking the star to the Earth. The redshift varies periodically in time due to the GW. This redshifting, applied to a distant pulsar, causes individual pulses to arrive at the Earth periodically early and late; it is this timing residual which is searched for by \textsc{pta}s.

As discussed following eq.~(\ref{eq:tetradcomponents_fourmomentum}), the star's astrometric position is given by \(n_{\Imath} \! \equiv \! p_{\Imath}/p_{\hat{0}}\) where \(p_{\hat{0}}\) is the negative of \(\Omega_{\textrm{obs}}\) in eq.~(\ref{eq:omegaobs}) and
\begin{align}
p_{\Imath}  = &\big(\eta_{\mu\nu} \! + \! h_{\mu\nu}(t_{0}, 0)\big) \big(p^{\mu} \!+ \delta p_{t_{0}}^{\mu}(0)\big)\big(\epsilon^{\nu}_{\Imath}\!+ \delta\epsilon^{\nu}_{\Imath}(t_{0})\big).
\end{align}
Combining the previous results gives the observed astrometric deflection of the star due to a plane \textsc{gw}, this is the same result as was found in \cite{Book:2010pf} (their eq.~(36)) with minor changes in notation;
\begin{widetext}
\vspace*{-10pt}
\begin{align}
\begin{split}
\delta n_{\Imath} = &\left[\Bigg(\!\left\{1 + \frac{\rmi (2 - q^{r} n_{r})}{\omega \lambda_{\mathsc{S}} \Omega (1 - q^{\ell} n_{\ell})}\,\Big(1 - \exp\left(-\rmi\omega\Omega\lambda_{\mathsc{S}}(1 - q^{s} n_{s})\right)\!\!\Big)\!\right\} n_{\Imath} \vphantom{\Bigg(\Bigg)^{2}}\right. \\[-5pt]
& \quad\quad\quad\quad \;\, - \left\{1 + \frac{\rmi}{\omega \lambda_{\mathsc{S}}\Omega (1 - q^{\ell} n_{\ell})}\,\Big(1 - \exp\left(-\rmi\omega\Omega\lambda_{\mathsc{S}} (1 - q^{s} n_{s})\right)\!\!\Big)\!\right\} q_{\Imath}\Bigg) \frac{H_{jk}n^{j}n^{k}}{2(1 - q^{\ell} n_{\ell})} \\
& \quad\quad\quad\quad\quad\quad\quad\quad \left.\vphantom{\Bigg(\Bigg)^{2}} - \left\{\frac{1}{2} + \frac{\rmi}{\omega \lambda_{\mathsc{S}}\Omega (1 - q^{\ell} n_{\ell})}\,\Big(1 - \exp\left(-\rmi\omega\Omega\lambda_{\mathsc{S}}(1 - q^{s} n_{s})\right)\!\!\Big)\!\right\} H_{\Imath j}n^{j}\right]\exp(-\rmi \omega t_{0})\,.
\end{split} \label{eq:astrometricsignalFULL}
\end{align}
\end{widetext}

As was found for the redshift, the deflection depends on the metric perturbations at the star and at the Earth; although not symmetrically. This loss of symmetry is because the deflection depends both on an integral along the null geodesic trajectory as per eq.~(\ref{eq:deltapi}), and an integral along the Earth's worldline as per eq.~(\ref{eq:partransbasistetrad}).

In the \textsc{pta} analysis of stochastic \textsc{gw} backgrounds it is common to drop the ``pulsar term'' (which is called the ``star term'' here). 
This is possible because in the limit where the \textsc{gw} wavelength is much shorter than the distance to the star, \(\omega\lambda_{\mathsc{S}} \Omega \gg 1\), the overlap reduction function (\textsc{orf}) tends to the result obtained by simply ignoring the ``pulsar term'' in eq.~(\ref{eq:formulaRedshift}) \cite{PhysRevD.79.062003}.

In the astrometric case the ``star terms'' can also sometimes be neglected, albeit for slightly different reasons. Consider eq.~(\ref{eq:astrometricsignalFULL}) in the distant source limit \(\omega\lambda_{\mathsc{S}}\Omega\gg 1\) (i.e.\ where the star is many \textsc{gw} wavelengths distant from the observer). At leading order, only the first term in each set of curly brackets remains, and the result becomes
\begin{align} \label{eq:AstroDefFinal}
\delta n_{\Imath} = \frac{1}{2} \left(\!\frac{n_{\Imath} - q_{\Imath}}{1 - q^{\ell} n_{\ell}}\,h_{\Jmath\hat{k}}(\textrm{E})\,n^{\Jmath} n^{\hat{k}} - h_{\Imath\Jmath}(\textrm{E})\,n^{\Jmath}\right).
\end{align}
Notice that in this limit all dependence on \(h_{ij}(\textrm{S})\) has been lost. 
Also notice that the distant source limit was taken for the astrometric response of a single star, not for the statistical response of a network (quantified via the \textsc{orf}) as in the \textsc{pta} case. In eq.~(\ref{eq:AstroDefFinal}) all contractions have been written using tetrad components; hereafter the hat notation denoting tetrad components will be dropped, and the astrometric deflections in eq.~(\ref{eq:AstroDefFinal}) will simply be denoted \(\delta n_{i}\).

The sensitivity of Gaia to \textsc{gw}s comes largely from the fact that it observes a large number of stars. These stars are generally well separated (by many gravitational wavelengths), therefore the small star terms will be uncorrelated between stars. At small angular separation many stars are at similar distances (star clusters); this needs consideration in any practical application. In contrast, the larger Earth term is correlated between all stars; it is this that Gaia will aim to detect. The independent star terms may be treated as an effective noise source in the experiment. Including all the individual ``star terms'' would slightly increase the sensitivity of Gaia to the lowest \textsc{gw} frequencies, however this would involve fitting for the distance to every observed star individually.

A problem occurs with the distant source limit of the astrometric deflection when the \textsc{gw} source is collinear with a star (i.e.\ when \(q^{i}n_{i}=1\)); the expression in eq.~(\ref{eq:AstroDefFinal}) will, in general, diverge.
In fact, this is usually not a problem, because in \textsc{gr} the \textsc{gw} polarizations are transverse (i.e.\ \(h_{ij}q^{j}=0\)) which ensures that eq.~(\ref{eq:AstroDefFinal}) has a smooth limit as \(n^{i}\rightarrow q^{i}\). 
However, the divergence is a problem when working with alternative polarizations with a longitudinal component (i.e.\ when \(h_{ij}q^{j}\neq 0\)). 
For the remainder of this paper we work with the distant source limit in eq.~(\ref{eq:AstroDefFinal}) whenever possible, and fall back on the full (non-divergent) expression in eq.~(\ref{eq:astrometricsignalFULL}) when the distant source limit approximation breaks down.

\begin{figure}[t]
\includestandalone[scale=1]{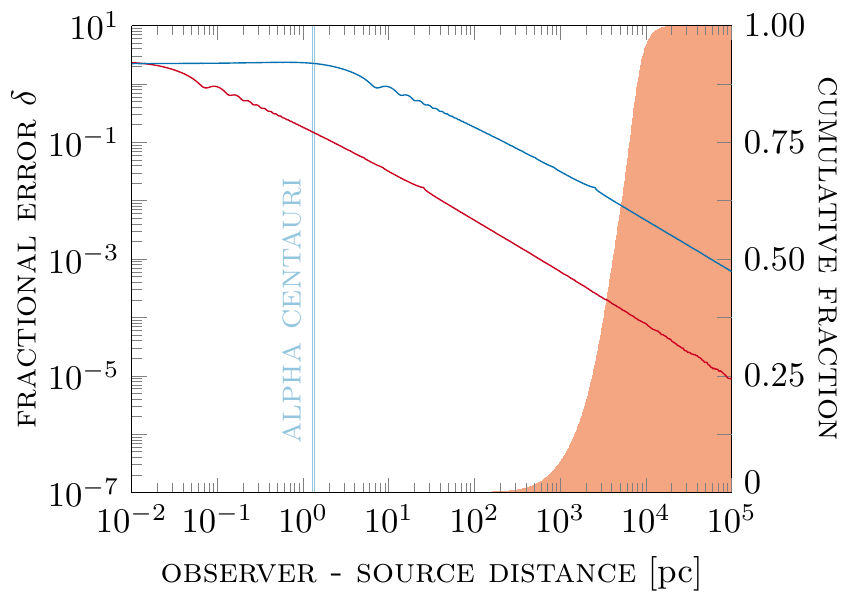}
\caption{The fractional error, \(\delta\) (see eq.~(\ref{eq:errorFormula})), in the distant source approximation as a function of the distance to the star. The red and blue curves correspond to frequencies of \(5\times 10^{-7}\,\hertz\) and \(5\times 10^{-9}\,\hertz\) respectively; these are the upper and lower frequency limits of Gaia's \textsc{gw} sensitivity \citep{PhysRevLett.119.261102}. Also shown on the secondary \(y\)-axis is the cumulative distribution of distances to objects in the Gaia DR1 catalog (simulated distances to stars taken from the \emph{Gaia Universe Model Snapshot} \citep{2012gums}). The distant-source limit is an good approximation for the majority of stars in the catalog.}
\label{fig:separationsStarCounts}
\end{figure}

For a transverse \textsc{gw} the distant source limit does not diverge and we may study the fractional error in the distant source approximation. The error is defined as
\begin{align}\label{eq:errorFormula}
\delta = \frac{|\delta n_{i}- \delta n_{i}^{\textrm{ds}}|}{|\delta n_{i}|},
\end{align}
where \(\delta n_{i}\) is given by eq.~(\ref{eq:astrometricsignalFULL}), and \(\delta n_{i}^{\textrm{ds}}\) by eq.~(\ref{eq:AstroDefFinal}). This quantity is plotted in Fig.~\ref{fig:separationsStarCounts} as a function of the distance to the source for two frequencies at the edges of the Gaia's \textsc{gw} bandwidth, \((10^{-8} - 3 \times 10^{-7})\,\hertz\) \citep{PhysRevLett.119.261102}. Also shown in Fig.~\ref{fig:separationsStarCounts} is the cumulative distribution of the distances to stars in the Gaia catalog. Fig.~\ref{fig:separationsStarCounts} shows that even for the longest \textsc{gw} wavelengths of interest to Gaia the error \(\delta < 10^{-2}\) for 90\% of stars in the Gaia catalog. This justifies the use of the distant source limit eq.~(\ref{eq:AstroDefFinal}) whenever it is not divergent.

\section{Gravitational wave polarizations}
\label{sec:gravitationalWavepolarizations}
\noindent
Whilst \textsc{gr} only allows for two \textsc{gw} polarization modes (the transverse and traceless \(+\) and \(\times\) modes), alternative theories can include up to 4 additional modes. Besides the two \textsc{gr} modes, there may be a transverse trace scalar mode, two vectorial modes with mixed transverse and longitudinal components, and a purely longitudinal scalar mode. For a detailed discussion the reader is referred to \cite{Will:2014kxa}.

\begin{figure*}
\centering\vspace{-10pt}
\subfloat{
	\includegraphics[scale=1]{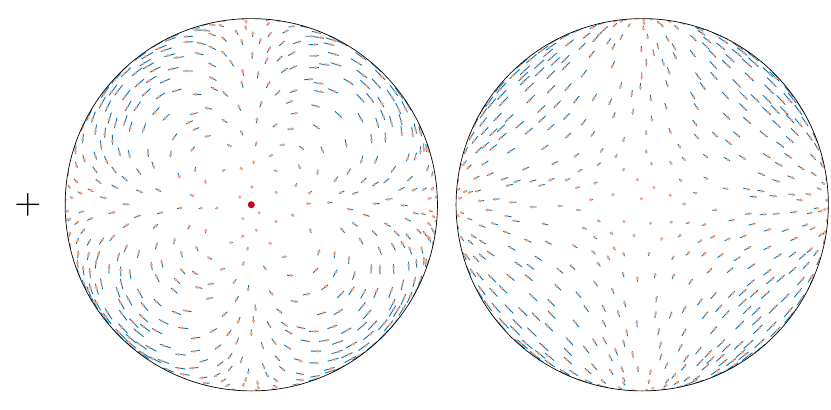}
}
\hspace{12pt}
\subfloat{
	\includegraphics[scale=1]{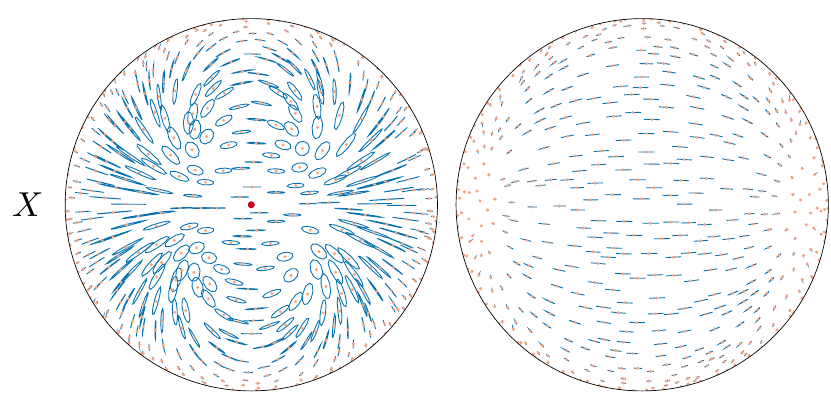}
}\\\vspace*{-10pt}
\subfloat{
	\includegraphics[scale=1]{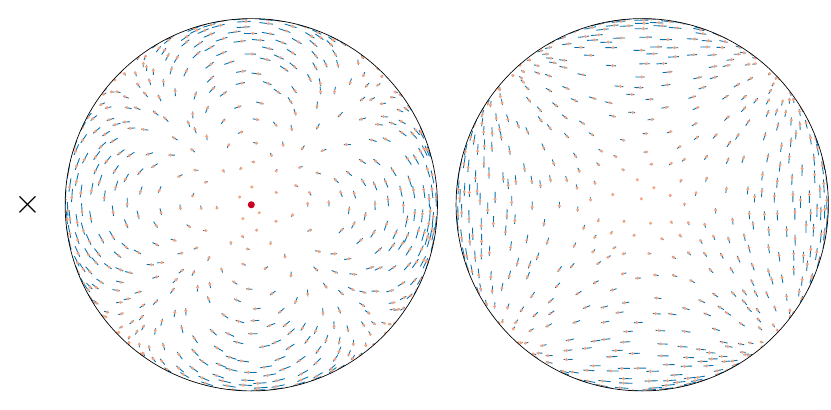}
}
\hspace{12pt}
\subfloat{
	\includegraphics[scale=1]{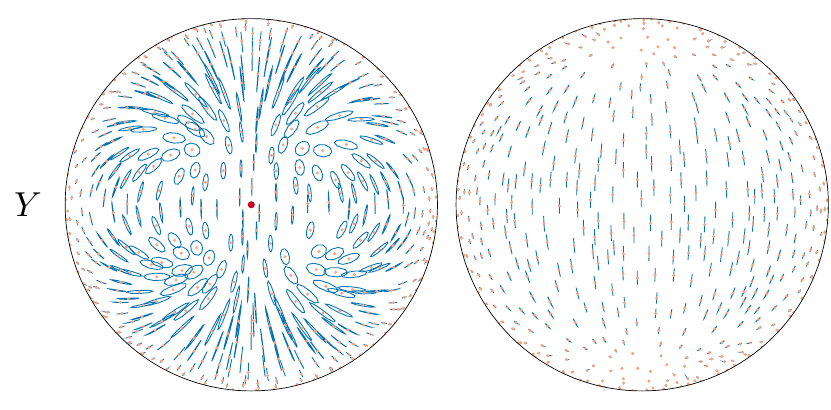}
}\\\vspace*{-10pt}
\subfloat{
	\includegraphics[scale=1]{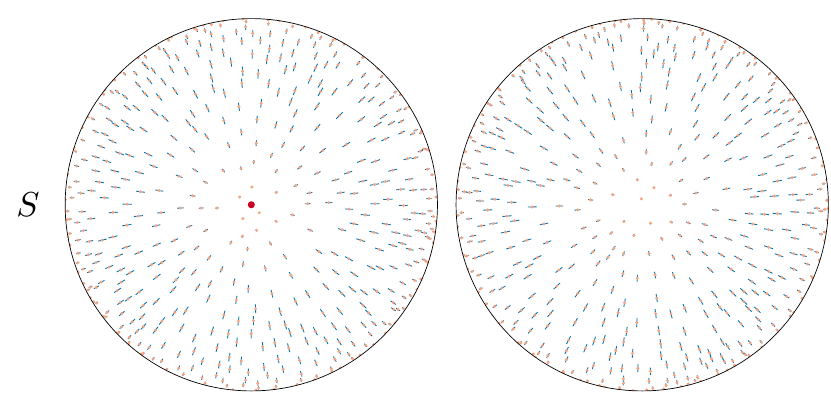}
}
\hspace{12pt}
\subfloat{
	\includegraphics[scale=1]{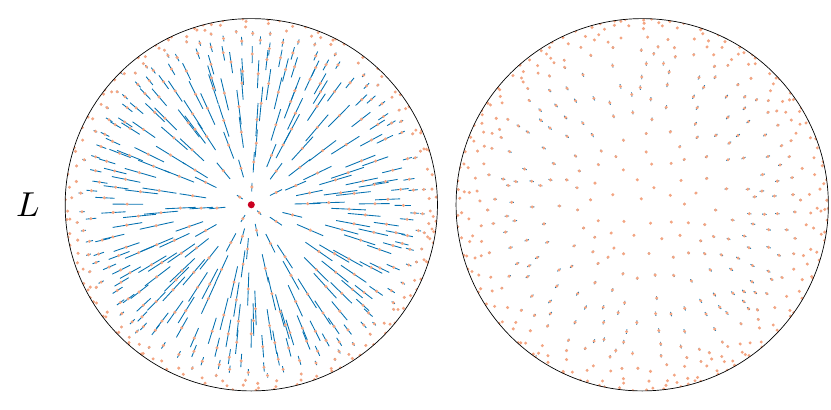}
}
\caption{Orthographic projections of the Northern (left) and Southern (right) hemispheres. Randomly placed on the sky are 1000 stars. A \textsc{gw} from a source located in the direction of the North pole (indicated by the red dots) is incident on the Earth causing the stars move periodically at the \textsc{gw} frequency. The blue lines show the traces which each star would leave as it moves on the sky. For clarity, the incident \textsc{gw} has the unphysically large characteristic strain amplitude \(A = 0.1\), and in the cases where the modes have longitudinal components (\(X, Y,\) and \(L\)), the star terms are included in the calculation of the astrometric deflection in order to avoid it becoming divergent near the North pole. In these latter three samples, all stars are placed at a distance of 10 gravitational wavelengths from the Earth. In reality, these neat elliptical patterns would appear more chaotic, since the distances to each star is generally different. The \(+\) and \(\times\) patterns are related to each other by a rotation through \(\sfrac{\pi}{4}\)) and the \(X\) and \(Y\) patterns are related by a rotation through \(\sfrac{\pi}{2}\). The astrometric patterns for the two \textsc{gr} states including the star terms are presented additionally in Appendix~\ref{app:astrometricPatterns}.}
\label{fig:starTraces}
\vspace*{-7pt}
\end{figure*}

In order to define the 6 \textsc{gw} polarization basis tensors we first introduce the orthonormal coordinate basis associated with the spherical polar coordinates \((r, \theta, \phi)\);
\begin{subequations}
\begin{align}
&\hat{e}_{i}^{r} = (\sin\theta \cos\phi,\,\sin\theta \sin\phi,\,\cos\theta), \\
&\hat{e}_{i}^{\theta} = (\cos\theta \cos\phi,\,\cos\theta \sin\phi,\,-\sin\theta), \\
&\hat{e}_{i}^{\phi} = (-\sin\phi,\,\cos\phi,\,0).
\end{align}
\end{subequations}
The symmetric spatial \textsc{gw} polarization tensor, \(H_{ij}\), for a \textsc{gw} traveling in direction \(q_{i} = -\,\hat{e}^{r}_{i}\) can then be decomposed in terms of basis tensors as
\begin{align}\label{eq:sumOfPolarizations}
H_{ij} = A_{+} \epsilon_{ij}^{+} &+ A_{\times} \epsilon_{ij}^{\times} + A_{\mathsc{S}}\,\epsilon_{ij}^{\mathsc{S}} \nonumber \\
&+ A_{\mathsc{X}} \epsilon_{ij}^{\mathsc{X}} + A_{\mathsc{Y}} \epsilon_{ij}^{\mathsc{Y}} + A_{\mathsc{L}} \epsilon_{ij}^{\mathsc{L}},
\end{align}
where the 6 \textsc{gw} basis tensors, \(\epsilon^{\mathsc{P}}_{ij}\), are defined as

\vspace{0.4cm}
\vspace{-\abovedisplayskip}
\noindent
\begin{subequations}%
\begin{minipage}[c]{0.495\linewidth}
\begin{align}
\epsilon_{ij}^{+}(q_{k}) &\!=\! \hat{e}_{i}^{\theta} \hat{e}_{j}^{\theta} \!-\! \hat{e}_{i}^{\phi} \hat{e}_{j}^{\phi}, \label{eq:tensorPlus} \\
\epsilon_{ij}^{\times}(q_{k}) &\!=\! \hat{e}_{i}^{\theta} \hat{e}_{j}^{\phi} \!+\! \hat{e}_{i}^{\phi} \hat{e}_{j}^{\theta}, \label{eq:tensorCross} \\
\epsilon_{ij}^{\mathsc{S}}(q_{k}) &\!=\! \hat{e}_{i}^{\theta} \hat{e}_{j}^{\theta} \!+\! \hat{e}_{i}^{\phi} \hat{e}_{j}^{\phi}, \label{eq:tensorScalar}
\end{align}
\end{minipage}
\begin{minipage}[c]{0.495\linewidth}
\begin{align}
\epsilon_{ij}^{\mathsc{X}} (q_{k})&\!=\! \hat{e}_{i}^{\theta} \hat{e}^{r}_{j} \!+\! \hat{e}^{r}_{i} \hat{e}_{j}^{\theta}, \label{eq:tensorX} \\
\epsilon_{ij}^{\mathsc{Y}} (q_{k})&\!=\! \hat{e}_{i}^{\phi} \hat{e}^{r}_{j} \!+\! \hat{e}^{r}_{i} \hat{e}_{j}^{\phi}, \label{eq:tensorY} \\
\epsilon_{ij}^{\mathsc{L}} (q_{k})&\!=\! \sqrt{2}\,\hat{e}^{r}_{i}\,\hat{e}^{r}_{j}. \label{eq:tensorLong}
\end{align}
\end{minipage}
\label{eq:polarizationModes}
\end{subequations}\vspace{0.2cm}\\
The factor of \(\sqrt{2}\) in the definition of \(\epsilon_{ij}^{\mathsc{L}} (q_{k})\) is for normalisation convenience and accounts for the fact that this tensor has only a single non-zero component, while all others have exactly 2. In the standard Cartesian coordinate system the generalized perturbation tensor takes the form
\begin{align}
H = \left(\begin{matrix}
A_{+} + A_{\mathsc{S}} & A_{\times} & A_{\mathsc{X}} \\
A_{\times} & -A_{+} + A_{\mathsc{S}} & A_{\mathsc{Y}} \\
A_{\mathsc{X}} & A_{\mathsc{Y}} & \sqrt{2}\,A_{\mathsc{L}}
\end{matrix}\right)\!.
\end{align}

Shown in Fig.~(\ref{fig:starTraces}) are the distant source limit astrometric deflection patterns for each of the 3 transverse \textsc{gw} polarization states, and the exact astrometric deflection patterns for the other 3 \textsc{gw} polarization modes with longitudinal components. The results in Fig.~(\ref{fig:starTraces}) were calculated using eq.~(\ref{eq:AstroDefFinal}) for the \(+, \times,\) and \(S\) polarization states, and using eq.~(\ref{eq:astrometricsignalFULL}) for the \(X, Y,\) and \(L\) states. Polarization tensors \(H_{ij}=A_{\mathsc{P}}\,\epsilon^{\mathsc{P}}_{ij}\) with \(A_{\mathsc{P}} = 0.1\) (unrealistically large for visualization purposes) were used throughout, and in the latter three cases, all stars were placed 10 gravitational wavelengths away from Earth. If the distant source limit were used for the three longitudinal modes, the plots would incorrectly show a divergence in the astrometric deflection pattern at the North pole due to the factor of \(1-q^{\ell}n_{\ell}\) in the denominator of eq.~(\ref{eq:AstroDefFinal}).
\vspace*{-5pt}

\section{Correlated Astrometric Deflections}
\label{sec:backgroundMapping}
\vspace*{-5pt}

\begin{figure}[t]
\includestandalone[scale=1]{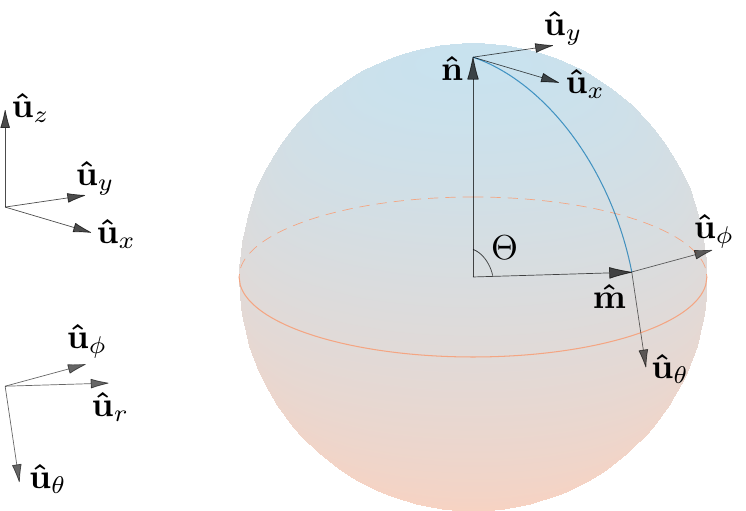}
\caption{Geometrical setup of the vectors involved in the calculation of the overlap reduction functions. A pair of stars, one of them nominally placed at the North pole, and a second one at angular separation \(\Theta\) (along the arc \(\phi = 0\)) are considered. Each of them experiences astrometric response due to a background of gravitational radiation. On the left-hand side of the Figure are shown the Cartesian triad associated with the point \(\hat{n}^{i}\) (top) and the curvilinear triad associated with the point \(\hat{m}^{i}\) (bottom). The astrometric deflection at point \(\hat{n}^{i}\) (\(\hat{m}^{i}\)) is a vector in the tangent plane to the sphere, and can be decomposed in terms of just \(\hat{\mathbf{e}}_{x}\) and \(\hat{\mathbf{e}}_{y}\) (\(\hat{\mathbf{e}}_{\theta}\) and \(\hat{\mathbf{e}}_{\phi}\)). We are interested in the correlations between these vector components.}
\label{fig:unitVectors}
\vspace*{-10pt}
\end{figure}

\noindent In Section~\ref{sec:astrometricResponse} of this article the astrometric response to a single monochromatic \textsc{gw} was derived. In this section the astrometric response to a stochastic background of \textsc{gw}s is considered. A stochastic background of \textsc{gw}s generates a stochastic pattern of astrometric deflections over the sky which is highly correlated at large angular scales. The pattern of this correlation depends on the polarization of the \textsc{gw}s which make up the background. The general framework for considering correlated vector fields on the sphere is introduced, and then in Sections~\ref{sec:tensorialModes} to \ref{sec:longitudinalModes} below, the correlation for several different combinations of the polarization states discussed in Section~\ref{sec:gravitationalWavepolarizations} are explicitly evaluated.

The discussion in this article is restricted to stochastic \textsc{gw} backgrounds which are Gaussian, stationary, isotropic and unpolarized (at this point we do not specify which polarization states comprise the background, just that they are uncorrelated with each other so the background is statistically unpolarized). The astrometric deflection is then a Gaussian random vector field on the sphere, and the statistical properties of this field are described by a correlation matrix. Adopting a convenient decomposition for this correlation matrix allows for an intuitive visualization of key features of the correlation over the sphere of the sky. This decomposition will also permit a clear comparison with the analogous calculations in the pulsar timing literature.

Consider the correlation between the astrometric deflections at two different points on the sky. The astrometric response is given by the distant source limit formula, eq.~(\ref{eq:AstroDefFinal}) (specific examples when this formula is not valid will be considered later in this Section). In this limit, the astrometric deflection depends only on the ``Earth term'' metric perturbation; this time-dependent metric perturbation can be Fourier decomposed as
\begin{align}
h_{ij}\!\left(t\right)\!=\!\mathfrak{R} \left\{\!\sum_{\mathsc{P}}\!\int_{0}^{\infty} \!\!\!\!\!\!\dd f\!\!\int_{S^{2}}\!\!\!\!\dd\Omega_{\vect{q}}\,A_{\mathsc{P}}\!\left(\vect{q}, f\right) e^{\textrm{--}2\pi\rmi f t}\epsilon_{ij}^{\mathsc{P}}\!\left(\vect{q}\right)\!\right\}\!,
\end{align}
where \(f = \omega / 2\pi\) is the linear \textsc{gw} frequency, \(A_{\mathsc{P}}\!\left(\vect{q}, f\right)\) are the (complex) Fourier coefficients, \(\dd\Omega_{\vect{q}}\) is the area element on the sphere, and \(\vect{q}\equiv q_{i}\) is the direction from which the \textsc{gw} originates. 
The label \(P\) indicates the \textsc{gw} polarization which, at this stage, may be any of the six states \(P \in \{+, \times, S, X, Y, L\}\) described in Section~\ref{sec:gravitationalWavepolarizations}.

The \textsc{gw} background is assumed to be Gaussian, zero-mean, stationary, isotropic, and unpolarized. The expectation values of the Fourier coefficients in such a background satisfy
\begin{subequations}\label{eq:amplitudeCorrs}
\begin{align}
&\left< A_{\mathsc{P}}\!\left(\vect{q}, f\right) A^{*}_{\mathsc{P}'}\!\left(\vect{q}', f'\right) \right> = C(f)\,\delta_{\mathsc{PP'}}\,\delta_{S^{2}}(\vect{q},\vect{q}')\,\delta(f-f') \label{eq:amplitudeCorrs1} \\
&\left<A_{\mathsc{P}}\!\left(\vect{q}, f\right) A_{\mathsc{P}'}\!\left(\vect{q}', f'\right)\right> = 0,
\end{align}
\end{subequations}
where the function \(C(f)\) is related to the spectral energy density in the \textsc{gw} background \citep{1993PhRvD..48.2389F}.

The astrometric deflection is linear in the metric perturbation, and the response to each Fourier mode may be calculated individually:
\begin{widetext}
\vspace*{-11pt}
\begin{align}
\delta n_{i} \left(\vect{n}, t\right) = \mathfrak{R} \left\{\sum_{\mathsc{P}}\!\int_{0}^{\infty} \!\!\!\!\!\!\dd f\,e^{-2\pi\rmi f t}\!\int_{S^{2}}\!\!\!\!\dd\Omega_{\vect{q}}\,A_{\mathsc{P}}\!\left(\vect{q}, f\right) \tensor{\Delta}{_{i}^{jk}}\!\left(\vect{n}, \vect{q}\right)\epsilon_{jk}^{\mathsc{P}}\!\left(\vect{q}\right)\!\right\}, \label{eq:CJMone}
\end{align}
\end{widetext}
\newpage where we have defined
\begin{align}\label{eq:defDelta}
\tensor{\Delta}{_{i}^{jk}} \left(\vect{n}, \vect{q}\right) = \frac{1}{2}\left(\!\frac{n_{i} - q_{i}}{1 - q^{\ell} n_{\ell}}\,n^{j} n^{k} - {\delta_{i}}^{j} n^{k}\!\right).
\end{align}

The linearity of the astrometric deflection with the metric perturbation also ensures that it will be a Gaussian random variable, and that its statistical properties depend only on the two-point correlation.  
Using eqs.~(\ref{eq:amplitudeCorrs}) and (\ref{eq:CJMone}), it is straightforward to show that the expectation of the product of astrometric responses \(\delta n_{i}\) and \(\delta m_{i}\) of two stars, situated at two different points on the sky, \(n_{i}\) and \(m_{i}\), separates into a factor depending on the measurement times and a factor depending on the locations of the stars on the sky,
\begin{equation}
\left<\delta n_{i} (\vect{n}, t)\,\delta m_{j} (\vect{m}, t')\right> = T(t,t')\,\Gamma_{ij} (\vect{n}, \vect{m})\,.
\end{equation}
Because both \(\delta n_{i} (\vect{n}, t)\) and \(\delta m_{j} (\vect{m}, t)\) have zero mean, this expectation is proportional to the correlation between the two quantities.
The temporal correlation factor is given by an integral over the \textsc{gw} frequency spectrum;
\begin{equation}
T (t,t') = \frac{1}{4}\!\int_{0}^{\infty} \!\!\!\!\!\!\dd f\; C(f)\!\left(e^{2\pi\rmi f (t-t')}+e^{-2\pi\rmi f (t-t')}\right).
\end{equation}
The spatial correlation factor is a sum of integrals of products of vectors over the sphere for each mode \(P\) (cf. the factor of \(\delta_{\mathsc{P P'}}\) in eq.~(\ref{eq:amplitudeCorrs1})):
\begin{equation}\label{eq:Gammasum}
\Gamma_{ij}(\vect{n}, \vect{m}) = \sum_{\mathsc{P}} \Gamma^{\mathsc{P}}_{ij} (\vect{n}, \vect{m})\,,
\end{equation}
where, for each polarization state, we have defined 
\begin{align} \label{eq:Gammadef}
&\Gamma_{ij}^{\mathsc{P}} (\vect{n}, \vect{m}) = \int_{S^{2}}\!\!\!\!\dd\Omega_{\vect{q}}\,\delta n_{i}^{\mathsc{P}}(\vect{n}, \vect{q})\,\delta m_{j}^{\mathsc{P}}(\vect{m}, \vect{q}), \\
&\mathrm{where}\,\,\delta n_{i}^{\mathsc{P}}(\vect{n}, \vect{q}) = \tensor{\Delta}{_{i}^{jk}} (\vect{n}, \vect{q})\,\epsilon^{\mathsc{P}}_{jk}(\vect{q})\,.
\end{align}
Only the spatial part of the correlation, \(\Gamma_{ij} (\vect{n}, \vect{m})\), depends on the polarization content of the \textsc{gw} background. Hereafter, only spatial correlations will be investigated.

Since the background is isotropic, the sky sphere can be rotated into the most convenient orientation. The first star, \(n_{i}\), is placed at the North Pole, and the second, \(m_{i}\), in the \(x\)-\(z\) plane (for anisotropic backgrounds, this transformation is still possible, though in that case the background needs to also be rotated into the new frame; see \citep{Gair:2015hra} for details). The stars have coordinates

\medskip
\vspace{-\abovedisplayskip}
\noindent
\begin{subequations}\label{eq:NEWEQLABEL_CJM}
\begin{minipage}[c]{0.42\linewidth}
\begin{align}
\vect{n} = (0, 0, 1),
\end{align}%
\end{minipage}
\begin{minipage}[c]{0.57\linewidth}
\begin{align}
\vect{m} = (\sin \Theta, 0, \cos \Theta).
\end{align}
\end{minipage}
\end{subequations}
\smallskip

The astrometric deflection vectors lie in the tangent plane of the sphere; it is now necessary to introduce a pair of basis vectors at both points on the sphere. The choice of basis is of course arbitrary; however, the following choice will prove to be convenient. For any pair of points \(n_{i}\) and \(m_{i}\) there is a unique (shortest) geodesic, \(\gamma\), linking \(n_{i}\) to \(m_{i}\); at both \(n_{i}\) and \(m_{i}\) the unit tangent vector to \(\gamma\) and the unit vector pointing to the left of \(\gamma\) form an orthonormal basis. For the values of \(n_{i}\) and \(m_{i}\) in eq.~(\ref{eq:NEWEQLABEL_CJM}) the coordinate expressions for these basis vectors are

\medskip
\vspace{-2pt}
\vspace{-\abovedisplayskip}
\noindent
\begin{minipage}[c]{0.4\linewidth}
\begin{subequations}\label{eq:tanBasisVectorsxy}
\begin{align}
& \hat{u}^{x} = (1, 0, 0), \label{eq:defux}\\
& \hat{u}^{y} = (0, 1, 0), \label{eq:defuy}
\end{align}%
\end{subequations}
\end{minipage}
\begin{minipage}[c]{0.59\linewidth}
\begin{subequations}\label{eq:tanBasisVectorstp}
\begin{align}
& \hat{u}^{\theta} = (\cos \Theta, 0, -\sin \Theta), \label{eq:defutheta} \\
& \hat{u}^{\phi} = (0, 1, 0), \label{eq:defuphi}
\end{align}%
\end{subequations}
\end{minipage}
\vspace{3pt}

\noindent whilst the general definitions for two arbitrary points on the sphere are

\medskip
\vspace{-2pt}
\vspace{-\abovedisplayskip}
\noindent
\begin{minipage}[c]{0.49\linewidth}
\begin{subequations}\label{eq:tanBasisVectorsxycoordinvar}
\begin{align}
& \hat{u}^{x} \!=\! {\frac{(\hat{n}\!\times\! \hat{m})\!\times\! \hat{n}}{\sqrt{1\,\textrm{--}\,(\hat{n}\cdot \hat{m})^{2}}}}, \\
& \hat{u}^{y} \!=\! {\frac{\hat{n}\!\times\! \hat{m}}{\sqrt{1\,\textrm{--}\,(\hat{n}\cdot \hat{m})^{2}}}} ,
\end{align}%
\end{subequations}
\end{minipage}
\begin{minipage}[c]{0.49\linewidth}
\begin{subequations}\label{eq:tanBasisVectorstpcoordinvar}
\begin{align}
& \hat{u}^{\theta} \!=\! {\frac{(\hat{n}\!\times\! \hat{m})\!\times\! \hat{n}}{\sqrt{1\,\textrm{--}\,(\hat{n}\cdot \hat{m})^{2}}}}, \\
& \hat{u}^{\phi} \!=\! {\frac{\hat{n}\!\times\! \hat{m}}{\sqrt{1\,\textrm{--}\,(\hat{n}\cdot \hat{m})^{2}}}} .
\end{align}%
\end{subequations}
\end{minipage}
\vspace{4pt}

\noindent The geometric setup is illustrated in Fig.~\ref{fig:unitVectors}.

The astrometric deflections may now be decomposed into this basis;
\begin{subequations}\label{eq:responseResolved}
\begin{align}
\delta n_{i}^{\mathsc{P}}(\vect{q}) &= \delta n_{x}^{\mathsc{P}}(\vect{q})\,\hat{u}_{i}^{x} + \delta n_{y}^{\mathsc{P}}(\vect{q})\,\hat{u}_{i}^{y}\,, \\
\delta m_{i}^{\mathsc{P}}(\vect{q}) &= \delta m_{\theta}^{\mathsc{P}}(\vect{q})\,\hat{u}_{i}^{\theta} + \delta m_{\phi}^{\mathsc{P}}(\vect{q})\,\hat{u}_{i}^{\phi}\,,
\end{align}
\end{subequations}
where the scalar coefficients are given by

\medskip
\vspace{-2pt}
\vspace{-\abovedisplayskip}
\noindent
\begin{minipage}[c]{0.48\linewidth}
\begin{subequations}\label{eq:tanBasisVectorsN}
\begin{align}
& \delta n_{x}^{\mathsc{P}}(\vect{q}) \!=\! \delta n_{i}^{\mathsc{P}}\!(\vect{q})\,\hat{u}_{x}^{i}, \label{eq:dnx} \\
& \delta n_{y}^{\mathsc{P}}(\vect{q}) \!=\! \delta n_{i}^{\mathsc{P}}\!(\vect{q})\,\hat{u}_{y}^{i}, \label{eq:dny}
\end{align}%
\end{subequations}
\end{minipage}
\begin{minipage}[c]{0.51\linewidth}
\begin{subequations}\label{eq:tanBasisVectorsM}
\begin{align}
& \delta m_{\theta}^{\mathsc{P}}(\vect{q}) \!=\! \delta m_{i}^{\mathsc{P}}\!(\vect{q})\,\hat{u}_{\theta}^{i}, \label{eq:dmtheta} \\
& \delta m_{\phi}^{\mathsc{P}}(\vect{q}) \!=\! \delta m_{i}^{\mathsc{P}}\!(\vect{q})\,\hat{u}_{\phi}^{i}. \label{eq:dmphi}
\end{align}%
\end{subequations}
\end{minipage}
\vspace{3pt}

\noindent Substituting eqs.~(\ref{eq:responseResolved}) into eq.~(\ref{eq:Gammadef}) and expanding gives an expression for the spatial correlation function as a sum of scalar integrals over the sphere,
\begin{subequations}
\begin{align}
\Gamma_{ij}^{\mathsc{P}} (\Theta) &= \hat{u}_{i}^{x} \hat{u}_{j}^{\theta}\;\int_{S^{2}} \!\!\!\dd\Omega_{\vect{q}}\;\delta n_{x}^{\mathsc{P}}(\vect{q})\,\delta m_{\theta}^{\mathsc{P}}(\vect{q}) \label{eq:defGammaxtheta}\\
&\quad + \hat{u}_{i}^{x} \hat{u}_{j}^{\phi}\;\int_{S^{2}} \!\!\!\dd\Omega_{\vect{q}}\;\delta n_{x}^{\mathsc{P}}(\vect{q})\,\delta m_{\phi}^{\mathsc{P}}(\vect{q}) \label{eq:defGammaxphi} \\
&\quad\quad + \hat{u}_{i}^{y} \hat{u}_{j}^{\theta}\;\int_{S^{2}} \!\!\!\dd\Omega_{\vect{q}}\;\delta n_{y}^{\mathsc{P}}(\vect{q})\,\delta m_{\theta}^{\mathsc{P}}(\vect{q}) \label{eq:defGammaytheta} \\
&\quad\quad\quad + \hat{u}_{i}^{y} \hat{u}_{j}^{\phi}\;\int_{S^{2}} \!\!\!\dd\Omega_{\vect{q}}\;\delta n_{y}^{\mathsc{P}}(\vect{q})\,\delta m_{\phi}^{\mathsc{P}}(\vect{q})\;. \label{eq:defGammayphi}
\end{align}
\end{subequations}

In fact, it can be shown that for any correlated vector field on the sphere which is statistically invariant under both rotations and parity transformations the \(y\)-\(\theta\) and \(x\)-\(\phi\) terms (i.e.\ (\ref{eq:defGammaxphi}) and (\ref{eq:defGammaytheta})) vanish.
These terms can also explicitly be shown to vanish for each \textsc{gw} polarization considered individually in Sections~\ref{sec:tensorialModes} to \ref{sec:longitudinalModes} below.
Therefore, remarkably, the full spatial correlation matrix is always fully specified by just two real-valued functions, and may be written as
\begin{align}\label{eq:fullCorrBasis}
    \Gamma_{ij}^{\mathsc{P}} (\vect{n}, \vect{m}) = \Gamma_{x\theta}^{\mathsc{P}} (\Theta)\,\hat{u}_{i}^{x} \hat{u}_{j}^{\theta} + \Gamma_{y\phi}^{\mathsc{P}} (\Theta)\,\hat{u}_{i}^{y} \hat{u}_{j}^{\phi}\,,
\end{align}
where \(\Theta= \arccos\!\left(\vect{n} \cdot \vect{m}\right)\), the unit vectors are defined in terms of \(n_{i}\) and \(m_{i}\) in eqs.~(\ref{eq:tanBasisVectorsxycoordinvar}) and (\ref{eq:tanBasisVectorstpcoordinvar}) and the functions \(\Gamma_{x\theta}^{\mathsc{P}}(\Theta)\) and \(\Gamma_{y\phi}^{\mathsc{P}}(\Theta)\) are defined as the integrals in terms (\ref{eq:defGammaxtheta}) and (\ref{eq:defGammayphi}) respectively. 
This result is the equivalent of eq.~(71) in \citep{Book:2010pf}.

One advantage of the decomposition of the spatial correlation matrix in eq.~(\ref{eq:fullCorrBasis}) is that the functions \(\Gamma_{x\theta}^{\mathsc{P}}(\Theta)\) and \(\Gamma_{y\phi}^{\mathsc{P}}(\Theta)\) have the clear interpretation as the \emph{scalar} correlations of the ``parallel'' and ``perpendicular'' components of the astrometric deflection. 
Here ``parallel'' means tangent to the geodesic linking the two points, and ``perpendicular'' means pointing to the left of this curve; see Fig.~\ref{fig:unitVectors}.

It is well known that any vector field \(v_{i}\) on the sphere admits a unique Helmholtz decomposition into gradient and curl parts; \({v_{i} = \nabla_{\!i}\,\phi+(\nabla \wedge)_{i}\,\psi}\), where \(\nabla\) is the surface gradient on the sphere, \({(\nabla\wedge) = \vect{r} \wedge\!\nabla}\) is the surface curl, and \(\phi\) and \(\psi\) are scalar fields on the sphere. Another advantage of the spatial correlation matrix in eq.~(\ref{eq:fullCorrBasis}) is that the functions \(\Gamma_{x\theta}^{\mathsc{P}}(\Theta)\) and \(\Gamma_{y\phi}^{\mathsc{P}}(\Theta)\) govern the statistical properties of the gradient and curl parts respectively. By comparing components, it can be seen that the \(\Gamma_{x\theta}^{\mathsc{P}}(\Theta)\) and \(\Gamma_{y\phi}^{\mathsc{P}}(\Theta)\) terms in eq.~(\ref{eq:fullCorrBasis}) are exactly the general divergence and curl kernels \(\Psi_{\textrm{div}}\) and \(\Psi_{\textrm{curl}}\) defined in \citep{FuselierWright2009}. A random vector field described by spatial correlation matrix with \(\Gamma_{y\phi}=0\) will be a pure divergence, and a vector field described by correlation matrix with \(\Gamma_{x\theta}=0\) will be a pure curl.

\begin{figure}[t]
\includestandalone[scale=1]{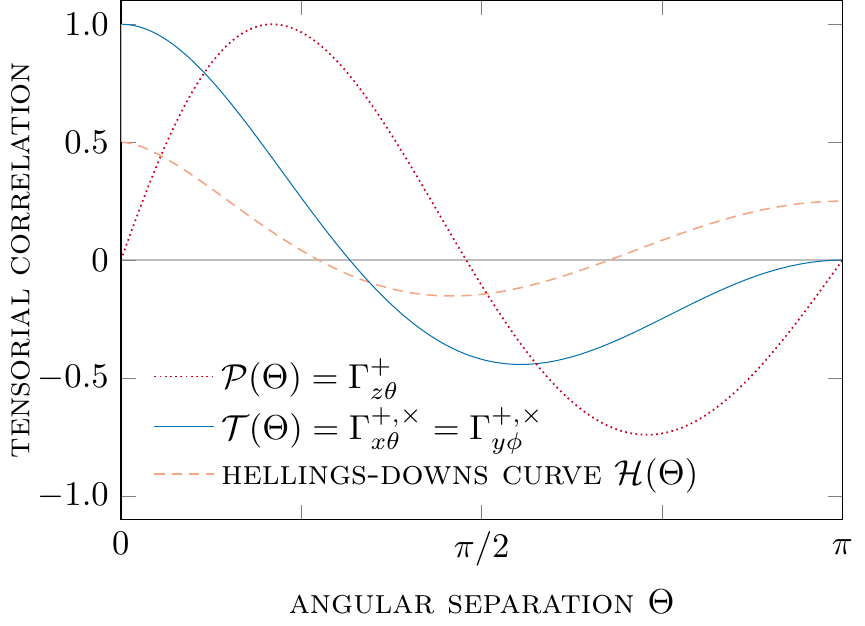}
\caption{The astrometric and redshift correlations as a function of angular separation on the sky in a background of tensorial, transverse-traceless \textsc{gw}s (i.e.\ \(+\) and \(\times\)). The well known Hellings--Downs curve, \(\mathcal{H} (\Theta)\), determines the redshift correlations and is shown here with the usual normalization \(\lim_{\Theta\rightarrow 0}H(\Theta)=\sfrac{1}{2}\) due to the presence of the pulsar term in eq.~(\ref{eq:formulaRedshift}). The astrometric correlations are similarly determined by a single function, \(\mathcal{T}(\Theta)\), which is shown with the normalization \(\mathcal{T}(0)=1\) as there is no star term in eq.~(\ref{eq:AstroDefFinal}). The function \(\mathcal{T}(\Theta)\) is the astrometric analog of the Hellings-Downs curve. The function \(\mathcal{P} (\Theta)\) is the redshift-astrometry analog of the Hellings-Downs curve and is introduced and discussed in Section~\ref{sec:redshiftAstroCorr}.}
\label{fig:plusCrossPlot}
\end{figure}

\subsection{Tensorial Transverse--Traceless Polarizations}
\label{sec:tensorialModes}

\noindent This section considers the astrometric correlations arising in a background of just the two \textsc{gr} polarization modes; i.e.\ \(P\in\{+,\times\}\). This calculation was considered previously in \citep{Book:2010pf}, here this result is reproduced within the framework outlined in the previous section.

In Appendix~\ref{app:PlusCrossInt} it is shown how to evaluate the integrals \(\Gamma_{x\theta}^{+}\), \(\Gamma_{y\phi}^{+}\), \(\Gamma_{x\theta}^{\times}\), and \(\Gamma_{y\phi}^{\times}\) defined in eq.~(\ref{eq:Gammadef}). Following eq.~(\ref{eq:Gammasum}), the spatial correlation matrix in a background with multiple polarizations is the sum of the individual spatial correlations, so \(\Gamma_{x\theta}^{+, \times}(\Theta)=\Gamma_{x\theta}^{+}(\Theta)+\Gamma_{x\theta}^{\times}(\Theta)\), and similarly for \(\Gamma_{y\phi}^{+,\times}(\Theta)\). Remarkably, these two functions turn out to be equal in this particular case,
\begin{widetext}
\begin{align}\label{eq:tensorialCorrelation}
\mathcal{T} (\Theta) = \Gamma_{x \theta}^{+, \times} (\Theta) = \Gamma_{y \phi}^{+, \times} (\Theta) = \frac{2\pi}{3} - \frac{14\pi}{3}\,\sin^{2}\!\left(\Theta/2\right) - 8\pi\,\frac{\sin^{4}\!\left(\Theta / 2\right)}{1 - \sin^{2}\!\left(\Theta / 2\right)}\,\ln\!\left(\sin\!\left(\Theta / 2\right)\right).
\end{align}
\end{widetext}
Throughout this section all correlation functions are written in terms of \(\sin\!\left(\Theta / 2\right)\). Therefore, the correlated astrometric deflection field generated by a Gaussian, stationary, isotropic, unpolarized \textsc{gw} background in \textsc{gr} is fully specified by a single real-valued function of the angular separation on the sphere, \(\mathcal{T}(\Theta)\).

\begin{figure}[t]
\includestandalone[scale=1]{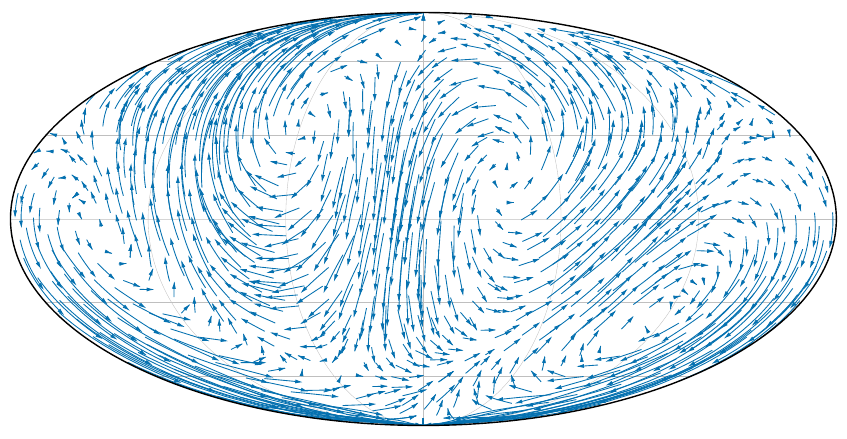}
\caption{A random realization of the astrometric deflection field for a background of tensorial \(+\) and \(\times\) waves. 
The position of each star is recorded twice, separated by a time \(\Delta t\). These two position are shown here (in Mollweide projection) at the foot and head respectively of each arrow. The length of each arrow is proportional to the total power in the \textsc{gw} background at frequencies \(f<1/\Delta t\). The length of the arrows has been greatly scaled up here for clarity.}
\label{fig:tensorialRealPlot}
\end{figure}

This function ought to be compared to the corresponding result for Pulsar Timing. The spatial correlation between the redshift at two different points on the sky is given by the well-known Hellings-and-Downs curve \citep{1983ApJ...265L..39H},
\begin{align}\label{eq:hellingsDowns}
\begin{split}
\mathcal{H}(\Theta) &= \frac{1}{2}\,(1 + \beta) - \frac{1}{4}\,\sin^{2}\!\left(\Theta / 2\right) \\[2pt]
& \quad\quad\quad\quad + 3 \sin^{2}\!\left(\Theta / 2\right) \ln\!\left(\sin\!\left(\Theta / 2\right)\right),
\end{split}
\end{align}
where \({\beta = 1}\) for co-located pulsars and is zero otherwise. The standard \textsc{pta} normalization is \(\lim_{\Theta\rightarrow 0} \mathcal{H}(\Theta) = \sfrac{1}{2}\); the \(\beta\) in eq.~(\ref{eq:hellingsDowns}) comes from the expectation of the pulsar terms in eq.~(\ref{eq:formulaRedshift}) which is non-zero only for the autocorrelation. In the astrometric case there are no star terms (see eq.~(\ref{eq:AstroDefFinal})), so in the case of a total time correlation (\(T(t, t') = 1\)), the normalization \(\mathcal{T}(0) = 1\) can be chosen. Regardless, it is the ``shape'' of these curves that is of most interest here.

The well known Hellings-Downs curve governs the spatial correlation of the redshift on the sky. Similarly, the function \(\mathcal{T}(\Theta)\) governs the spatial correlation of the astrometric deflection on the sky.  The function \(\mathcal{T} (\Theta)\) can therefore be considered as the astrometric analog of the Hellings-Downs curve. Both \(\mathcal{T}(\Theta)\) and \(\mathcal{H}(\Theta)\) are shown in Fig.~\ref{fig:plusCrossPlot}. 

In order to gain a better understanding of what this vector field correlation over the sky means it is useful to draw a realization of this random process and to plot the result. The results are shown in Fig.~\ref{fig:tensorialRealPlot}, and an overview of the procedure used to produce the data in this plot can be found in Appendix~\ref{app:RandomReal}.

\subsection{Scalar ``Breathing" Polarization}

\begin{figure}[b]
\includestandalone[scale=1]{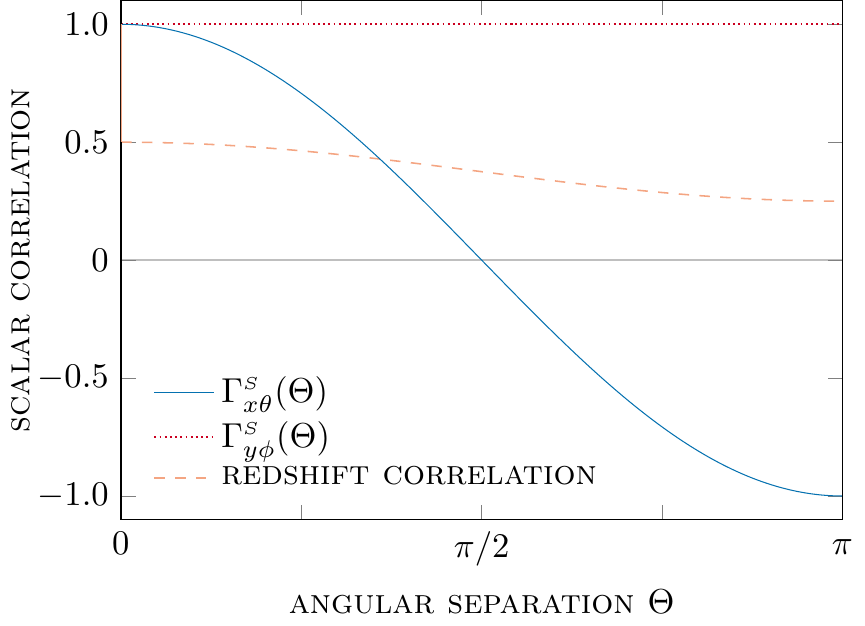}
\caption{The astrometric and redshift correlations as a function of angular separation on the sky in a background of scalar, ``breathing'' \textsc{gw}s (i.e.\ \(S\)). The functions which determine the astrometric correlations (\(\Gamma_{x\theta}^{\mathsc{S}}(\Theta)\) and \(\Gamma_{y\phi}^{\mathsc{S}}(\Theta)\), see eq.~(\ref{eq:scalarCorrelation})) are normalized so that their maximum is unity. The \textsc{pta} result for the correlated redshift in eq.~(\ref{eq:ptascalarcorr}) is plotted, normalized to \(\sfrac{1}{2}\) at \(\Theta=0\).}
\label{fig:scalarPlot}
\end{figure}

\noindent The astrometric correlations arising in a background of transverse scalar \textsc{gw}s (i.e.\ \(P\in\{S\}\)) is considered here. Appendix~\ref{app:SInt} shows how to evaluate the integrals \(\Gamma_{x\theta}^{\mathsc{S}}\) and \(\Gamma_{y\phi}^{\mathsc{S}}\) defined in eq.~(\ref{eq:Gammadef}); here only the results of these integrals are presented.
\begin{subequations}\label{eq:scalarCorrelation}
\begin{align}
\Gamma_{x \theta}^{\mathsc{S}} (\Theta) &= \frac{\pi}{3}\,\cos \Theta \equiv \frac{\pi}{3} - \frac{2\pi}{3}\,\sin^{2}\!\left(\Theta / 2\right), \label{eq:scalarCorr1}\\
\Gamma_{y \phi}^{\mathsc{S}} (\Theta) &= \frac{\pi}{3}. \label{eq:scalarCorr2}
\end{align}
\end{subequations}
Again, this should be compared to the \textsc{pta} result for the redshift correlation in a stochastic background of ``breathing" \textsc{gw}s. This was derived by \citep{0004-637X-685-2-1304} as
\begin{align}\label{eq:ptascalarcorr}
\textrm{corr} (\Theta) = \frac{1}{2}\,(1 + \beta) + \frac{1}{2} - \frac{1}{4}\,\sin^{2}\!\left(\Theta / 2\right),
\end{align}
the variable \(\beta\) is defined just after eq.~(\ref{eq:hellingsDowns}). All three of these functions are plotted in Fig.~\ref{fig:scalarPlot}.

\begin{figure}[b]
\includestandalone[scale=1]{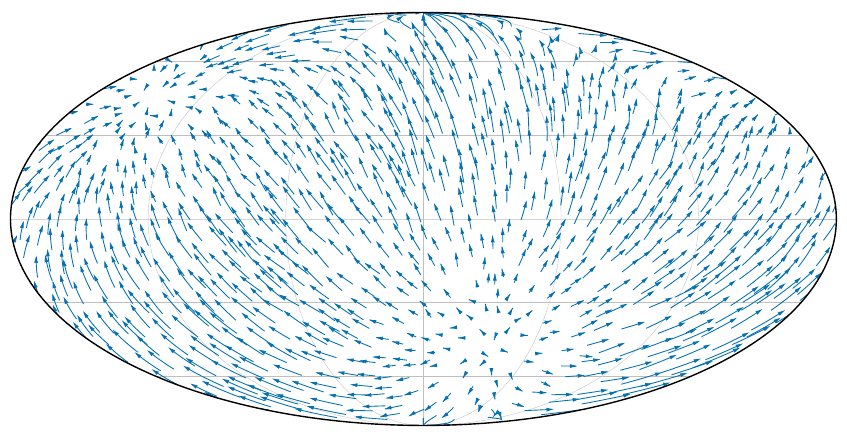}
\caption{A random realization of the astrometric deflection field for a background of scalar ``breathing'' \(S\) waves. This was produced in the same way as Fig.~\ref{fig:tensorialRealPlot}. It is clear from the plot that the astrometric deflection vector field has a random dipole-like structure on the sphere; the origin of this behavior is the fact that \(\Gamma_{y\phi}^{S} (\Theta) \equiv \textrm{constant}\), and the astrometric deflection at any two points on the sky are perfectly correlated.}
\label{fig:scalarRealPlot}
\end{figure}

The most surprising aspect of astrometric correlation is the result for \(\Gamma_{y\phi}^{\mathsc{S}} (\Theta)\); the ``perpendicular'' components of the astrometric deflection at any two points on the sky are always \emph{perfectly} correlated. This is an extremely strong constraint which any allowed realizations of the vector field must obey. The interpretation of this becomes clearer when a random realization of the correlation is drawn; this is shown in Fig.~\ref{fig:scalarRealPlot}.

The random realizations of the astrometric deflections plotted in Figs.~\ref{fig:tensorialRealPlot} and \ref{fig:scalarRealPlot} are qualitatively different. The transverse traceless polarisations of \textsc{gr} produce a distinctive curl-like pattern at large angular scales, whereas the transverse-trace (or scalar) mode generates a dipole-like structure on the sky. The polarization content of the stochastic \textsc{gw} background determines the spatial correlations among the astrometric deflections. If \emph{Gaia}, or some other future astrometry mission, is able to measure the stochastic pattern of astrometric deflections due to a background of \textsc{gw}s, the measured correlations will encode details of the polarization content of the background and thereby enable a test of \textsc{gr}.

\subsection{Vectorial Polarizations}
\label{sec:vectorialModes}

\begin{figure}[t]
\includestandalone[scale=1]{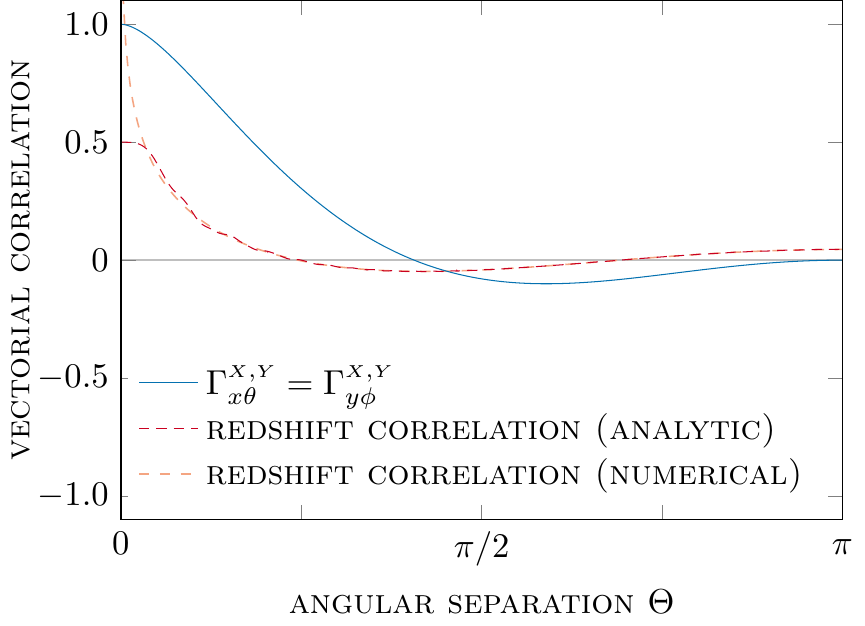}
\caption{The astrometric and redshift correlations as a function of angular separation on the sky in a background of vectorial \textsc{gw}s (i.e.\ \(P \in \{X,Y\}\)). The function which determines the astrometric correlation (\(\Gamma_{x\theta}^{\mathsc{X}, \mathsc{Y}} (\Theta) = \Gamma_{y\phi}^{\mathsc{X}, \mathsc{Y}} (\Theta)\), see eq.~(\ref{eq:vectorialAstrCorr})) is normalized so that its maximum is unity. The numerical redshift result (for pulsars at distances \((100 \lambda_{\mathrm{GW}}, 200\lambda_{\mathrm{GW}})\)) for the correlated redshift is plotted (normalized to \(\sfrac{1}{2}\) at \(\Theta=0\)) along with the divergent result from eq.~(\ref{eq:vectorialPtaCorrDiv}).}
\label{fig:vectorPlot}
\end{figure}

\noindent After analysing the tensorial modes and the scalar ``breathing" mode, it is interesting to consider the astrometric correlations arising in a background of just the two vectorial polarization modes; i.e.\ \(P \in \{X,Y\}\). These calculations have an additional complication over those in the preceding sections as the vectorial polarizations have a longitudinal component which introduces a singularity into the ``Earth term''-only redshift and astrometric responses (see eqs.~(\ref{eq:formulaRedshift}) and (\ref{eq:AstroDefFinal})).
In the case of the redshift correlation, as was found in \cite{0004-637X-685-2-1304}, this means the correlation curve diverges at \(\Theta=0\);
\begin{align}\label{eq:vectorialPtaCorrDiv}
\Gamma_{z}^{\mathsc{X}, \mathsc{Y}} (\Theta) = - \frac{28 \pi}{3} + \frac{32 \pi}{3}\,\sin^{2}\left(\Theta/2\right) - 8 \pi \ln\!\left(\sin\!\left(\Theta/2\right)\right).
\end{align}
This result is plotted in Fig.~\ref{fig:vectorPlot}. The divergence at the origin is a result of the use of the ``Earth term''-only redshift response. If the ``star term'' is included the result becomes finite, and the correlation depends on the distance to the star. When including the star term the integration must be performed numerically; the results of this numerical integration are also shown in Fig.~\ref{fig:vectorPlot} for two pulsars at distances of 100 and 200 gravitational wavelengths respectively.

In contrast, the divergence in the astrometric response is of a logarithmic nature (i.e., of the type \(\int \dd x\,f(x) / x\)), and is regularized by the integral over the sky. This means that the resulting correlation curve is non-divergent, even though the two individual astrometric responses do diverge. Appendix~\ref{app:VectorialInt} discusses how to evaluate the integrals \(\Gamma_{x\theta}^{\mathsc{X}}\), \(\Gamma_{y\phi}^{\mathsc{X}}\), \(\Gamma_{x\theta}^{\mathsc{Y}}\), and \(\Gamma_{y\phi}^{\mathsc{Y}}\) analytically. Following eq.~(\ref{eq:Gammasum}), the spatial correlation matrix in a background with multiple polarizations is the sum of the individual spatial correlations: \(\Gamma_{x\theta}^{\mathsc{X}, \mathsc{Y}}(\Theta)=\Gamma_{x\theta}^{\mathsc{X}}(\Theta)+\Gamma_{x\theta}^{\mathsc{Y}}(\Theta)\), and similarly for \(\Gamma_{y\phi}^{\mathsc{X}, \mathsc{Y}}(\Theta)\). Again, these two functions turn out to be equal:
\begin{align}\label{eq:vectorialAstrCorr}
\begin{split}
&\Gamma_{x \theta}^{\mathsc{X}, \mathsc{Y}} (\Theta) = \Gamma_{y \phi}^{\mathsc{X}, \mathsc{Y}} (\Theta) = \\
&\quad \frac{4\pi}{3} + \frac{8\pi}{3}\,\sin^{2}\!\left(\Theta / 2\right) + 8\pi\,\frac{\sin^{2}\!\left(\Theta / 2\right)}{1 - \sin^{2}\!\left(\Theta / 2\right)}\,\ln\!\left(\sin\!\left(\Theta / 2\right)\right).
\end{split}
\end{align}
This vectorial astrometric correlation function is also plotted in Fig.~\ref{fig:vectorPlot}.

\subsection{Scalar ``Longitudinal'' Polarization}
\label{sec:longitudinalModes}

\begin{figure}[t]
\includestandalone[scale=1]{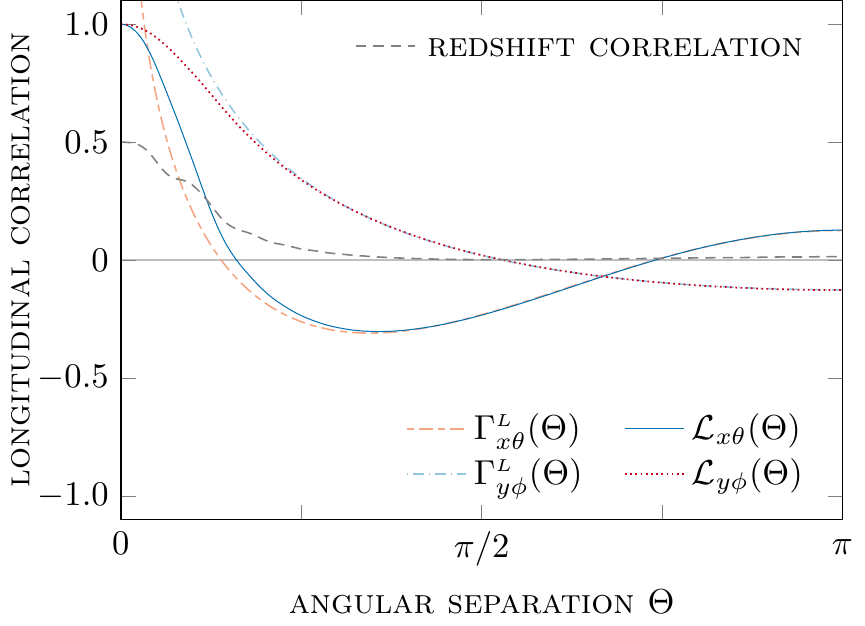}
\caption{The astrometric and redshift correlations as a function of angular separation on the sky in a background of scalar longitudinal \textsc{gw}s (i.e.\ \(L\)). The numerical curves which determine the astrometric correlation (\(\mathcal{L}_{x\theta} (\Theta)\) and \(\mathcal{L}_{y\phi} (\Theta)\), see Appendix~\ref{app:numericalintegral}) are normalized so that their maximum is unity; the two analytical divergent curves \(\Gamma_{x\theta}^{\mathsc{L}} (\Theta)\) and \(\Gamma_{y\phi}^{\mathsc{L}} (\Theta)\) (see eq.~(\ref{eq:longCorrDiv})) are also plotted. The numerical redshift result for the correlated redshift is plotted, normalized to \(\sfrac{1}{2}\) at \(\Theta=0\).}
\label{fig:longitudinalPlot}
\end{figure}

\noindent The scalar longitudinal mode, on the other hand, is more interesting, as in this case the ``Earth term''-only astrometric correlation curves do diverge at \(\Theta = 0\). These functions are given by (see Appendix~\ref{app:longIntegralDer} for details of the evaluation of the relevant integrals, and Fig.~\ref{fig:longitudinalPlot} for plots of the two functions)
\begin{subequations}\label{eq:longCorrDiv}
\begin{align}
\Gamma_{x \theta}^{\mathsc{L}} (\Theta) &= -\,\frac{10\pi}{3} + \frac{8\pi}{3}\,\sin^{2}\!\left(\Theta / 2\right) - 2\pi\,\frac{\ln\!\left(\sin\!\left(\Theta / 2\right)\right)}{1 - \sin^{2}\!\left(\Theta/2\right)}, \label{eq:longCorr1}\\
\Gamma_{y \phi}^{\mathsc{L}} (\Theta) &= -\,\frac{4\pi}{3} - 2\pi\,\frac{\ln\!\left(\sin\!\left(\Theta / 2\right)\right)}{1 - \sin^{2}\!\left(\Theta/2\right)}. \label{eq:longCorr2}
\end{align}
\end{subequations}
Similarly to vectorial redshift correlation, the divergence is a result of using the ``Earth term''-only astrometric response in eq.~(\ref{eq:AstroDefFinal}). If instead, the full astrometric response in eq.~(\ref{eq:astrometricsignalFULL}) is used the correlation is finite, although the integrals need to be evaluated numerically in this case (see Appendix~\ref{app:numericalintegral} for details). When using the full astrometric response, the correlation curves depend on the distance to the stars; two such curves are shown in Fig.~\ref{fig:longitudinalPlot} for stars at distances of 100 and 200 gravitational wavelengths.

\begin{figure}[b]
\includestandalone[scale=1]{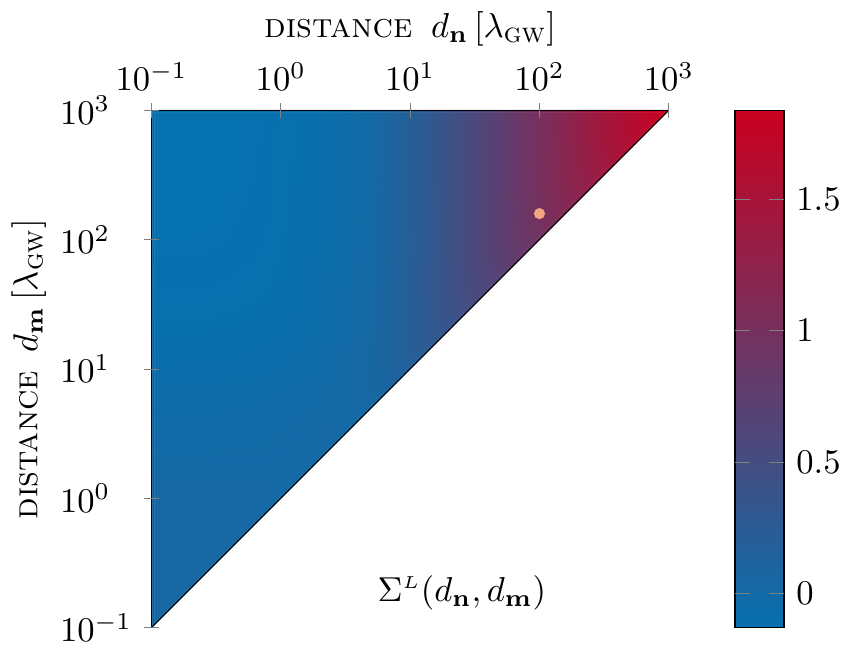}
\caption{Surface plot of the astrometric longitudinal correlation at \(\Theta = 0\), given by \(\Sigma^{\mathsc{L}} (d_{\vect{n}}, d_{\vect{m}})\) for \(d_{\vect{m}} \ge d_{\vect{n}}\) (see eq.~(\ref{eq:longSurf}) in Appendix~\ref{app:corrSurfaces} for the precise expression. The two distances are expressed in terms of gravitational wavelengths. A dot marks the point \((d_{\vect{n}}, d_{\vect{m}}) = (100, 200)\), which is used for computing the numerical integral in Fig.~\ref{fig:longitudinalPlot}.}
\label{fig:longSurfPlot}
\end{figure}

One conclusion which can be drawn from the curves in Fig.~\ref{fig:longitudinalPlot} is that that for a longitudinally-polarized \textsc{gw} background there are only strong astrometric correlations between stars at small angular separations. This is in marked contrast to the \textsc{gr} case of a tensorial \{\(+\),\(\times\)\} background, where correlations of order unity persist at all angular scales.

In order to better understand the behaviour at \(\Theta=0\), it is useful to consider the full (non-divergent) correlation including the distances to the stars at the point \(\Theta = 0\). 
At this point the full integral can be evaluated analytically, giving a correlation which is a function of just the distances \(d_{n}\) and \(d_{m}\) to the two stars. This function quantifies the cross-correlation of the deflections of two stars which appear at the same point on the sky, but are separated in distance by many gravitational wavelengths. A plot of this function is given in Fig.~\ref{fig:longSurfPlot} and an explicit expression for it is given in Appendix~\ref{app:corrSurfaces}.

\section{Redshift-Astrometry correlations}
\label{sec:redshiftAstroCorr}
\noindent
As discussed above, a stochastic background of \textsc{gw}s can, in principle, be detected by both pulsar timing or astrometric measurements. However, improved sensitivities can be obtained by combining these two techniques. In addition to the straightforward increase in signal to noise that comes with the increased amount of data, there is an additional benefit that comes from now being able to search for \textsc{gw}s in the cross-correlation between the two data sets. It is this redshift-astrometric correlation which is considered here.

As described above, an isotropic, unpolarized background of \textsc{gw}s causes a correlated redshift pattern on the sky. The correlation at two points on the sky is described by a single real-valued function of the angular separation (\(\mathcal{H}(\Theta)\), see eq.~(\ref{eq:hellingsDowns}) and Fig.~\ref{fig:plusCrossPlot}) known as the Hellings-Downs curve~\cite{1983ApJ...265L..39H}. This result is a very robust prediction within \textsc{gr}; it depends only on the existence of the two polarizations predicted by the theory and the homogeneity and isotropy of the universe on scales comparable to the distance to the \textsc{gw} sources. It does not depend on the dynamics of the individual sources generating the background. Detection of a Hellings-Downs correlated redshift pattern via pulsar timing would be clear evidence for a stochastic \textsc{gw} background.

Similarly, a background of \textsc{gw}s causes a correlated astrometric deflection pattern on the sky. This correlation pattern is also fully specified by a single real-valued function of the angular separation (\(\mathcal{T}(\Theta)\), see eq.~(\ref{eq:tensorialCorrelation}) and Fig.~\ref{fig:plusCrossPlot}) which was first derived in \cite{Book:2010pf}, although not in the current form. This is the astrometric analog of the Hellings-Downs curve; it is a similarly robust prediction of \textsc{gr} and detection of this pattern via astrometric measurements would provide similarly clear evidence for a stochastic \textsc{gw} background.

\begin{figure}[t]
\includestandalone[scale=1]{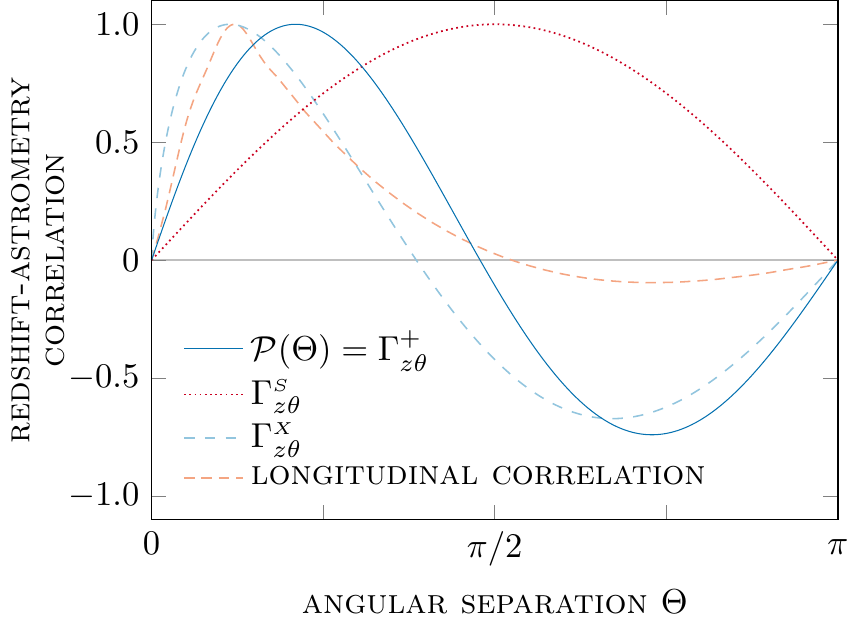}
\caption{The redshift-astrometric correlations as a function of angular separation on the sky, given by eqs.~(\ref{eq:ptaAstrGR}), (\ref{eq:redshiftAstrS}), and (\ref{eq:redshiftAstrX}) for different polarizations. The numerical result for the scalar longitudinal correlation is plotted too (see Appendix~\ref{app:numericalintegral}). All functions are normalized so that their maximum is unity.}
\label{fig:ptaAstrCorr}
\end{figure}

Additionally, there is a correlation between the redshift and astrometric deflection. The redshift of a pulsar in direction \(\vect{n}\) is correlated with the astrometric deflection, \({\delta m_{i} = \delta m_{\theta}\,\hat{u}^{\theta}_{i}+\delta m_{\phi}\,\hat{u}^{\phi}_{i}}\), of a star in direction \(\vect{m}\) via
\begin{align}
\big< z(\vect{n}) \, \delta m_{i}(\vect{m})\big> \propto &\,\int\textrm{d}\Omega_{\mathbf{q}}\,z(\vect{n})\,\delta m_{i}(\vect{m})\,,\nonumber \\ 
\propto & \,\int\textrm{d}\Omega_{\mathbf{q}}\,z(\vect{n}) \big(\delta m_{\theta}\,\hat{u}^{\theta}_{i}+\delta m_{\phi}\,\hat{u}^{\phi}_{i}\big) \,,\nonumber \\
\propto & \;\Gamma_{z\theta} (\Theta)\,\hat{u}^{\theta}_{i} + \Gamma_{z\phi} (\Theta)\,\hat{u}^{\theta}_{i}\,, \label{eq:important}
\end{align}
where \(\Theta=\arccos(\vect{n} \cdot \vect{m})\). General considerations again show that \(\Gamma_{z\phi}\) always vanishes. For a \textsc{gw} background of \(+\) and \(\times\) waves the remaining function evaluates to
\begin{align}\label{eq:ptaAstrGR}
\begin{split}
&\mathcal{P} (\Theta) = \Gamma_{z\theta}^{+} (\Theta) = \frac{8\pi}{3}\,\sin\!\left(\Theta/2\right)\cos\!\left(\Theta/2\right) \\
&\quad\quad\quad\quad\quad\quad\quad\quad + 8\pi\,\frac{\sin^{3}\!\left(\Theta/2\right)}{\cos\!\left(\Theta/2\right)}\,\ln\!\left(\sin\!\left(\Theta/2\right)\right).
\end{split}
\end{align}
The basis vectors \(\hat{u}^{\theta}_{i}\) and \(\hat{u}^{\phi}_{i}\) are defined in Section~\ref{sec:backgroundMapping} and illustrated in Fig.~\ref{fig:unitVectors}. From eq.~(\ref{eq:important}) it follows that the redshift of a pulsar is correlated with the ``parallel component'' of the astrometric deflection of a star, and is uncorrelated with the ``perpendicular component'' (see Fig.~\ref{fig:unitVectors} for an illustration of the geometric setup).

Furthermore, the correlation between a pulsar and the response from a \(\times\)-polarized \textsc{gw} vanishes,
\begin{align}
\Gamma_{z\theta}^{\times} (\Theta) = \Gamma_{z\phi}^{\times} (\Theta) = 0;
\end{align}
therefore the entire correlation between the redshift of a pulsars and the astrometric deflection of a star for \(P \in \{+, \times\}\) is governed by \(\mathcal{P} (\Theta)\) in eq.~(\ref{eq:ptaAstrGR}). This is shown, together with the other curves relevant to \(\textsc{gr}\), in Fig.~\ref{fig:plusCrossPlot}, and with the other redshift-astrometry curves in Fig.~\ref{fig:ptaAstrCorr}.

\textsc{pta}-astrometry correlations like the one in eq.~(\ref{eq:ptaAstrGR}) can be found for the other polarization states, too. In particular, the curves can be derived explicitly in the case of scalar ``breathing" and vectorial modes:
\begin{align}
\begin{split}
&\Gamma_{z\theta}^{\mathsc{S}} (\Theta) = -\frac{20\pi}{3}\,\sin\!\left(\Theta/2\right)\cos\!\left(\Theta/2\right) \\
&\quad\quad\quad\quad\quad\quad\quad\quad - 8\pi\,\frac{\sin\!\left(\Theta/2\right)}{\cos\!\left(\Theta/2\right)}\,\ln\!\left(\sin\!\left(\Theta/2\right)\right),
\end{split} \label{eq:redshiftAstrS}\\
&\Gamma_{z\theta}^{\mathsc{X}} (\Theta) = \frac{2\pi}{3}\,\sin\!\left(\Theta/2\right)\cos\!\left(\Theta/2\right), \label{eq:redshiftAstrX}
\end{align}
while the correlations \(\Gamma_{z\phi}^{\mathsc{S}}\), \(\Gamma_{z\phi}^{\mathsc{X}}\), \(\Gamma_{z\theta}^{\mathsc{Y}}\) and \(\Gamma_{z\phi}^{\mathsc{Y}}\) all vanish. The scalar longitudinal mode curve cannot be derived explicitly using this method, and has therefore been computed numerically (again, only the \(x-\theta\) correlation is non-zero). Plots of all 4 curves can be found in Fig.~\ref{fig:ptaAstrCorr}.

Again, the redshift-astrometric correlation pattern is fully described by a single real-valued function of the angular separation, \(\mathcal{P} (\Theta)\). This is the redshift-astrometric analog of the Hellings-Downs curve, and is a similarly robust prediction within \textsc{gr}. The three functions \(\mathcal{H}(\Theta)\), \(\mathcal{T}(\Theta)\) and \(\mathcal{P} (\Theta)\) provide a starting point for searching for a stochastic \textsc{gw} background using a combination of pulsar timing and astrometric data; \(\mathcal{H}(\Theta)\) describes the spatial correlations of the redshift over the sky, \(\mathcal{T}(\Theta)\) describes the astrometric correlations, and \(\mathcal{P}(\Theta)\) describes the cross-correlation between the redshifts and astrometric deflections.

\section{Conclusions}
\noindent
The change in the apparent position of a star, or the astrometric response, caused by an arbitrarily-polarized gravitational wave has been considered. This astrometric response depends, in general, on both the metric perturbation at the photon emission and absorption events at the star and Earth respectively. If the star is many gravitational wavelengths distant from the Earth, and if the \textsc{gw} is transverse (i.e., \(+\), \(\times\), and \(S\)), then the star terms can be neglected and the astrometric deflection depends only on the metric perturbation near the Earth.

This paper considered the correlated astrometric signal due to a stationary and isotropic \textsc{gw} background. If the background is unpolarized, then the correlation separates into a sum of an integral over the sphere of the sky for each polarization state. These integrals we evaluated for each of the 6 possible \textsc{gw} polarization states; although in one case the integration could be performed only numerically. A new decomposition for the astrometric correlation matrix has been introduced. This decomposition has a clear geometric interpretation and enables a nice comparison to be drawn with existing results for the redshift correlations from the pulsar timing literature. In the special case of a tensorially-polarized \textsc{gw} background, the astrometric correlation is governed by a single function which can be considered as the astrometric analogue of the Hellings--Downs curve.

The cross-correlation between the redshift and astrometric signals has also been derived for all six \textsc{gw} polarizations. This may form the basis for a joint pulsar timing and astrometric search for the low-frequency stochastic gravitational wave background.

Future work on this topic should include establishing how the absolute sensitivity of \emph{Gaia} to stochastic \textsc{gw} backgrounds depends on the polarization content. It would also be possible to consider how the astrometric correlations are changed in the presence of an anisotropic \textsc{gw} background; this would require a more detailed analysis similar that already performed for redshift correlations \citep{2014PhRvD..90h2001G, Gair:2015hra}. Another interesting possibility is to consider how the astrometric and redshift correlations change if the graviton has a mass; the correlations derived in this paper depend on the assumption that \textsc{gw}s propagate at the speed of light.

\section*{Acknowledgements}
\noindent
DM is funded by the \textsc{stfc}. CM acknowledges financial support provided under the European Union's H2020 ERC Consolidator Grant ``Matter and strong-field gravity: New frontiers in Einstein’s theory'' grant agreement no. MaGRaTh–646597. This project has received funding from the European Union's Horizon 2020 research and innovation programme under the Marie Sklodowska-Curie grant agreement No 690904 and CM would like to acknowledge networking support by the COST Action CA16104. The authors are grateful to Steve Taylor and Michalis Agathos for various discussions which contributed to the contents of this article.

\bibliographystyle{apsrev4-1.bst}
\bibliography{main}

\begin{thebibliography}{38}%
\makeatletter
\providecommand \@ifxundefined [1]{%
 \@ifx{#1\undefined}
}%
\providecommand \@ifnum [1]{%
 \ifnum #1\expandafter \@firstoftwo
 \else \expandafter \@secondoftwo
 \fi
}%
\providecommand \@ifx [1]{%
 \ifx #1\expandafter \@firstoftwo
 \else \expandafter \@secondoftwo
 \fi
}%
\providecommand \natexlab [1]{#1}%
\providecommand \enquote  [1]{``#1''}%
\providecommand \bibnamefont  [1]{#1}%
\providecommand \bibfnamefont [1]{#1}%
\providecommand \citenamefont [1]{#1}%
\providecommand \href@noop [0]{\@secondoftwo}%
\providecommand \href [0]{\begingroup \@sanitize@url \@href}%
\providecommand \@href[1]{\@@startlink{#1}\@@href}%
\providecommand \@@href[1]{\endgroup#1\@@endlink}%
\providecommand \@sanitize@url [0]{\catcode `\\12\catcode `\$12\catcode
  `\&12\catcode `\#12\catcode `\^12\catcode `\_12\catcode `\%12\relax}%
\providecommand \@@startlink[1]{}%
\providecommand \@@endlink[0]{}%
\providecommand \url  [0]{\begingroup\@sanitize@url \@url }%
\providecommand \@url [1]{\endgroup\@href {#1}{\urlprefix }}%
\providecommand \urlprefix  [0]{URL }%
\providecommand \Eprint [0]{\href }%
\providecommand \doibase [0]{http://dx.doi.org/}%
\providecommand \selectlanguage [0]{\@gobble}%
\providecommand \bibinfo  [0]{\@secondoftwo}%
\providecommand \bibfield  [0]{\@secondoftwo}%
\providecommand \translation [1]{[#1]}%
\providecommand \BibitemOpen [0]{}%
\providecommand \bibitemStop [0]{}%
\providecommand \bibitemNoStop [0]{.\EOS\space}%
\providecommand \EOS [0]{\spacefactor3000\relax}%
\providecommand \BibitemShut  [1]{\csname bibitem#1\endcsname}%
\let\auto@bib@innerbib\@empty
\bibitem [{\citenamefont {{Abbott}}\ \emph
  {et~al.}(2016{\natexlab{a}})\citenamefont {{Abbott}}, \citenamefont
  {{Abbott}}, \citenamefont {{Abbott}}, \citenamefont {{Abernathy}} \emph
  {et~al.}}]{2016PhRvL.116f1102A}%
  \BibitemOpen
  \bibfield  {author} {\bibinfo {author} {\bibfnamefont {B.~P.}\ \bibnamefont
  {{Abbott}}}, \bibinfo {author} {\bibfnamefont {R.}~\bibnamefont {{Abbott}}},
  \bibinfo {author} {\bibfnamefont {T.~D.}\ \bibnamefont {{Abbott}}}, \bibinfo
  {author} {\bibfnamefont {M.~R.}\ \bibnamefont {{Abernathy}}},  \emph
  {et~al.},\ }\href {\doibase 10.1103/PhysRevLett.116.061102} {\bibfield
  {journal} {\bibinfo  {journal} {Phys.\ Rev.\ Lett.}\ }\textbf {\bibinfo
  {volume} {116}},\ \bibinfo {eid} {061102} (\bibinfo {year}
  {2016}{\natexlab{a}})},\ \Eprint {http://arxiv.org/abs/1602.03837}
  {arXiv:1602.03837 [gr-qc]} \BibitemShut {NoStop}%
\bibitem [{\citenamefont {{Abbott}}\ \emph
  {et~al.}(2016{\natexlab{b}})\citenamefont {{Abbott}}, \citenamefont
  {{Abbott}}, \citenamefont {{Abbott}}, \citenamefont {{Abernathy}},
  \citenamefont {{Acernese}} \emph {et~al.}}]{2016PhRvL.116x1103A}%
  \BibitemOpen
  \bibfield  {author} {\bibinfo {author} {\bibfnamefont {B.~P.}\ \bibnamefont
  {{Abbott}}}, \bibinfo {author} {\bibfnamefont {R.}~\bibnamefont {{Abbott}}},
  \bibinfo {author} {\bibfnamefont {T.~D.}\ \bibnamefont {{Abbott}}}, \bibinfo
  {author} {\bibfnamefont {M.~R.}\ \bibnamefont {{Abernathy}}}, \bibinfo
  {author} {\bibfnamefont {F.}~\bibnamefont {{Acernese}}},  \emph {et~al.},\
  }\href {\doibase 10.1103/PhysRevLett.116.241103} {\bibfield  {journal}
  {\bibinfo  {journal} {Phys.\ Rev.\ Lett.}\ }\textbf {\bibinfo {volume}
  {116}},\ \bibinfo {eid} {241103} (\bibinfo {year} {2016}{\natexlab{b}})},\
  \Eprint {http://arxiv.org/abs/1606.04855} {arXiv:1606.04855 [gr-qc]}
  \BibitemShut {NoStop}%
\bibitem [{\citenamefont {{Abbott}}\ \emph
  {et~al.}(2017{\natexlab{a}})\citenamefont {{Abbott}}, \citenamefont
  {{Abbott}}, \citenamefont {{Abbott}}, \citenamefont {{Acernese}},
  \citenamefont {{Ackley}} \emph {et~al.}}]{2017PhRvL.118v1101A}%
  \BibitemOpen
  \bibfield  {author} {\bibinfo {author} {\bibfnamefont {B.~P.}\ \bibnamefont
  {{Abbott}}}, \bibinfo {author} {\bibfnamefont {R.}~\bibnamefont {{Abbott}}},
  \bibinfo {author} {\bibfnamefont {T.~D.}\ \bibnamefont {{Abbott}}}, \bibinfo
  {author} {\bibfnamefont {F.}~\bibnamefont {{Acernese}}}, \bibinfo {author}
  {\bibfnamefont {K.}~\bibnamefont {{Ackley}}},  \emph {et~al.},\ }\href
  {\doibase 10.1103/PhysRevLett.118.221101} {\bibfield  {journal} {\bibinfo
  {journal} {Phys.\ Rev.\ Lett.}\ }\textbf {\bibinfo {volume} {118}},\ \bibinfo
  {eid} {221101} (\bibinfo {year} {2017}{\natexlab{a}})},\ \Eprint
  {http://arxiv.org/abs/1706.01812} {arXiv:1706.01812 [gr-qc]} \BibitemShut
  {NoStop}%
\bibitem [{\citenamefont {{Abbott}}\ \emph
  {et~al.}(2017{\natexlab{b}})\citenamefont {{Abbott}}, \citenamefont
  {{Abbott}}, \citenamefont {{Abbott}}, \citenamefont {{Acernese}},
  \citenamefont {{Ackley}} \emph {et~al.}}]{2017PhRvL.119n1101A}%
  \BibitemOpen
  \bibfield  {author} {\bibinfo {author} {\bibfnamefont {B.~P.}\ \bibnamefont
  {{Abbott}}}, \bibinfo {author} {\bibfnamefont {R.}~\bibnamefont {{Abbott}}},
  \bibinfo {author} {\bibfnamefont {T.~D.}\ \bibnamefont {{Abbott}}}, \bibinfo
  {author} {\bibfnamefont {F.}~\bibnamefont {{Acernese}}}, \bibinfo {author}
  {\bibfnamefont {K.}~\bibnamefont {{Ackley}}},  \emph {et~al.},\ }\href
  {\doibase 10.1103/PhysRevLett.119.141101} {\bibfield  {journal} {\bibinfo
  {journal} {Phys.\ Rev.\ Lett.}\ }\textbf {\bibinfo {volume} {119}},\ \bibinfo
  {eid} {141101} (\bibinfo {year} {2017}{\natexlab{b}})},\ \Eprint
  {http://arxiv.org/abs/1709.09660} {arXiv:1709.09660 [gr-qc]} \BibitemShut
  {NoStop}%
\bibitem [{\citenamefont {{The LIGO Scientific Collaboration}}\ \emph
  {et~al.}(2017)\citenamefont {{The LIGO Scientific Collaboration}},
  \citenamefont {{the Virgo Collaboration}}, \citenamefont {{Abbott}},
  \citenamefont {{Abbott}}, \citenamefont {{Abbott}}, \citenamefont
  {{Acernese}}, \citenamefont {{Ackley}} \emph {et~al.}}]{2017arXiv171105578T}%
  \BibitemOpen
  \bibfield  {author} {\bibinfo {author} {\bibnamefont {{The LIGO Scientific
  Collaboration}}}, \bibinfo {author} {\bibnamefont {{the Virgo
  Collaboration}}}, \bibinfo {author} {\bibfnamefont {B.~P.}\ \bibnamefont
  {{Abbott}}}, \bibinfo {author} {\bibfnamefont {R.}~\bibnamefont {{Abbott}}},
  \bibinfo {author} {\bibfnamefont {T.~D.}\ \bibnamefont {{Abbott}}}, \bibinfo
  {author} {\bibfnamefont {F.}~\bibnamefont {{Acernese}}}, \bibinfo {author}
  {\bibfnamefont {K.}~\bibnamefont {{Ackley}}},  \emph {et~al.},\ }\href@noop
  {} {\bibfield  {journal} {\bibinfo  {journal} {ArXiv e-prints}\ } (\bibinfo
  {year} {2017})},\ \Eprint {http://arxiv.org/abs/1711.05578} {arXiv:1711.05578
  [astro-ph.HE]} \BibitemShut {NoStop}%
\bibitem [{\citenamefont {{Abbott}}\ \emph
  {et~al.}(2017{\natexlab{c}})\citenamefont {{Abbott}}, \citenamefont
  {{Abbott}}, \citenamefont {{Abbott}}, \citenamefont {{Acernese}},
  \citenamefont {{Ackley}} \emph {et~al.}}]{2017PhRvL.119p1101A}%
  \BibitemOpen
  \bibfield  {author} {\bibinfo {author} {\bibfnamefont {B.~P.}\ \bibnamefont
  {{Abbott}}}, \bibinfo {author} {\bibfnamefont {R.}~\bibnamefont {{Abbott}}},
  \bibinfo {author} {\bibfnamefont {T.~D.}\ \bibnamefont {{Abbott}}}, \bibinfo
  {author} {\bibfnamefont {F.}~\bibnamefont {{Acernese}}}, \bibinfo {author}
  {\bibfnamefont {K.}~\bibnamefont {{Ackley}}},  \emph {et~al.},\ }\href
  {\doibase 10.1103/PhysRevLett.119.161101} {\bibfield  {journal} {\bibinfo
  {journal} {Phys.\ Rev.\ Lett.}\ }\textbf {\bibinfo {volume} {119}},\ \bibinfo
  {eid} {161101} (\bibinfo {year} {2017}{\natexlab{c}})},\ \Eprint
  {http://arxiv.org/abs/1710.05832} {arXiv:1710.05832 [gr-qc]} \BibitemShut
  {NoStop}%
\bibitem [{\citenamefont {Blair}\ \emph {et~al.}(2015)\citenamefont {Blair}
  \emph {et~al.}}]{Blair:2016idv}%
  \BibitemOpen
  \bibfield  {author} {\bibinfo {author} {\bibfnamefont {D.}~\bibnamefont
  {Blair}} \emph {et~al.},\ }\href {\doibase 10.1007/s11433-015-5748-6}
  {\bibfield  {journal} {\bibinfo  {journal} {Sci.\ China\ Phys.\ Mech.\
  Astron.}\ }\textbf {\bibinfo {volume} {58}},\ \bibinfo {pages} {120402}
  (\bibinfo {year} {2015})},\ \Eprint {http://arxiv.org/abs/1602.02872}
  {arXiv:1602.02872 [physics.ins-det]} \BibitemShut {NoStop}%
\bibitem [{\citenamefont {Hughes}\ \emph {et~al.}(2001)\citenamefont {Hughes},
  \citenamefont {Marka}, \citenamefont {Bender},\ and\ \citenamefont
  {Hogan}}]{Hughes:2001ch}%
  \BibitemOpen
  \bibfield  {author} {\bibinfo {author} {\bibfnamefont {S.~A.}\ \bibnamefont
  {Hughes}}, \bibinfo {author} {\bibfnamefont {S.}~\bibnamefont {Marka}},
  \bibinfo {author} {\bibfnamefont {P.~L.}\ \bibnamefont {Bender}}, \ and\
  \bibinfo {author} {\bibfnamefont {C.~J.}\ \bibnamefont {Hogan}},\ }\bibfield
  {booktitle} {\emph {\bibinfo {booktitle} {{Proceedings, APS / DPF / DPB
  Summer Study on the Future of Particle Physics (Snowmass 2001), Snowmass,
  Colorado, 30 Jun - 21 Jul 2001}}},\ }\href@noop {} {\bibfield  {journal}
  {\bibinfo  {journal} {eConf}\ }\textbf {\bibinfo {volume} {C010630}},\
  \bibinfo {pages} {P402} (\bibinfo {year} {2001})},\ \Eprint
  {http://arxiv.org/abs/astro-ph/0110349} {arXiv:astro-ph/0110349 [astro-ph]}
  \BibitemShut {NoStop}%
\bibitem [{\citenamefont {Abbott}\ \emph {et~al.}(2016)\citenamefont {Abbott}
  \emph {et~al.}}]{TheLIGOScientific:2016htt}%
  \BibitemOpen
  \bibfield  {author} {\bibinfo {author} {\bibfnamefont {B.~P.}\ \bibnamefont
  {Abbott}} \emph {et~al.} (\bibinfo {collaboration} {Virgo, LIGO
  Scientific}),\ }\href {\doibase 10.3847/2041-8205/818/2/L22} {\bibfield
  {journal} {\bibinfo  {journal} {Astrophys.\ J.}\ }\textbf {\bibinfo {volume}
  {818}},\ \bibinfo {pages} {L22} (\bibinfo {year} {2016})},\ \Eprint
  {http://arxiv.org/abs/1602.03846} {arXiv:1602.03846 [astro-ph.HE]}
  \BibitemShut {NoStop}%
\bibitem [{\citenamefont {Abbott}\ \emph {et~al.}(2017)\citenamefont {Abbott}
  \emph {et~al.}}]{Abbott:2017xzu}%
  \BibitemOpen
  \bibfield  {author} {\bibinfo {author} {\bibfnamefont {B.~P.}\ \bibnamefont
  {Abbott}} \emph {et~al.} (\bibinfo {collaboration} {LIGO Scientific,
  VINROUGE, Las Cumbres Observatory, DES, DLT40, Virgo, 1M2H, Dark Energy
  Camera GW-E, MASTER}),\ }\href {\doibase 10.1038/nature24471} {\bibfield
  {journal} {\bibinfo  {journal} {Nature}\ }\textbf {\bibinfo {volume} {551}},\
  \bibinfo {pages} {85} (\bibinfo {year} {2017})},\ \Eprint
  {http://arxiv.org/abs/1710.05835} {arXiv:1710.05835 [astro-ph.CO]}
  \BibitemShut {NoStop}%
\bibitem [{\citenamefont {{Abbott}}\ \emph
  {et~al.}(2016{\natexlab{a}})\citenamefont {{Abbott}}, \citenamefont
  {{Abbott}}, \citenamefont {{Abbott}}, \citenamefont {{Abernathy}},
  \citenamefont {{Acernese}} \emph {et~al.}}]{2016PhRvL.116v1101A}%
  \BibitemOpen
  \bibfield  {author} {\bibinfo {author} {\bibfnamefont {B.~P.}\ \bibnamefont
  {{Abbott}}}, \bibinfo {author} {\bibfnamefont {R.}~\bibnamefont {{Abbott}}},
  \bibinfo {author} {\bibfnamefont {T.~D.}\ \bibnamefont {{Abbott}}}, \bibinfo
  {author} {\bibfnamefont {M.~R.}\ \bibnamefont {{Abernathy}}}, \bibinfo
  {author} {\bibfnamefont {F.}~\bibnamefont {{Acernese}}},  \emph {et~al.},\
  }\href {\doibase 10.1103/PhysRevLett.116.221101} {\bibfield  {journal}
  {\bibinfo  {journal} {Phys.\ Rev.\ Lett.}\ }\textbf {\bibinfo {volume}
  {116}},\ \bibinfo {eid} {221101} (\bibinfo {year} {2016}{\natexlab{a}})},\
  \Eprint {http://arxiv.org/abs/1602.03841} {arXiv:1602.03841 [gr-qc]}
  \BibitemShut {NoStop}%
\bibitem [{\citenamefont {{Abbott}}\ \emph
  {et~al.}(2016{\natexlab{b}})\citenamefont {{Abbott}}, \citenamefont
  {{Abbott}}, \citenamefont {{Abbott}}, \citenamefont {{Abernathy}},
  \citenamefont {{Acernese}} \emph {et~al.}}]{2016PhRvX...6d1015A}%
  \BibitemOpen
  \bibfield  {author} {\bibinfo {author} {\bibfnamefont {B.~P.}\ \bibnamefont
  {{Abbott}}}, \bibinfo {author} {\bibfnamefont {R.}~\bibnamefont {{Abbott}}},
  \bibinfo {author} {\bibfnamefont {T.~D.}\ \bibnamefont {{Abbott}}}, \bibinfo
  {author} {\bibfnamefont {M.~R.}\ \bibnamefont {{Abernathy}}}, \bibinfo
  {author} {\bibfnamefont {F.}~\bibnamefont {{Acernese}}},  \emph {et~al.},\
  }\href {\doibase 10.1103/PhysRevX.6.041015} {\bibfield  {journal} {\bibinfo
  {journal} {Phys.\ Rev.\ X}\ }\textbf {\bibinfo {volume} {6}},\ \bibinfo {eid}
  {041015} (\bibinfo {year} {2016}{\natexlab{b}})},\ \Eprint
  {http://arxiv.org/abs/1606.04856} {arXiv:1606.04856 [gr-qc]} \BibitemShut
  {NoStop}%
\bibitem [{\citenamefont {{Abbott}}\ \emph {et~al.}(2017)\citenamefont
  {{Abbott}}, \citenamefont {{Abbott}}, \citenamefont {{Abbott}}, \citenamefont
  {{Acernese}}, \citenamefont {{Ackley}} \emph {et~al.}}]{2017ApJ...848L..13A}%
  \BibitemOpen
  \bibfield  {author} {\bibinfo {author} {\bibfnamefont {B.~P.}\ \bibnamefont
  {{Abbott}}}, \bibinfo {author} {\bibfnamefont {R.}~\bibnamefont {{Abbott}}},
  \bibinfo {author} {\bibfnamefont {T.~D.}\ \bibnamefont {{Abbott}}}, \bibinfo
  {author} {\bibfnamefont {F.}~\bibnamefont {{Acernese}}}, \bibinfo {author}
  {\bibfnamefont {K.}~\bibnamefont {{Ackley}}},  \emph {et~al.},\ }\href
  {\doibase 10.3847/2041-8213/aa920c} {\bibfield  {journal} {\bibinfo
  {journal} {Astrophys.\ J.\ Lett.}\ }\textbf {\bibinfo {volume} {848}},\
  \bibinfo {eid} {L13} (\bibinfo {year} {2017})},\ \Eprint
  {http://arxiv.org/abs/1710.05834} {arXiv:1710.05834 [astro-ph.HE]}
  \BibitemShut {NoStop}%
\bibitem [{\citenamefont {{Hou}}\ \emph {et~al.}(2017)\citenamefont {{Hou}},
  \citenamefont {{Gong}},\ and\ \citenamefont {{Liu}}}]{2017arXiv170401899H}%
  \BibitemOpen
  \bibfield  {author} {\bibinfo {author} {\bibfnamefont {S.}~\bibnamefont
  {{Hou}}}, \bibinfo {author} {\bibfnamefont {Y.}~\bibnamefont {{Gong}}}, \
  and\ \bibinfo {author} {\bibfnamefont {Y.}~\bibnamefont {{Liu}}},\
  }\href@noop {} {\bibfield  {journal} {\bibinfo  {journal} {ArXiv e-prints}\ }
  (\bibinfo {year} {2017})},\ \Eprint {http://arxiv.org/abs/1704.01899}
  {arXiv:1704.01899 [gr-qc]} \BibitemShut {NoStop}%
\bibitem [{\citenamefont {Will}(2014)}]{Will:2014kxa}%
  \BibitemOpen
  \bibfield  {author} {\bibinfo {author} {\bibfnamefont {C.~M.}\ \bibnamefont
  {Will}},\ }\href {\doibase 10.12942/lrr-2014-4} {\bibfield  {journal}
  {\bibinfo  {journal} {Living\ Rev.\ Rel.}\ }\textbf {\bibinfo {volume}
  {17}},\ \bibinfo {pages} {4} (\bibinfo {year} {2014})},\ \Eprint
  {http://arxiv.org/abs/1403.7377} {arXiv:1403.7377 [gr-qc]} \BibitemShut
  {NoStop}%
\bibitem [{\citenamefont {Lee}\ \emph {et~al.}(2008)\citenamefont {Lee},
  \citenamefont {Jenet},\ and\ \citenamefont {Price}}]{0004-637X-685-2-1304}%
  \BibitemOpen
  \bibfield  {author} {\bibinfo {author} {\bibfnamefont {K.~J.}\ \bibnamefont
  {Lee}}, \bibinfo {author} {\bibfnamefont {F.~A.}\ \bibnamefont {Jenet}}, \
  and\ \bibinfo {author} {\bibfnamefont {R.~H.}\ \bibnamefont {Price}},\ }\href
  {http://stacks.iop.org/0004-637X/685/i=2/a=1304} {\bibfield  {journal}
  {\bibinfo  {journal} {Astrophys.\ J.}\ }\textbf {\bibinfo {volume} {685}},\
  \bibinfo {pages} {1304} (\bibinfo {year} {2008})}\BibitemShut {NoStop}%
\bibitem [{\citenamefont {{Gaia Collaboration}}\ \emph
  {et~al.}(2016)\citenamefont {{Gaia Collaboration}}, \citenamefont {{Prusti}},
  \citenamefont {{de Bruijne}}, \citenamefont {{Brown}}, \citenamefont
  {{Vallenari}}, \citenamefont {{Babusiaux}} \emph
  {et~al.}}]{2016A&A...595A...1G}%
  \BibitemOpen
  \bibfield  {author} {\bibinfo {author} {\bibnamefont {{Gaia Collaboration}}},
  \bibinfo {author} {\bibfnamefont {T.}~\bibnamefont {{Prusti}}}, \bibinfo
  {author} {\bibfnamefont {J.~H.~J.}\ \bibnamefont {{de Bruijne}}}, \bibinfo
  {author} {\bibfnamefont {A.~G.~A.}\ \bibnamefont {{Brown}}}, \bibinfo
  {author} {\bibfnamefont {A.}~\bibnamefont {{Vallenari}}}, \bibinfo {author}
  {\bibfnamefont {C.}~\bibnamefont {{Babusiaux}}},  \emph {et~al.},\ }\href
  {\doibase 10.1051/0004-6361/201629272} {\bibfield  {journal} {\bibinfo
  {journal} {Astron.\ Astrophys.}\ }\textbf {\bibinfo {volume} {595}},\
  \bibinfo {eid} {A1} (\bibinfo {year} {2016})},\ \Eprint
  {http://arxiv.org/abs/1609.04153} {arXiv:1609.04153 [astro-ph.IM]}
  \BibitemShut {NoStop}%
\bibitem [{\citenamefont {{Gilmore}}\ \emph {et~al.}(1998)\citenamefont
  {{Gilmore}}, \citenamefont {{Perryman}}, \citenamefont {{Lindegren}},
  \citenamefont {{Favata}}, \citenamefont {{Hoeg}} \emph
  {et~al.}}]{1998SPIE.3350..541G}%
  \BibitemOpen
  \bibfield  {author} {\bibinfo {author} {\bibfnamefont {G.~F.}\ \bibnamefont
  {{Gilmore}}}, \bibinfo {author} {\bibfnamefont {M.~A.}\ \bibnamefont
  {{Perryman}}}, \bibinfo {author} {\bibfnamefont {L.}~\bibnamefont
  {{Lindegren}}}, \bibinfo {author} {\bibfnamefont {F.}~\bibnamefont
  {{Favata}}}, \bibinfo {author} {\bibfnamefont {E.}~\bibnamefont {{Hoeg}}},
  \emph {et~al.},\ }in\ \href {\doibase 10.1117/12.317134} {\emph {\bibinfo
  {booktitle} {Astronomical Interferometry}}},\ \bibinfo {series} {Proc. SPIE},
  Vol.\ \bibinfo {volume} {3350},\ \bibinfo {editor} {edited by\ \bibinfo
  {editor} {\bibfnamefont {R.~D.}\ \bibnamefont {{Reasenberg}}}}\ (\bibinfo
  {year} {1998})\ pp.\ \bibinfo {pages} {541--550},\ \Eprint
  {http://arxiv.org/abs/astro-ph/9805180} {astro-ph/9805180} \BibitemShut
  {NoStop}%
\bibitem [{\citenamefont {{Braginsky}}\ \emph {et~al.}(1990)\citenamefont
  {{Braginsky}}, \citenamefont {{Kardashev}}, \citenamefont {{Polnarev}},\ and\
  \citenamefont {{Novikov}}}]{1990NCimB.105.1141B}%
  \BibitemOpen
  \bibfield  {author} {\bibinfo {author} {\bibfnamefont {V.~B.}\ \bibnamefont
  {{Braginsky}}}, \bibinfo {author} {\bibfnamefont {N.~S.}\ \bibnamefont
  {{Kardashev}}}, \bibinfo {author} {\bibfnamefont {A.~G.}\ \bibnamefont
  {{Polnarev}}}, \ and\ \bibinfo {author} {\bibfnamefont {I.~D.}\ \bibnamefont
  {{Novikov}}},\ }\href@noop {} {\bibfield  {journal} {\bibinfo  {journal}
  {Nuovo Cimento B}\ }\textbf {\bibinfo {volume} {105}},\ \bibinfo {pages}
  {1141} (\bibinfo {year} {1990})}\BibitemShut {NoStop}%
\bibitem [{\citenamefont {Gwinn}\ \emph {et~al.}(1997)\citenamefont {Gwinn},
  \citenamefont {Eubanks}, \citenamefont {Pyne}, \citenamefont {Birkinshaw},\
  and\ \citenamefont {Matsakis}}]{Gwinn:1996gv}%
  \BibitemOpen
  \bibfield  {author} {\bibinfo {author} {\bibfnamefont {C.~R.}\ \bibnamefont
  {Gwinn}}, \bibinfo {author} {\bibfnamefont {T.~M.}\ \bibnamefont {Eubanks}},
  \bibinfo {author} {\bibfnamefont {T.}~\bibnamefont {Pyne}}, \bibinfo {author}
  {\bibfnamefont {M.}~\bibnamefont {Birkinshaw}}, \ and\ \bibinfo {author}
  {\bibfnamefont {D.~N.}\ \bibnamefont {Matsakis}},\ }\href {\doibase
  10.1086/304424} {\bibfield  {journal} {\bibinfo  {journal} {Astrophys.\ J.}\
  }\textbf {\bibinfo {volume} {485}},\ \bibinfo {pages} {87} (\bibinfo {year}
  {1997})},\ \Eprint {http://arxiv.org/abs/astro-ph/9610086}
  {arXiv:astro-ph/9610086 [astro-ph]} \BibitemShut {NoStop}%
\bibitem [{\citenamefont {Book}\ and\ \citenamefont
  {Flanagan}(2011)}]{Book:2010pf}%
  \BibitemOpen
  \bibfield  {author} {\bibinfo {author} {\bibfnamefont {L.~G.}\ \bibnamefont
  {Book}}\ and\ \bibinfo {author} {\bibfnamefont {E.~E.}\ \bibnamefont
  {Flanagan}},\ }\href {\doibase 10.1103/PhysRevD.83.024024} {\bibfield
  {journal} {\bibinfo  {journal} {Phys.\ Rev.}\ }\textbf {\bibinfo {volume}
  {D83}},\ \bibinfo {pages} {024024} (\bibinfo {year} {2011})},\ \Eprint
  {http://arxiv.org/abs/1009.4192} {arXiv:1009.4192 [astro-ph.CO]} \BibitemShut
  {NoStop}%
\bibitem [{\citenamefont {{Mignard}}\ and\ \citenamefont
  {{Klioner}}(2012)}]{2012A&A...547A..59M}%
  \BibitemOpen
  \bibfield  {author} {\bibinfo {author} {\bibfnamefont {F.}~\bibnamefont
  {{Mignard}}}\ and\ \bibinfo {author} {\bibfnamefont {S.}~\bibnamefont
  {{Klioner}}},\ }\href {\doibase 10.1051/0004-6361/201219927} {\ \textbf
  {\bibinfo {volume} {547}},\ \bibinfo {eid} {A59} (\bibinfo {year} {2012})},\
  \Eprint {http://arxiv.org/abs/1207.0025} {arXiv:1207.0025 [astro-ph.IM]}
  \BibitemShut {NoStop}%
\bibitem [{\citenamefont {Klioner}(2018)}]{Klioner:2017asb}%
  \BibitemOpen
  \bibfield  {author} {\bibinfo {author} {\bibfnamefont {S.~A.}\ \bibnamefont
  {Klioner}},\ }\href {\doibase 10.1088/1361-6382/aa9f57} {\bibfield  {journal}
  {\bibinfo  {journal} {Class.\ Quantum\ Grav.}\ }\textbf {\bibinfo {volume}
  {35}},\ \bibinfo {pages} {045005} (\bibinfo {year} {2018})},\ \Eprint
  {http://arxiv.org/abs/1710.11474} {arXiv:1710.11474 [astro-ph.HE]}
  \BibitemShut {NoStop}%
\bibitem [{\citenamefont {Moore}\ \emph {et~al.}(2017)\citenamefont {Moore},
  \citenamefont {Mihaylov}, \citenamefont {Lasenby},\ and\ \citenamefont
  {Gilmore}}]{PhysRevLett.119.261102}%
  \BibitemOpen
  \bibfield  {author} {\bibinfo {author} {\bibfnamefont {C.~J.}\ \bibnamefont
  {Moore}}, \bibinfo {author} {\bibfnamefont {D.~P.}\ \bibnamefont {Mihaylov}},
  \bibinfo {author} {\bibfnamefont {A.}~\bibnamefont {Lasenby}}, \ and\
  \bibinfo {author} {\bibfnamefont {G.}~\bibnamefont {Gilmore}},\ }\href
  {\doibase 10.1103/PhysRevLett.119.261102} {\bibfield  {journal} {\bibinfo
  {journal} {Phys.\ Rev.\ Lett.}\ }\textbf {\bibinfo {volume} {119}},\ \bibinfo
  {pages} {261102} (\bibinfo {year} {2017})}\BibitemShut {NoStop}%
\bibitem [{\citenamefont {{Flanagan}}\ and\ \citenamefont
  {{Hughes}}(2005)}]{2005NJPh....7..204F}%
  \BibitemOpen
  \bibfield  {author} {\bibinfo {author} {\bibfnamefont {{\'E}.~{\'E}.}\
  \bibnamefont {{Flanagan}}}\ and\ \bibinfo {author} {\bibfnamefont {S.~A.}\
  \bibnamefont {{Hughes}}},\ }\href {\doibase 10.1088/1367-2630/7/1/204}
  {\bibfield  {journal} {\bibinfo  {journal} {New\ J.\ Phys.}\ }\textbf
  {\bibinfo {volume} {7}},\ \bibinfo {pages} {204} (\bibinfo {year} {2005})},\
  \Eprint {http://arxiv.org/abs/gr-qc/0501041} {gr-qc/0501041} \BibitemShut
  {NoStop}%
\bibitem [{\citenamefont {Perrodin}\ and\ \citenamefont
  {Sesana}(2017)}]{Perrodin:2017bxr}%
  \BibitemOpen
  \bibfield  {author} {\bibinfo {author} {\bibfnamefont {D.}~\bibnamefont
  {Perrodin}}\ and\ \bibinfo {author} {\bibfnamefont {A.}~\bibnamefont
  {Sesana}},\ }\href@noop {} {\  (\bibinfo {year} {2017})},\ \Eprint
  {http://arxiv.org/abs/1709.02816} {arXiv:1709.02816 [astro-ph.HE]}
  \BibitemShut {NoStop}%
\bibitem [{\citenamefont {{Arzoumanian}}\ \emph {et~al.}(2016)\citenamefont
  {{Arzoumanian}}, \citenamefont {{Brazier}}, \citenamefont {{Burke-Spolaor}},
  \citenamefont {{Chamberlin}}, \citenamefont {{Chatterjee}} \emph
  {et~al.}}]{2016ApJ...821...13A}%
  \BibitemOpen
  \bibfield  {author} {\bibinfo {author} {\bibfnamefont {Z.}~\bibnamefont
  {{Arzoumanian}}}, \bibinfo {author} {\bibfnamefont {A.}~\bibnamefont
  {{Brazier}}}, \bibinfo {author} {\bibfnamefont {S.}~\bibnamefont
  {{Burke-Spolaor}}}, \bibinfo {author} {\bibfnamefont {S.~J.}\ \bibnamefont
  {{Chamberlin}}}, \bibinfo {author} {\bibfnamefont {S.}~\bibnamefont
  {{Chatterjee}}},  \emph {et~al.},\ }\href {\doibase
  10.3847/0004-637X/821/1/13} {\bibfield  {journal} {\bibinfo  {journal}
  {Astrophys.\ J.}\ }\textbf {\bibinfo {volume} {821}},\ \bibinfo {eid} {13}
  (\bibinfo {year} {2016})},\ \Eprint {http://arxiv.org/abs/1508.03024}
  {arXiv:1508.03024} \BibitemShut {NoStop}%
\bibitem [{\citenamefont {{Lentati}}\ \emph {et~al.}(2015)\citenamefont
  {{Lentati}}, \citenamefont {{Taylor}}, \citenamefont {{Mingarelli}},
  \citenamefont {{Sesana}}, \citenamefont {{Sanidas}} \emph
  {et~al.}}]{2015MNRAS.453.2576L}%
  \BibitemOpen
  \bibfield  {author} {\bibinfo {author} {\bibfnamefont {L.}~\bibnamefont
  {{Lentati}}}, \bibinfo {author} {\bibfnamefont {S.~R.}\ \bibnamefont
  {{Taylor}}}, \bibinfo {author} {\bibfnamefont {C.~M.~F.}\ \bibnamefont
  {{Mingarelli}}}, \bibinfo {author} {\bibfnamefont {A.}~\bibnamefont
  {{Sesana}}}, \bibinfo {author} {\bibfnamefont {S.~A.}\ \bibnamefont
  {{Sanidas}}},  \emph {et~al.},\ }\href {\doibase 10.1093/mnras/stv1538}
  {\bibfield  {journal} {\bibinfo  {journal} {Mon.\ Notices\ Royal\ Astron.\
  Soc.}\ }\textbf {\bibinfo {volume} {453}},\ \bibinfo {pages} {2576} (\bibinfo
  {year} {2015})},\ \Eprint {http://arxiv.org/abs/1504.03692}
  {arXiv:1504.03692} \BibitemShut {NoStop}%
\bibitem [{\citenamefont {{Shannon}}\ \emph {et~al.}(2013)\citenamefont
  {{Shannon}}, \citenamefont {{Ravi}}, \citenamefont {{Coles}}, \citenamefont
  {{Hobbs}}, \citenamefont {{Keith}} \emph {et~al.}}]{2013Sci...342..334S}%
  \BibitemOpen
  \bibfield  {author} {\bibinfo {author} {\bibfnamefont {R.~M.}\ \bibnamefont
  {{Shannon}}}, \bibinfo {author} {\bibfnamefont {V.}~\bibnamefont {{Ravi}}},
  \bibinfo {author} {\bibfnamefont {W.~A.}\ \bibnamefont {{Coles}}}, \bibinfo
  {author} {\bibfnamefont {G.}~\bibnamefont {{Hobbs}}}, \bibinfo {author}
  {\bibfnamefont {M.~J.}\ \bibnamefont {{Keith}}},  \emph {et~al.},\
  }\href@noop {} {\bibfield  {journal} {\bibinfo  {journal} {Science}\ }\textbf
  {\bibinfo {volume} {342}},\ \bibinfo {pages} {334} (\bibinfo {year}
  {2013})},\ \Eprint {http://arxiv.org/abs/1310.4569} {arXiv:1310.4569}
  \BibitemShut {NoStop}%
\bibitem [{\citenamefont {{Verbiest}}\ \emph {et~al.}(2016)\citenamefont
  {{Verbiest}}, \citenamefont {{Lentati}}, \citenamefont {{Hobbs}},
  \citenamefont {{van Haasteren}}, \citenamefont {{Demorest}} \emph
  {et~al.}}]{2016MNRAS.458.1267V}%
  \BibitemOpen
  \bibfield  {author} {\bibinfo {author} {\bibfnamefont {J.~P.~W.}\
  \bibnamefont {{Verbiest}}}, \bibinfo {author} {\bibfnamefont
  {L.}~\bibnamefont {{Lentati}}}, \bibinfo {author} {\bibfnamefont
  {G.}~\bibnamefont {{Hobbs}}}, \bibinfo {author} {\bibfnamefont
  {R.}~\bibnamefont {{van Haasteren}}}, \bibinfo {author} {\bibfnamefont
  {P.~B.}\ \bibnamefont {{Demorest}}},  \emph {et~al.},\ }\href {\doibase
  10.1093/mnras/stw347} {\bibfield  {journal} {\bibinfo  {journal} {Mon.\
  Notices\ Royal\ Astron.\ Soc.}\ }\textbf {\bibinfo {volume} {458}},\ \bibinfo
  {pages} {1267} (\bibinfo {year} {2016})},\ \Eprint
  {http://arxiv.org/abs/1602.03640} {arXiv:1602.03640 [astro-ph.IM]}
  \BibitemShut {NoStop}%
\bibitem [{\citenamefont {Gair}\ \emph {et~al.}(2015)\citenamefont {Gair},
  \citenamefont {Romano},\ and\ \citenamefont {Taylor}}]{Gair:2015hra}%
  \BibitemOpen
  \bibfield  {author} {\bibinfo {author} {\bibfnamefont {J.~R.}\ \bibnamefont
  {Gair}}, \bibinfo {author} {\bibfnamefont {J.~D.}\ \bibnamefont {Romano}}, \
  and\ \bibinfo {author} {\bibfnamefont {S.~R.}\ \bibnamefont {Taylor}},\
  }\href {\doibase 10.1103/PhysRevD.92.102003} {\bibfield  {journal} {\bibinfo
  {journal} {Phys.\ Rev.}\ }\textbf {\bibinfo {volume} {D92}},\ \bibinfo
  {pages} {102003} (\bibinfo {year} {2015})},\ \Eprint
  {http://arxiv.org/abs/1506.08668} {arXiv:1506.08668 [gr-qc]} \BibitemShut
  {NoStop}%
\bibitem [{\citenamefont {Schutz}(2010)}]{Schutz}%
  \BibitemOpen
  \bibfield  {author} {\bibinfo {author} {\bibfnamefont {B.~F.}\ \bibnamefont
  {Schutz}},\ }\href {\doibase 10.1017/S1743921309990457} {\bibfield  {journal}
  {\bibinfo  {journal} {IAU Symp.}\ }\textbf {\bibinfo {volume} {261}},\
  \bibinfo {pages} {234} (\bibinfo {year} {2010})}\BibitemShut {NoStop}%
\bibitem [{\citenamefont {Finn}\ \emph {et~al.}(2009)\citenamefont {Finn},
  \citenamefont {Larson},\ and\ \citenamefont {Romano}}]{PhysRevD.79.062003}%
  \BibitemOpen
  \bibfield  {author} {\bibinfo {author} {\bibfnamefont {L.~S.}\ \bibnamefont
  {Finn}}, \bibinfo {author} {\bibfnamefont {S.~L.}\ \bibnamefont {Larson}}, \
  and\ \bibinfo {author} {\bibfnamefont {J.~D.}\ \bibnamefont {Romano}},\
  }\href {\doibase 10.1103/PhysRevD.79.062003} {\bibfield  {journal} {\bibinfo
  {journal} {Phys.\ Rev.\ D}\ }\textbf {\bibinfo {volume} {79}},\ \bibinfo
  {pages} {062003} (\bibinfo {year} {2009})}\BibitemShut {NoStop}%
\bibitem [{\citenamefont {{Robin}}\ \emph {et~al.}(2012)\citenamefont
  {{Robin}}, \citenamefont {{Luri}}, \citenamefont {{Reyl{\'e}}}, \citenamefont
  {{Isasi}}, \citenamefont {{Grux}} \emph {et~al.}}]{2012gums}%
  \BibitemOpen
  \bibfield  {author} {\bibinfo {author} {\bibfnamefont {A.~C.}\ \bibnamefont
  {{Robin}}}, \bibinfo {author} {\bibfnamefont {X.}~\bibnamefont {{Luri}}},
  \bibinfo {author} {\bibfnamefont {C.}~\bibnamefont {{Reyl{\'e}}}}, \bibinfo
  {author} {\bibfnamefont {Y.}~\bibnamefont {{Isasi}}}, \bibinfo {author}
  {\bibfnamefont {E.}~\bibnamefont {{Grux}}},  \emph {et~al.},\ }\href
  {\doibase 10.1051/0004-6361/201118646} {\bibfield  {journal} {\bibinfo
  {journal} {Astron.\ Astrophys.}\ }\textbf {\bibinfo {volume} {543}},\
  \bibinfo {eid} {A100} (\bibinfo {year} {2012})},\ \Eprint
  {http://arxiv.org/abs/1202.0132} {arXiv:1202.0132} \BibitemShut {NoStop}%
\bibitem [{\citenamefont {{Flanagan}}(1993)}]{1993PhRvD..48.2389F}%
  \BibitemOpen
  \bibfield  {author} {\bibinfo {author} {\bibfnamefont {E.~E.}\ \bibnamefont
  {{Flanagan}}},\ }\href {\doibase 10.1103/PhysRevD.48.2389} {\bibfield
  {journal} {\bibinfo  {journal} {Phys.\ Rev.\ D}\ }\textbf {\bibinfo {volume}
  {48}},\ \bibinfo {pages} {2389} (\bibinfo {year} {1993})},\ \Eprint
  {http://arxiv.org/abs/astro-ph/9305029} {astro-ph/9305029} \BibitemShut
  {NoStop}%
\bibitem [{\citenamefont {Fuselier}\ and\ \citenamefont
  {Wright}(2009)}]{FuselierWright2009}%
  \BibitemOpen
  \bibfield  {author} {\bibinfo {author} {\bibfnamefont {E.~J.}\ \bibnamefont
  {Fuselier}}\ and\ \bibinfo {author} {\bibfnamefont {G.~B.}\ \bibnamefont
  {Wright}},\ }\href@noop {} {\bibfield  {journal} {\bibinfo  {journal} {\ J.\
  Num.\ Anal.}\ }\textbf {\bibinfo {volume} {47}},\ \bibinfo {pages} {3213}
  (\bibinfo {year} {2009})}\BibitemShut {NoStop}%
\bibitem [{\citenamefont {{Hellings}}\ and\ \citenamefont
  {{Downs}}(1983)}]{1983ApJ...265L..39H}%
  \BibitemOpen
  \bibfield  {author} {\bibinfo {author} {\bibfnamefont {R.~W.}\ \bibnamefont
  {{Hellings}}}\ and\ \bibinfo {author} {\bibfnamefont {G.~S.}\ \bibnamefont
  {{Downs}}},\ }\href {\doibase 10.1086/183954} {\bibfield  {journal} {\bibinfo
   {journal} {Astrophys.\ J.\ Lett.}\ }\textbf {\bibinfo {volume} {265}},\
  \bibinfo {pages} {L39} (\bibinfo {year} {1983})}\BibitemShut {NoStop}%
\bibitem [{\citenamefont {{Gair}}\ \emph {et~al.}(2014)\citenamefont {{Gair}},
  \citenamefont {{Romano}}, \citenamefont {{Taylor}},\ and\ \citenamefont
  {{Mingarelli}}}]{2014PhRvD..90h2001G}%
  \BibitemOpen
  \bibfield  {author} {\bibinfo {author} {\bibfnamefont {J.}~\bibnamefont
  {{Gair}}}, \bibinfo {author} {\bibfnamefont {J.~D.}\ \bibnamefont
  {{Romano}}}, \bibinfo {author} {\bibfnamefont {S.}~\bibnamefont {{Taylor}}},
  \ and\ \bibinfo {author} {\bibfnamefont {C.~M.~F.}\ \bibnamefont
  {{Mingarelli}}},\ }\href {\doibase 10.1103/PhysRevD.90.082001} {\bibfield
  {journal} {\bibinfo  {journal} {Phys.\ Rev.\ D}\ }\textbf {\bibinfo {volume}
  {90}},\ \bibinfo {eid} {082001} (\bibinfo {year} {2014})},\ \Eprint
  {http://arxiv.org/abs/1406.4664} {arXiv:1406.4664 [gr-qc]} \BibitemShut
  {NoStop}%
\end{thebibliography}%

\newpage

\onecolumngrid

\begin{appendices}
\numberwithin{equation}{section}
\renewcommand{\theequation}{\Alph{section}\arabic{equation}}
\renewcommand{\thesubsection}{\Roman{subsection}}

\section{Additional Astrometric Deflection Patterns}
\label{app:astrometricPatterns}
\begin{figure}[h!]
\centering
\subfloat{
	\includegraphics[scale=1]{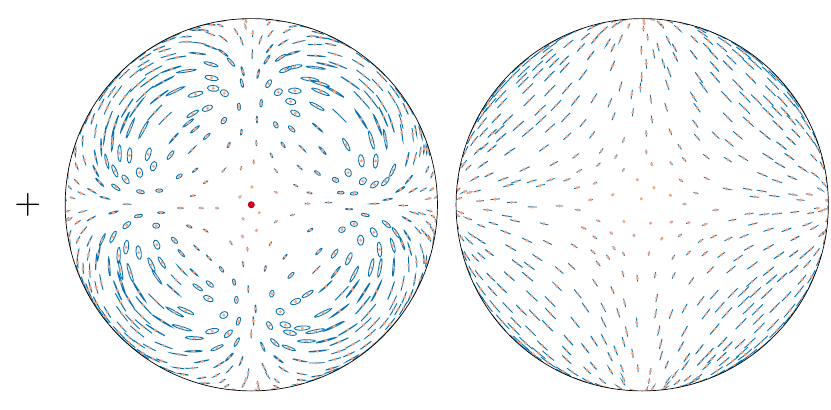}
}
\hspace{12pt}
\subfloat{
	\includegraphics[scale=1]{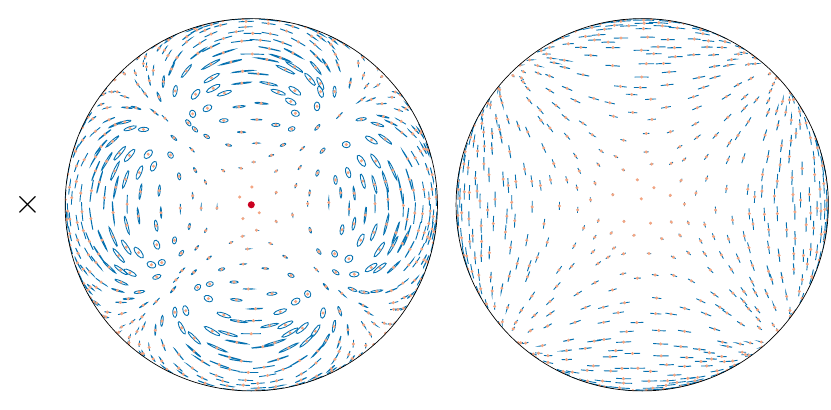}
}
\caption{Orthographic projections of the Northern (left) and Southern (right) hemispheres. On the sky are chosen 1000 stars. A \textsc{gw} from a source located at the North pole (indicated by the red dots) is incident on the Earth causing the stars move periodically at the \textsc{gw} frequency, according to eq.~(\ref{eq:astrometricsignalFULL}). All stars are placed at a distance of 10 gravitational wavelengths from the Earth; in reality, these neat elliptical patterns would appear more chaotic, since the distances to each star is generally different. The blue lines show the resulting movement tracks. The incident \textsc{gw} has the unphysically large characteristic strain amplitude \(A = 0.1\).}
\label{fig:starTracesApp}
\end{figure}

\noindent
In Fig.~\ref{fig:starTraces} of Section~\ref{sec:gravitationalWavepolarizations} the astrometric deflection patterns were plotted for all 6 \textsc{gw} polarization states; for the three transverse modes (\(+\), \(\times\), and \(S\)) the distant source limit formula (\ref{eq:AstroDefFinal}) was used, whilst for the three modes with longitudinal components (\(X\), \(Y\), and \(L\)) the full formula (including the star terms) eq.~(\ref{eq:astrometricsignalFULL}) was used. In this appendix the effect of including the star terms on the deflection patterns for the two transverse \textsc{gr} modes (\(+\) and \(\times\)) is illustrated. In Fig.~\ref{fig:starTracesApp} the new astrometric deflection patterns are plotted; these plots were produced in the same way as the top row of Fig.~\ref{fig:starTraces} except eq.~(\ref{eq:astrometricsignalFULL}) was used in place of eq.~(\ref{eq:AstroDefFinal}) (all stars are placed 10 gravitational wavelengths away from the Earth). The extra terms in the full expression for the astrometric deflection introduce an additional oscillatory deflection pattern which is out of phase with the Earth term pattern plotted in Fig.~\ref{fig:starTraces}, this causes each star to trace out a small ellipse on the sky. If the stars are further away from the Earth then the phase difference between the two oscillations changes and the amplitude of the additional oscillation is reduced (the ellipses appear rotated and their eccentricity is increased); in the limit of infinite distance the patterns in Fig.~\ref{fig:starTraces} are recovered. In these figures the ellipses are aligned in a regular pattern on the sky because the stars are all the same distance from the Earth. If all the distances were different, the ellipses would all be misaligned and the extra motion from the full formula in eq.~(\ref{eq:astrometricsignalFULL}) would appear to be a random noise superposed on the regular Earth term pattern plotted in Fig.~\ref{fig:starTraces}.

\section{The Tensorial Astrometric Integrals}
\label{app:PlusCrossInt}
\noindent In Sections~\ref{sec:tensorialModes} to \ref{sec:longitudinalModes} the details of the evaluations of the spatial correlation integrals for the different polarization modes were omitted for brevity; the details are presented in this Appendix~\ref{app:PlusCrossInt}, and in Appendices~\ref{app:SInt}, \ref{app:VectorialInt}, and \ref{app:longIntegralDer}.

In the main text the astrometric response at each of the two star positions was resolved in the tangent plane along a pair of basis vectors. It was then shown how the spatial correlation matrix can be written in terms of just two scalar integrals over the sky; one involving the \(x\) and \(\theta\) components, and one involving the \(y\) and \(\phi\) components. In this Appendix these two integrals will be evaluated, first for the \(+\) mode and then for the \(\times\) mode.

Firstly, the \(x\)-\(\theta\) term for the \(+\) polarized \textsc{gw} state is considered; the correlation integral is defined in eq.~(\ref{eq:defGammaxtheta}) as
\begin{align}
\Gamma^{+}_{x\theta}(\Theta) = \int_{S^{2}}\!\!\!\dd\Omega_{\vect{q}}\;\delta n_{x}^{+}(\vect{q})\,\delta m_{\theta}^{+}(\vect{q}) = \int_{0}^{2\pi}\!\!\!\!\dd\phi\int_{0}^{\pi}\!\!\!\dd\theta\,\sin\theta \,\delta n_{x}^{+}(\theta,\phi)\,\delta m_{\theta}^{+}(\theta,\phi) \,, \label{eq:appintONE}
\end{align}
where the vector \(\vect{q}=(\sin\theta\cos\phi,\sin\theta\sin\phi,\cos\theta)\) is the direction on the sky from which the \textsc{gw} originates. The components of the astrometric deflections, \(\delta n_{x}^{+}(\theta,\phi)\) and \(\delta m_{\theta}^{+}(\theta,\phi)\), may be evaluated from the definitions in eqs.~(\ref{eq:dnx}) and (\ref{eq:dmtheta}), using the formula for the astrometric deflection in eq.~(\ref{eq:AstroDefFinal}), the expression for the \textsc{gw} basis tensor in eq.~(\ref{eq:tensorPlus}), and the expressions for the basis vectors tangent to the sphere in eqs.~(\ref{eq:defux}) and (\ref{eq:defutheta}):
\begin{subequations}
\begin{align}
\delta n_{x}^{+}(\theta,\phi) = &\, -\frac{1}{2}\,\sin\theta\cos\phi \,, \\
\delta m_{\theta}^{+}(\theta,\phi) = &\, \frac{1}{1 - \cos\Theta \cos\theta - \sin\Theta \sin\theta \cos\phi} \left(\frac{3}{16}\,\sin\Theta\,(1 - \cos\!\left(2\theta\right)) + \frac{1}{4} \left(\cos\!\left(2\Theta\right) \sin\!\left(2\theta\right) - 2\cos\Theta \sin\theta\right) \cos\phi \right. \nonumber \\ & \left. \quad + \frac{1}{16} \left(8 \sin\Theta\,\cos\theta - 3\sin\!\left(2\Theta\right) - \sin\!\left(2\Theta\right)\cos\!\left(2\theta\right)\right) \cos\!\left(2\phi\right)\!\right) \,.
\end{align}
\end{subequations}
The azimuthal integral over \(\phi\) in eq.~(\ref{eq:appintONE}) may be evaluated using the result derived in Appendix~\ref{app:azimuthalIntegral} (previously published by \citep{2014PhRvD..90h2001G} with a sign error). Using this result the double integral in eq.~(\ref{eq:appintONE}) becomes the single integral
\begin{align}
\begin{split}
\Gamma^{+}_{x\theta}(\Theta) =& - 2\pi\int_{0}^{\Theta}\!\!\!\dd\theta \left(\cos^{4}\!\left(\theta/2\right) + \cos^{2}\!\left(\theta/2\right) - 1\right) \frac{\sin^{3}\!\left(\theta/2\right)}{\cos\!\left(\theta/2\right)} + 2\pi\int_{\Theta}^{\pi}\!\!\!\dd\theta \left(\sin^{4}\!\left(\theta/2\right) + \sin^{2}\!\left(\theta/2\right) - 1\right) \frac{\cos^{3}\!\left(\theta/2\right)}{\sin\!\left(\theta/2\right)} \\
&+ 2\pi \cos^{2}\!\left(\Theta/2\right)\int_{0}^{\Theta}\!\!\!\dd\theta \left(2\cos^{4}\!\left(\theta/2\right) + 1\right) \frac{\sin^{3}\!\left(\theta/2\right)}{\cos\!\left(\theta/2\right)} - 2\pi \sin^{2}\!\left(\Theta/2\right)\int_{\Theta}^{\pi}\!\!\!\dd\theta \left(2\sin^{4}\!\left(\theta/2\right) + 1\right) \frac{\cos^{3}\!\left(\theta/2\right)}{\sin\!\left(\theta/2\right)} \\
&- \frac{2\pi}{\sin^{2}\!\left(\Theta/2\right)} \int_{0}^{\Theta}\!\!\!\dd\theta\,\frac{\sin^{5}\!\left(\theta/2\right)}{\cos\!\left(\theta/2\right)} + \frac{2\pi}{\cos^{2}\!\left(\Theta/2\right)} \int_{\Theta}^{\pi}\!\!\!\dd\theta\,\frac{\cos^{5}\!\left(\theta/2\right)}{\sin\!\left(\theta/2\right)}\,.
\end{split}
\end{align}
This integral may now be evaluated using standard techniques to give
\begin{align}
\Gamma_{x \theta}^{+} (\Theta) = \frac{7\pi}{3} - \frac{14\pi}{3}\,\sin^{2}\!\left(\Theta / 2\right) - 4\pi\,\frac{\sin^{4}\!\left(\Theta / 2\right)}{1 - \sin^{2}\!\left(\Theta / 2\right)}\,\ln\!\left(\sin\!\left(\Theta / 2\right)\right) + 4\pi\,\frac{\cos^{4}\!\left(\Theta / 2\right)}{1 - \cos^{2}\!\left(\Theta / 2\right)}\,\ln\!\left(\cos\!\left(\Theta / 2\right)\right). \label{eq:RES1}
\end{align}

Secondly, the \(y\)-\(\phi\) component term for the \(+\) polarised \textsc{gw} state is considered; the relevant correlation integral was defined in eq.~(\ref{eq:defGammayphi}) as
\begin{equation}
\Gamma^{+}_{y\phi}(\Theta) = \int_{0}^{2\pi}\!\!\!\!\dd\phi\int_{0}^{\pi}\!\!\!\dd\theta\,\sin\theta \,\delta n_{y}^{+}(\theta,\phi)\,\delta m_{\phi}^{+}(\theta,\phi) \,. \label{eq:appintTWO}
\end{equation}
The components of the astrometric deflections may be evaluated from the definitions in eqs.~(\ref{eq:dny}) and (\ref{eq:dmphi}), using the formula for the astrometric deflection in eq.~(\ref{eq:AstroDefFinal}), the expression for the \textsc{gw} basis tensor in eq.~(\ref{eq:tensorPlus}), and the expressions for the basis vectors in eqs.~(\ref{eq:defuy}) and (\ref{eq:defuphi}):
\begin{subequations}
\begin{align}
&\delta n_{y}^{+}(\theta,\phi) = -\frac{1}{2}\,\sin\theta \sin\phi \,, \\
\begin{split}
&\delta m_{\phi}^{+}(\theta,\phi) = \frac{1}{1 - \cos\Theta \cos\theta - \sin\Theta \sin\theta \cos\phi} \left(\frac{1}{2}\left(\cos\Theta\,\sin\theta\,\cos\theta - \sin\!\left(2\Theta\right)\,\sin\theta\right)\,\sin\phi \right. \\
&\quad\quad\quad\quad\quad\quad\quad\quad\quad\quad\quad\quad\quad\quad\quad\quad\quad\quad\quad\quad\quad\quad \left. + \frac{1}{8} \left(2 \sin\!\left(2\Theta\right) \cos\theta - 3\sin\Theta - \sin\Theta\cos\!\left(2\theta\right)\right)\,\sin\!\left(2\phi\right)\!\right).
\end{split}
\end{align}
\end{subequations}
The integral for \(\Gamma^{+}_{y\phi}(\Theta)\) in eq.~(\ref{eq:appintTWO}) may be evaluated in the same way as that for \(\Gamma^{+}_{x\theta}(\Theta)\) above to give
\begin{align}
\Gamma_{y \phi}^{+} (\Theta) = - \frac{5\pi}{3} - 4\pi\,\frac{\sin^{4}\!\left(\Theta / 2\right)}{1 - \sin^{2}\!\left(\Theta / 2\right)}\,\ln\!\left(\sin\!\left(\Theta / 2\right)\right) - 4\pi\,\frac{\cos^{4}\!\left(\Theta / 2\right)}{1 - \cos^{2}\!\left(\Theta / 2\right)}\,\ln\!\left(\cos\!\left(\Theta / 2\right)\right). \label{eq:RES2}
\end{align}
Thirdly, the two analogous integrals for the \(\times\) \textsc{gr} polarization state are considered; these correlation integrals are defined in eqs.~(\ref{eq:defGammaxtheta}) and (\ref{eq:defGammayphi}):

\medskip
\vspace{-\abovedisplayskip}
\noindent
\begin{subequations}
\begin{minipage}[c]{0.495\linewidth}
\begin{align}
\Gamma^{\times}_{x\theta}(\Theta) = \int_{0}^{\pi}\!\!\dd\theta\int_{0}^{2\pi}\!\!\!\dd\phi\;\sin\theta \;\delta n_{x}^{\times}(\theta,\phi)\,\delta m_{\theta}^{\times}(\theta,\phi), \label{eq:appintTHREE}
\end{align}%
\end{minipage}
\begin{minipage}[c]{0.50\linewidth}
\begin{align}
\Gamma^{\times}_{y\phi}(\Theta) = \int_{0}^{\pi}\!\!\dd\theta\int_{0}^{2\pi}\!\!\!\dd\phi\;\sin\theta \;\delta n_{y}^{\times}(\theta,\phi)\,\delta m_{\phi}^{\times}(\theta,\phi). \label{eq:appintFOUR}
\end{align}
\end{minipage}
\end{subequations}
\smallskip

\noindent The components of the astrometric deflection may be evaluated from the definitions in eqs.~(\ref{eq:dnx}), (\ref{eq:dmtheta}), (\ref{eq:dny}), and (\ref{eq:dmphi}), using the formula for the astrometric deflection in eq.~(\ref{eq:AstroDefFinal}), the expression for the \textsc{gw} basis tensor in eq.~(\ref{eq:tensorCross}), and the expressions for the basis vectors in eqs.~(\ref{eq:defux}), (\ref{eq:defutheta}), (\ref{eq:defuy}), and (\ref{eq:defuphi}). Due to the symmtetry between the \(+\) and \(\times\) modes, these components are closely related to those found above for the \(+\) components:

\begin{subequations}
\noindent\begin{minipage}[l]{0.495\linewidth}
\begin{align}
\delta n_{x}^{\times} (\theta,\phi) = \delta n_{y}^{+} (\theta,\phi)\,,
\end{align}
\end{minipage}\hspace*{0.03\linewidth}
\begin{minipage}[l]{0.47\linewidth}
\begin{align}
\delta n_{y}^{\times} (\theta,\phi) = -\,\delta n_{x}^{+} (\theta,\phi)\,,
\end{align}
\end{minipage}\\[-2pt]
\noindent\begin{minipage}[l]{0.495\linewidth}
\begin{align}
\delta m_{\theta}^{\times} (\theta,\phi) = \delta m_{\phi}^{+} (\theta,\phi)\,,
\end{align}
\end{minipage}\hspace*{0.03\linewidth}
\begin{minipage}[l]{0.47\linewidth}
\begin{align}
\delta m_{\phi}^{\times} (\theta,\phi) = -\,\delta m_{\theta}^{+} (\theta,\phi)\,.
\end{align}
\end{minipage}
\end{subequations}\\[5pt]

\noindent The two integrals for \(\Gamma^{\times}_{x\theta}(\Theta)\) and \(\Gamma^{\times}_{y\phi}(\Theta)\) in eqs.~(\ref{eq:appintTHREE}) and (\ref{eq:appintFOUR}) may be evaluated in the same way as those for \(\Gamma^{+}_{x\theta}(\Theta)\) and \(\Gamma^{+}_{y\phi}(\Theta)\) above to give

\medskip
\vspace{-\abovedisplayskip}
\noindent
\begin{subequations}
\begin{minipage}[c]{0.495\linewidth}
\begin{align}
\Gamma^{\times}_{x\theta}(\Theta) = &\, \Gamma^{+}_{y\phi}(\Theta) \,, \label{eq:RES3}
\end{align}%
\end{minipage}
\begin{minipage}[c]{0.50\linewidth}
\begin{align}
\Gamma^{\times}_{y\phi}(\Theta) = &\,  \Gamma^{+}_{x\theta}(\Theta)\,. \label{eq:RES4}
\end{align}
\end{minipage}
\end{subequations}
\smallskip

\begin{figure}[t]
\centering
\begin{minipage}[c]{0.48\linewidth}
	\centering
	\includegraphics[scale=1]{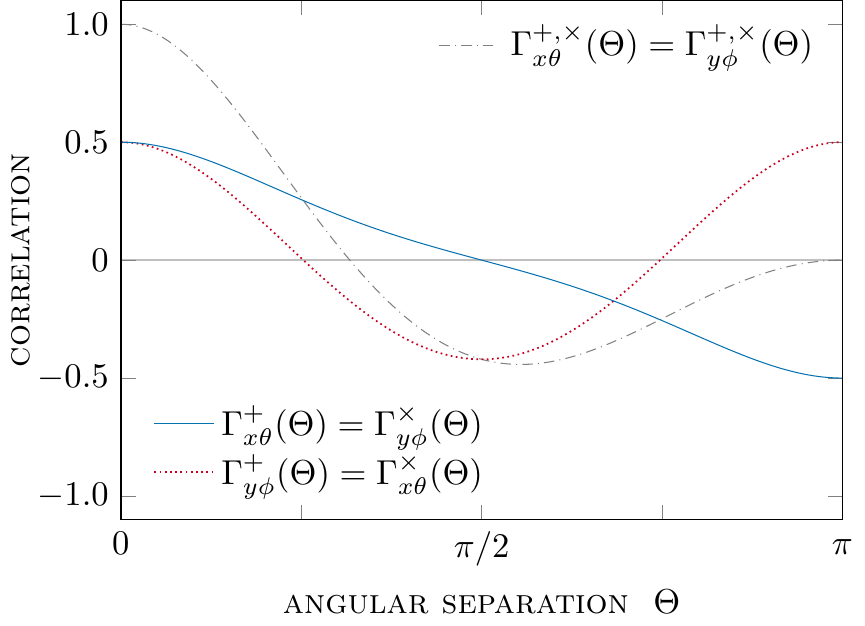}
	\caption{Plot of the tensorial correlation functions in eq.~(\ref{eq:RES5}), normalized so that their absolute maximum is unity. Also shown are the functions for each of the two \textsc{gr} modes, given by eqs.~(\ref{eq:RES1}), (\ref{eq:RES2}), (\ref{eq:RES3}), and (\ref{eq:RES4}), rescaled by the same normalization constant.}
	\label{fig:tensorialExtraPlot}
\end{minipage}%
\hspace*{0.03\linewidth}
\begin{minipage}[c]{0.48\linewidth}
	\centering
	\includegraphics[scale=1]{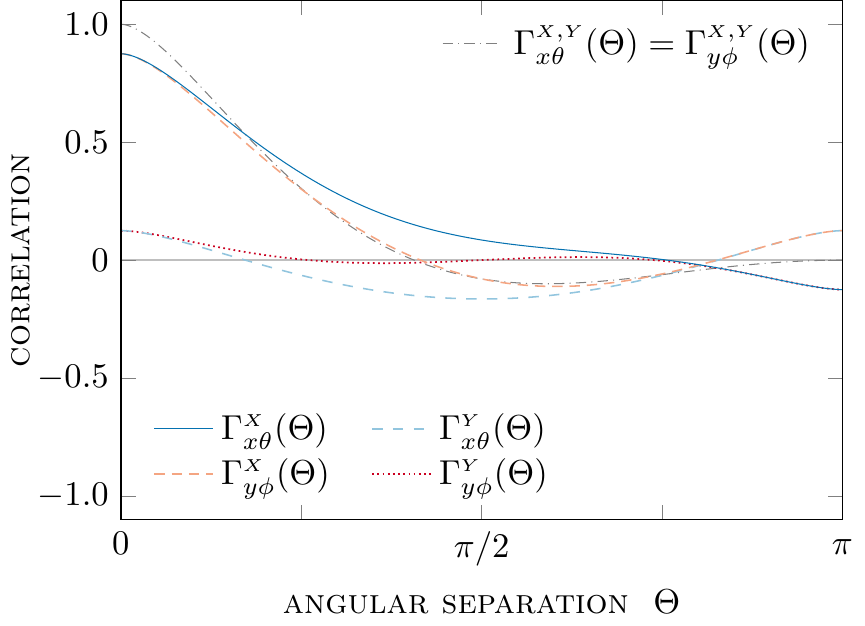}
	\caption{Plot of the vectorial correlation functions in eq.~(\ref{eq:vectorialAstrCorr}), normalized so that their absolute maximum is unity. Also shown are the functions for each of the two \textsc{gr} modes, given by eqs.~(\ref{eq:Vres1}), (\ref{eq:Vres2}), (\ref{eq:Vres3}), and (\ref{eq:Vres4}), rescaled by the same normalization constant.}
	\label{fig:vectorialExtraPlot}
\end{minipage}
\end{figure}

As was described in the main text, for an unpolarised background containing equal power of both \(+\) and \(\times\) polarization states we may define the combined spatial correlation functions \({\Gamma^{+,\times}_{x\theta}(\Theta)=\Gamma^{+}_{x\theta}(\Theta)+\Gamma^{\times}_{x\theta}(\Theta)}\) and \({\Gamma^{+,\times}_{y\phi}(\Theta)=\Gamma^{+}_{y\phi}(\Theta)+\Gamma^{\times}_{y\phi}(\Theta)}\). 
These new functions may be evaluated by taking the sums of the expression in eqs.~(\ref{eq:RES1}), (\ref{eq:RES2}), (\ref{eq:RES3}), and (\ref{eq:RES4}) to give the result which appeared in the main text,
\begin{align}\label{eq:RES5}
\Gamma_{x \theta}^{+, \times} (\Theta) = \Gamma_{y \phi}^{+, \times} (\Theta) = \frac{2\pi}{3} - \frac{14\pi}{3}\,\sin^{2}\!\left(\Theta/2\right) - 8\pi\,\frac{\sin^{4}\!\left(\Theta / 2\right)}{1 - \sin^{2}\!\left(\Theta / 2\right)}\,\ln\!\left(\sin\!\left(\Theta / 2\right)\right).
\end{align}
The three functions \({\Gamma^{+}_{x\theta}(\Theta)=\Gamma^{\times}_{y\phi}(\Theta)}\), \({\Gamma^{+}_{y\phi}(\Theta)=\Gamma^{\times}_{x\theta}(\Theta)}\), and \({\Gamma^{+,\times}_{x\theta}(\Theta)=\Gamma^{+,\times}_{y\phi}(\Theta)}\), are plotted in Fig.~\ref{fig:tensorialExtraPlot}.

\subsection{Azimuthal Integral}
\label{app:azimuthalIntegral}
\noindent
The following integral appears in most of the spatial correlation integrals;
\begin{align}
I_{n} (\theta, \Theta) =\!\int_{0}^{2\pi} \!\!\!\!\! \dd\phi\,\frac{\cos(n\phi)}{1 - \cos\Theta \cos\theta - \sin\Theta \sin\theta \cos\phi} = \mathfrak{R}\left\{\int_{0}^{2\pi} \!\!\!\!\! \dd\phi\,\frac{e^{\rmi n \phi}}{1 - \cos\Theta \cos\theta - \sin\Theta \sin\theta \cos\phi}\right\}.
\end{align}
The integral is tidied up by denoting the new variables

\medskip
\vspace{-\abovedisplayskip}
\noindent
\begin{subequations}
\begin{minipage}[c]{0.495\linewidth}
\begin{align}
\mathfrak{a} (\theta, \Theta) = 1 - \cos\Theta \cos\theta\,,
\end{align}%
\end{minipage}
\begin{minipage}[c]{0.50\linewidth}
\begin{align}
\mathfrak{b} (\theta, \Theta) = - \frac{\sin\Theta \sin\theta}{1 - \cos\Theta \cos\theta}\,.
\end{align}
\end{minipage}
\end{subequations}
\smallskip

If \(z = e^{\rmi\phi}\), then \(\dd\phi = \dd z/(\rmi z)\) and from \(\bar{z} = 1 / z = \cos\phi - \rmi \sin\phi\), it can be found that \(\cos\phi = \sfrac{1}{2}\!\left(z + z^{-1}\right)\) and finally \(I_{n} (\theta, \Theta)\) is expressed as a complex integral over the circle \(\gamma = \{|z| = - 1 / \mathfrak{b}\}\)
\begin{align}\label{eq:contourIntegral}
I_{n} (\theta, \Theta) = \mathfrak{R}\left\{\frac{2}{\rmi \mathfrak{a}}\,\oint_{\gamma} \dd z\,\frac{z^{|n|}}{\mathfrak{b} z^{2} + 2 z + \mathfrak{b}}\right\}.
\end{align}

\noindent The modulus sign is needed since the cosine function is even. The integrand has 2 distinct poles, each at \(z_{\pm} = \left(-1 \pm \sqrt{1 - \mathfrak{b}^{2}}\right)\!/\mathfrak{b}\).
To positions of those poles are determined by the range of the function \(\mathfrak{b} (\theta, \Theta)\). Both \(\theta\) and \(\Theta\) range from 0 to \(\pi\); the substitution \(x = \cos\theta\) and \(y = \cos\Theta\) leads to
\begin{align}
\mathfrak{b} (x, y) = - \frac{\sqrt{1 - x^{2}} \sqrt{1 - y^{2}}}{1 - x y}.
\end{align}
By investigating the derivatives of \(\mathfrak{b} (x, y)\), it can be established that the function ranges between \(-1\) and \(0\), which corresponds to \(z_{-} \in [-1/ \mathfrak{b}, +\infty)\) and \(z_{+} \in (0, - 1/\mathfrak{b}]\). The value of the contour integral is determined by the residue of the root which lies inside the contour \(\gamma\);
\begin{align}
2 \pi \rmi\,\mathrm{Res}_{\,z = z_{+}}\!\left\{\frac{z^{n}}{\mathfrak{b} z^{2} + 2 z + \mathfrak{b}}\right\} = \pi \rmi\,\frac{z_{+}^{n}}{\mathfrak{b} z_{+} + 1}.
\end{align}
Plugging this back into eq.~(\ref{eq:contourIntegral}) yields a result for \(I_{n}\).
\vspace{5pt}
\vspace{-\abovedisplayskip}
\noindent
\begin{subequations}
\begin{align}
I_{n} (\theta, \Theta) = \frac{2 \pi}{\mathfrak{a}\sqrt{1-\mathfrak{b}^{2}}}\left(\!\frac{-1 + \sqrt{1 - \mathfrak{b}^{2}}}{\mathfrak{b}}\right)\np{|n|} &= \frac{2 \pi}{\left|\cos\theta - \cos\Theta\right|}\,\Bigg(\!\frac{1 - \cos\Theta \cos\theta + \left|\cos\theta - \cos\Theta\right|}{\sin\Theta \sin\theta}\!\Bigg)\np{|n|} \\
&= \begin{cases}
\dfrac{2 \pi}{\cos\theta - \cos\Theta}\,\Bigg(\!\dfrac{(1 + \cos\Theta)(1 - \cos\theta)}{(1 - \cos\Theta)(1 + \cos\theta)}\!\Bigg)\np{|n|/2},\quad 0 \leq \theta \leq \Theta, \\
\dfrac{2 \pi}{\cos\Theta - \cos\theta}\,\Bigg(\!\dfrac{(1 - \cos\Theta)(1 + \cos\theta)}{(1 + \cos\Theta)(1 - \cos\theta)}\!\Bigg)\np{|n|/2},\quad \Theta < \theta \leq \pi.
\end{cases}
\end{align}
\end{subequations}
\vspace{-10pt}

\section{Random Realisations of the Astrometric Deflections on the Sky}
\label{app:RandomReal}
\noindent
The Figs.~\ref{fig:tensorialRealPlot} and \ref{fig:scalarRealPlot} show one possible realization of the \textsc{gw} background for the \textsc{gr} and transverse scalar polarizations respectively. 
In this appendix the method for producing such realizations is described.

Pick \(N\) distinct arbitrary points (or ``stars'') on the sky. The Cartesian coordinates of the \(i^{\textrm{th}}\) star are \(\vect{n}_{i}=(x_{i}, y_{i}, z_{i})\) and satisfy \(x_{i}^{2}+y_{i}^{2}+z_{i}^{2} = 1\). Consider the following vector of length \(3N\) formed from the Cartesian coordinates,
\begin{align}
\vect{x} = (x_{1},y_{1},z_{1},x_{2},y_{2},z_{2},\ldots,x_{\cn},y_{\cn},z_{\cn}).
\end{align}
The quantity one needs to find to find is the change in these coordinates, \(\bm{\delta}\vect{x}\).
Formally, this is distributed as a zero mean Gaussian random variable,
\begin{align}
\bm{\delta}\vect{x} \sim \mathcal{N}(\mathbf{0},\mathbf{C}) ,
\end{align}
with the \(3N\times 3N\) correlation matrix formed block-by-block from the \(3 \times 3\) spatial matrices defined in eq.~\ref{eq:Gammasum};
\begin{align}\label{eq:corrMatrixCart}
\mathbf{C} = \left(\begin{array}{@{}c|c|c|c@{}}
\mathbf{\Gamma}(\vect{n}_{1},\vect{n}_{1}) & \mathbf{\Gamma}(\vect{n}_{1},\vect{n}_{2}) & \ldots & \mathbf{\Gamma}(\vect{n}_{1},\vect{n}_{\cn}) \\\hline
\mathbf{\Gamma}(\vect{n}_{2},\vect{n}_{1}) & \mathbf{\Gamma}(\vect{n}_{2},\vect{n}_{2}) & \ldots & \mathbf{\Gamma}(\vect{n}_{2},\vect{n}_{\cn}) \\\hline
\vdots & \vdots & \ddots & \vdots \\\hline
\mathbf{\Gamma}(\vect{n}_{\cn},\vect{n}_{1}) & \mathbf{\Gamma}(\vect{n}_{\cn},\vect{n}_{2}) & \ldots & \mathbf{\Gamma}(\vect{n}_{\cn},\vect{n}_{\cn}) \\
\end{array}\right).
\end{align}
This is only valid formally, because the matrix \(\mathbf{C}\) is not positive definite; in fact, it has \(N\) zero eigenvalues. The origin of this behavior is that the 3 Cartesian coordinates are being used to describe an intrinsically a 2-dimensional process on the sphere. This can be rectified by instead considering the changes in the polar coordinates of each star;
\begin{align}
\begin{pmatrix} \delta\theta_{i} \\ \delta\phi_{i} \end{pmatrix} = \mathbf{P}_{i}\cdot\!\begin{pmatrix} \delta x_{i} \\ \delta y_{i} \\ \delta z_{i} \end{pmatrix},
\quad\text{where}\quad\;
\mathbf{P}_{i} = \begin{pmatrix}
0 & 0 & -\dfrac{1}{\sqrt{1 - z_{i}^{2}}} \\
-\dfrac{y_{i}}{x_{i}^{2} + y_{i}^{2}} & \dfrac{x_{i}}{x_{i}^{2} + y_{i}^{2}} & 0
\end{pmatrix}.
\end{align}
\begin{align}
\Rightarrow \;\; \bm{\delta}\vect{\theta} \equiv (\delta \theta_{1} , \delta \phi_{1} ,  \delta \theta_{2} , \delta \phi_{2} , \ldots , \delta \theta_{\cn} , \delta \phi_{\cn} ) = \mathbf{\Pi}\cdot\bm{\delta}\vect{x} \quad\textrm{where}\quad \mathbf{\Pi} = \left(\begin{array}{@{}c|c|c|c@{}}
\mathbf{P}_{1} & \mathbf{0} & \ldots & \mathbf{0} \\ \hline
\mathbf{0} & \mathbf{P}_{2} & \ldots & \mathbf{0} \\ \hline
\vdots & \vdots & \ddots & \vdots \\ \hline
\mathbf{0} & \mathbf{0} & \ldots & \mathbf{P}_{\cn} \\
\end{array}\right).
\end{align}
The matrix \(\mathbf{\Pi}\) has shape \(2N \times 3N\). The vector \(\bm{\delta}\vect{\theta}\) is also distributed as a Gaussian random variable, \(\bm{\delta}\vect{\theta} \sim \mathcal{N}(\mathbf{0},\mathbf{C}')\), where the new covariance matrix \(\mathbf{C}' = \mathbf{\Pi}\cdot\mathbf{C}\cdot\mathbf{\Pi}^{\textrm{T}}\) is strictly positive definite. The parameter \(A\) is some overall amplitude chosen in Figs.~\ref{fig:tensorialRealPlot} and \ref{fig:scalarRealPlot} to be \(A=0.01\). A random realization of \(\bm{\delta}\vect{\theta}\) may now be obtained without an obstacle, and the vector \(\bm{\delta}\vect{x}\) can be obtained by simple geometry.

The plots in Figs.~\ref{fig:tensorialRealPlot} and \ref{fig:scalarRealPlot} were produced using \(N = 1000\) stars placed regularly on the sky, and show the original star positions, \(\vect{x}\), and the new positions, \(\vect{x}+\bm{\delta}\vect{x}\), joined by a smooth curves using the Mollweide projection.

\section{The Transverse Scalar ``Breathing" Mode Astrometric Integrals}
\label{app:SInt}
\noindent
In this appendix the evaluation of spatial correlation integrals for the transverse scalar \textsc{gw} polarization state, \(S\), is briefly described. The integration is very similar to those for the \(+\) and \(\times\) states described in Appendix~\ref{app:PlusCrossInt}. The relevant correlation integrals were defined in eqs.~(\ref{eq:defGammaxtheta}) and (\ref{eq:defGammayphi}) as
\begin{subequations}
\begin{align}
\Gamma^{\mathsc{S}}_{x\theta}(\Theta) = \int_{S^{2}}\!\!\!\dd\Omega_{\vect{q}}\;\delta n_{x}^{\mathsc{S}} (\vect{q})\,\delta m_{\theta}^{\mathsc{S}} (\vect{q}) = \int_{0}^{2\pi}\!\!\!\!\dd\phi\int_{0}^{\pi}\!\!\!\dd\theta\,\sin\theta \,\delta n_{x}^{\mathsc{S}} (\theta,\phi)\,\delta m_{\theta}^{\mathsc{S}} (\theta,\phi) \,, \label{eq:appSintONE} \\
\Gamma^{\mathsc{S}}_{y\phi}(\Theta) = \int_{S^{2}}\!\!\!\dd\Omega_{\vect{q}}\;\delta n_{y}^{\mathsc{S}} (\vect{q})\,\delta m_{\phi}^{\mathsc{S}} (\vect{q}) = \int_{0}^{2\pi}\!\!\!\!\dd\phi\int_{0}^{\pi}\!\!\!\dd\theta\,\sin\theta \,\delta n_{y}^{\mathsc{S}} (\theta,\phi)\,\delta m_{\phi}^{\mathsc{S}} (\theta,\phi) \,, \label{eq:appSintTWO}
\end{align}
\end{subequations}
where the vector \(\vect{q}=(\sin\theta\cos\phi,\sin\theta\sin\phi,\cos\theta)\) is the direction on the sky from which the \textsc{gw} originates. The components of the astrometric deflections may be evaluated from the definitions in eqs.~(\ref{eq:dnx}), (\ref{eq:dmtheta}), (\ref{eq:dny}), and (\ref{eq:dmphi}), using the formula for the astrometric deflection in eq.~(\ref{eq:AstroDefFinal}), the expression for the \textsc{gw} basis tensor in eq.~(\ref{eq:tensorScalar}), and the expressions for the basis vectors tangent to the sphere in eqs.~(\ref{eq:defux}), (\ref{eq:defutheta}), (\ref{eq:defuy}), and (\ref{eq:defuphi}):

\begin{subequations}
\noindent\begin{minipage}[l]{0.485\linewidth}
\begin{flalign}
&\hspace*{11pt}\delta n_{x}^{\mathsc{S}} (\theta,\phi) = -\frac{1}{2}\,\sin\theta \cos\phi,&
\end{flalign}
\end{minipage}\hspace*{0.03\linewidth}
\begin{minipage}[l]{0.48\linewidth}
\begin{flalign}
&\delta n_{y}^{\mathsc{S}} (\theta,\phi) = \delta m_{\phi}^{\mathsc{S}} (\theta,\phi) = -\frac{1}{2}\,\sin\theta \sin\phi,&
\end{flalign}
\end{minipage}\\
\begin{flalign}
&\hspace*{11pt}\delta m_{\theta}^{\mathsc{S}} (\theta,\phi) = \frac{1}{2}\,\sin\Theta \cos\theta - \frac{1}{2}\,\cos\Theta \sin\theta \cos\phi.&
\end{flalign}
\end{subequations}
The azimuthal and polar integrals for \(\Gamma^{\mathsc{S}}_{x\theta}(\Theta)\) and \(\Gamma^{\mathsc{S}}_{y\phi}(\Theta)\) in eqs.~(\ref{eq:appSintONE}) and (\ref{eq:appSintTWO}) may be evaluated in the same way as those for \(\Gamma^{+,\times}_{x\theta}(\Theta)\) and \(\Gamma^{+,\times}_{y\phi}(\Theta)\) in Appendix~\ref{app:PlusCrossInt} to give the results in eqs.~(\ref{eq:scalarCorr1}) and (\ref{eq:scalarCorr2}) of the main text:

\begin{subequations}
\noindent\begin{minipage}[l]{0.485\linewidth}
\begin{flalign}
&\hspace*{11pt}\Gamma_{x \theta}^{\mathsc{S}} (\Theta) = \frac{\pi}{3}\,\cos \Theta \equiv \frac{\pi}{3} - \frac{2\pi}{3}\,\sin^{2}\!\left(\Theta / 2\right), \tag{\ref{eq:scalarCorr1}}&
\end{flalign}
\end{minipage}\hspace*{0.03\linewidth}
\begin{minipage}[l]{0.48\linewidth}
\begin{flalign}
&\Gamma_{y \phi}^{\mathsc{S}} (\Theta) = \frac{\pi}{3}. \tag{\ref{eq:scalarCorr2}}&
\end{flalign}
\end{minipage}
\end{subequations}

\section{The Vectorial Astrometric Integrals}
\label{app:VectorialInt}
\noindent
In this Appendix the evaluation of spatial correlation integrals for the vectorial \textsc{gw} polarization states, \(X\) and \(Y\), is briefly described. The derivation is very similar to those for the \(+\) and \(\times\) states described in Appendix~\ref{app:PlusCrossInt}. What is interesting in this case is that even though the standalone astrometric response in both vectorial polarizations is divergent at the origin (in Fig.~\ref{fig:starTraces} the ``star terms" were added to remove this divergence), the correlation functions are perfectly regular and finite for all relevant values of \(\Theta\).

Firstly, the \(X\) mode is considered; the relevant correlation integrals were defined in eqs.~(\ref{eq:defGammaxtheta}) and (\ref{eq:defGammayphi}) as 
\begin{subequations}
\begin{align}
\Gamma^{\mathsc{X}}_{x\theta}(\Theta) = \int_{S^{2}}\!\!\!\dd\Omega_{\vect{q}}\;\delta n_{x}^{\mathsc{X}} (\vect{q})\,\delta m_{\theta}^{\mathsc{X}} (\vect{q}) = \int_{0}^{2\pi}\!\!\!\!\dd\phi\int_{0}^{\pi}\!\!\!\dd\theta\,\sin\theta \,\delta n_{x}^{\mathsc{X}} (\theta,\phi)\,\delta m_{\theta}^{\mathsc{X}} (\theta,\phi) \,, \label{eq:appVxintONE} \\
\Gamma^{\mathsc{X}}_{y\phi}(\Theta) = \int_{S^{2}}\!\!\!\dd\Omega_{\vect{q}}\;\delta n_{y}^{\mathsc{X}} (\vect{q})\,\delta m_{\phi}^{\mathsc{X}} (\vect{q}) = \int_{0}^{2\pi}\!\!\!\!\dd\phi\int_{0}^{\pi}\!\!\!\dd\theta\,\sin\theta \,\delta n_{y}^{\mathsc{X}} (\theta,\phi)\,\delta m_{\phi}^{\mathsc{X}} (\theta,\phi) \,, \label{eq:appVxintTWO}
\end{align}
\end{subequations}
where the vector \(\vect{q}=(\sin\theta\cos\phi,\sin\theta\sin\phi,\cos\theta)\) is the direction on the sky from which the \textsc{gw} originates. The components of the astrometric deflections may be evaluated from the definitions in eqs.~(\ref{eq:dnx}), (\ref{eq:dmtheta}), (\ref{eq:dny}), and (\ref{eq:dmphi}), using the formula for the astrometric deflection in eq.~(\ref{eq:AstroDefFinal}), the expression for the \textsc{gw} basis tensor in eq.~(\ref{eq:tensorX}), and the expressions for the basis vectors tangent to the sphere in eqs.~(\ref{eq:defux}), (\ref{eq:defutheta}), (\ref{eq:defuy}), and (\ref{eq:defuphi}):

\begin{subequations}
\noindent\begin{minipage}[l]{0.485\linewidth}
\begin{flalign}
&\hspace*{12pt}\delta n_{x}^{\mathsc{X}} (\theta,\phi) = \frac{1}{2}\,(1 + 2\cos\theta) \cos\phi\,,&
\end{flalign}
\end{minipage}\hspace*{0.03\linewidth}
\begin{minipage}[l]{0.48\linewidth}
\begin{flalign}
&\delta n_{y}^{\mathsc{X}} (\theta,\phi) = \frac{1}{2}\,(1 + 2\cos\theta) \sin\phi\,,&
\end{flalign}
\end{minipage}\\[2pt]
\begin{flalign}
\begin{split}
&\delta m_{\theta}^{\mathsc{X}} (\theta,\phi) = \frac{1}{1 - \cos\Theta \cos\theta - \sin\Theta \sin\theta \cos\phi} \left(\frac{1}{8}\left(2 \sin\Theta \sin\theta - 3 \sin\!\left(2\Theta\right) \sin\!\left(2\theta\right)\right) \right. \\
&\quad\quad\quad\quad\quad\quad\quad\quad\quad\;\;\; \left. + \frac{1}{2} \left(\cos\Theta \cos\theta - \cos\!\left(2\Theta\right) \cos\!\left(2\theta\right)\right)\cos\phi + \frac{1}{4} \left(\sin\Theta\,\sin\theta - \sin\!\left(2\Theta\right) \sin\!\left(2\theta\right)\right) \cos\!\left(2\phi\right)\!\right),\;
\end{split}\\[5pt]
\begin{split}
&\delta m_{\phi}^{\mathsc{X}} (\theta,\phi) = \frac{1}{1 - \cos\Theta \cos\theta - \sin\Theta \sin\theta \cos\phi} \left(\frac{1}{2}\,\cos\Theta \left(\cos\theta - \cos\!\left(2\theta\right)\right)\,\sin\phi \right. \\
&\quad\quad\quad\quad\quad\quad\quad\quad\quad\quad\quad\quad\quad\quad\quad\quad\quad\quad\quad\quad\quad\quad\quad\quad\quad\quad\quad\; \left. + \frac{1}{4} \left(\sin\!\left(2\Theta\right) \sin\theta - \sin\Theta \sin\!\left(2\theta\right)\right)\,\sin\!\left(2\phi\right)\!\right).
\end{split}
\end{flalign}
\end{subequations}
The equations for \(\delta m_{\theta}^{\mathsc{X}} (\theta,\phi)\) and \(\delta m_{\phi}^{\mathsc{X}} (\theta,\phi)\) diverge when \(\Theta = 0\). Nevertheless, the azimuthal and polar integrals for \(\Gamma^{\mathsc{X}}_{y\phi} (\Theta)\) and \(\Gamma^{\mathsc{X}}_{y\phi} (\Theta)\) in eqs.~(\ref{eq:appVxintONE}) and (\ref{eq:appVxintTWO}) may be evaluated in the same way as those for \(\Gamma^{+,\times}_{x\theta} (\Theta)\) and \(\Gamma^{+,\times}_{y\phi} (\Theta)\) in Appendix~\ref{app:PlusCrossInt} to give two results that are regular for all values of \(\Theta\):
\begin{subequations}
\begin{align}
\Gamma_{x \theta}^{\mathsc{X}} (\Theta) = &\, \frac{\pi}{6} + \frac{2\pi}{3}\,\sin^{2}\!\left(\Theta / 2\right) + 6\pi\,\frac{\sin^{2}\!\left(\Theta / 2\right)}{1 - \sin^{2}\!\left(\Theta / 2\right)}\,\ln\!\left(\sin\!\left(\Theta / 2\right)\right) - 2\pi\,\frac{\cos^{2}\!\left(\Theta / 2\right)}{1 - \cos^{2}\!\left(\Theta / 2\right)}\,\ln\!\left(\cos\!\left(\Theta / 2\right)\right),  \label{eq:Vres1}\\
\Gamma_{y \phi}^{\mathsc{X}} (\Theta) = &\, \frac{13\pi}{6} + \pi \sin^{2}\!\left(\Theta / 2\right) + 6\pi\,\frac{\sin^{2}\!\left(\Theta / 2\right)}{1 - \sin^{2}\!\left(\Theta / 2\right)}\,\ln\!\left(\sin\!\left(\Theta / 2\right)\right) + 2\pi\,\frac{\cos^{2}\!\left(\Theta / 2\right)}{1 - \cos^{2}\!\left(\Theta / 2\right)}\,\ln\!\left(\cos\!\left(\Theta / 2\right)\right). \label{eq:Vres2}
\end{align}
\end{subequations}

Secondly, the \(Y\) mode is considered; the relevant correlation integrals were defined in eqs.~(\ref{eq:defGammaxtheta}) and (\ref{eq:defGammayphi}) as

\medskip
\vspace{-\abovedisplayskip}
\noindent
\begin{subequations}
\begin{minipage}[c]{0.495\linewidth}
\begin{align}
\Gamma^{\mathsc{Y}}_{x\theta} (\Theta) = \int_{0}^{2\pi}\!\!\!\!\dd\phi\int_{0}^{\pi}\!\!\!\dd\theta\,\sin\theta \,\delta n_{x}^{\mathsc{Y}}(\theta,\phi)\,\delta m_{\theta}^{\mathsc{X}} (\theta,\phi) \,, \label{eq:appVyintONE}
\end{align}%
\end{minipage}
\begin{minipage}[c]{0.50\linewidth}
\begin{align}
\Gamma^{\mathsc{Y}}_{y\phi} (\Theta) = \int_{0}^{2\pi}\!\!\!\!\dd\phi\int_{0}^{\pi}\!\!\!\dd\theta\,\sin\theta \,\delta n_{y}^{\mathsc{Y}}(\theta,\phi)\,\delta m_{\phi}^{\mathsc{X}} (\theta,\phi) \,. \label{eq:appVyintTWO}
\end{align}
\end{minipage}
\end{subequations}
\smallskip

The components of the astrometric deflections may be evaluated from the definitions in eqs.~(\ref{eq:dnx}), (\ref{eq:dmtheta}), (\ref{eq:dny}), and (\ref{eq:dmphi}), using the formula for the astrometric deflection in eq.~(\ref{eq:AstroDefFinal}), the expression for the \textsc{gw} basis tensor in eq.~(\ref{eq:tensorY}), and the expressions for the basis vectors tangent to the sphere in eqs.~(\ref{eq:defux}), (\ref{eq:defutheta}), (\ref{eq:defuy}), and (\ref{eq:defuphi}):

\begin{subequations}
\noindent\begin{minipage}[l]{0.485\linewidth}
\begin{flalign}
&\hspace*{13pt}\delta n_{x}^{\mathsc{Y}} (\theta,\phi) = \frac{1}{2}\,\cos\theta \sin\phi\,,&
\end{flalign}
\end{minipage}\hspace*{0.03\linewidth}
\begin{minipage}[l]{0.48\linewidth}
\begin{flalign}
&\delta n_{y}^{\mathsc{Y}} (\theta,\phi) = - \frac{1}{2}\,\cos\theta \cos\phi\,,&
\end{flalign}
\end{minipage}\\[2pt]
\begin{flalign}
\begin{split}
&\delta m_{\theta}^{\mathsc{Y}} (\theta,\phi) = \frac{1}{1 - \cos\Theta \cos\theta - \sin\Theta \sin\theta \cos\phi} \left(\frac{1}{2}\left(\cos\!\left(2\Theta\right) - \cos\Theta \cos\theta\right)\cos\theta \sin\phi \right. \\
&\quad\quad\quad\quad\quad\quad\quad\quad\quad\quad\quad\quad\quad\quad\quad\quad\quad\quad\quad\quad\quad\quad\quad\quad\quad\quad\quad\quad\;\; \left. + \frac{1}{4}\,\sin\Theta \left(2 \cos\Theta - \cos\theta\right)\sin\theta \sin\!\left(2\phi\right)\!\right),
\end{split}\\[5pt]
\begin{split}
&\delta m_{\phi}^{\mathsc{Y}} (\theta,\phi) = \frac{1}{1 - \cos\Theta \cos\theta - \sin\Theta \sin\theta \cos\phi} \left(\frac{3}{16}\,\sin\!\left(2\Theta\right) \sin\!\left(2\theta\right) + \frac{1}{4} \left(1 - 2\cos\Theta \cos\theta + \cos\!\left(2\Theta\right) \cos\!\left(2\theta\right)\right)\,\cos\phi \right. \\
&\quad\quad\quad\quad\quad\quad\quad\quad\quad\quad\quad\quad\quad\quad\quad\quad\quad\quad\quad\quad\quad\quad\quad\quad\quad\quad \left. + \frac{1}{16} \left(\sin\!\left(2\Theta\right) \sin\!\left(2\theta\right) - 8 \sin\Theta \sin\theta\right)\,\cos\!\left(2\phi\right)\!\right).
\end{split}
\end{flalign}
\end{subequations}
The integrals for \(\Gamma^{\mathsc{Y}}_{y\phi} (\Theta)\) and \(\Gamma^{\mathsc{Y}}_{y\phi} (\Theta)\) in eqs.~(\ref{eq:appVyintONE}) and (\ref{eq:appVyintTWO}) may be evaluated in the same way as the integrals in Appendix~\ref{app:PlusCrossInt}:
\begin{subequations}
\begin{align}
\Gamma_{x \theta}^{\mathsc{Y}} (\Theta) = &\, \frac{7\pi}{6} + 2\pi\,\frac{\sin^{2}\!\left(\Theta / 2\right)}{1 - \sin^{2}\!\left(\Theta / 2\right)}\,\ln\!\left(\sin\!\left(\Theta / 2\right)\right) + 2\pi\,\frac{\cos^{2}\!\left(\Theta / 2\right)}{1 - \cos^{2}\!\left(\Theta / 2\right)}\,\ln\!\left(\cos\!\left(\Theta / 2\right)\right), \label{eq:Vres3}\\
\Gamma_{y \phi}^{\mathsc{Y}} (\Theta) = &\, - \frac{5\pi}{6} + \frac{5\pi}{3}\,\sin^{2}\!\left(\Theta / 2\right) + 2\pi\,\frac{\sin^{2}\!\left(\Theta / 2\right)}{1 - \sin^{2}\!\left(\Theta / 2\right)}\,\ln\!\left(\sin\!\left(\Theta / 2\right)\right) - 2\pi\,\frac{\cos^{2}\!\left(\Theta / 2\right)}{1 - \cos^{2}\!\left(\Theta / 2\right)}\,\ln\!\left(\cos\!\left(\Theta / 2\right)\right). \label{eq:Vres4}
\end{align}
\end{subequations}
As was described in the main text, for a unpolarised background containing equal power of both \(X\) and \(Y\) polarization states we may define the combined spatial correlation functions \({\Gamma_{x\theta}^{\mathsc{X},\mathsc{Y}}(\Theta) = \Gamma_{x\theta}(\Theta)^{\mathsc{X}} + \Gamma_{x\theta}^{\mathsc{Y}} (\Theta)}\) and \({\Gamma_{y\phi}(\Theta)^{\mathsc{X},\mathsc{Y}} = \Gamma_{y\phi}(\Theta)^{\mathsc{X}} + \Gamma_{y\phi}^{\mathsc{Y}} (\Theta)}\). These new functions may be evaluated by taking the sums of the expression in eqs.~(\ref{eq:Vres1}), (\ref{eq:Vres2}), (\ref{eq:Vres3}), and (\ref{eq:Vres4}) to give the results which appeared in eq.~(\ref{eq:vectorialAstrCorr}) of the main text,
\begin{align}
\Gamma_{x \theta}^{\mathsc{X}, \mathsc{Y}} (\Theta) = \Gamma_{y \phi}^{\mathsc{X}, \mathsc{Y}} (\Theta) = \frac{4\pi}{3} + \frac{8\pi}{3}\,\sin^{2}\!\left(\Theta / 2\right) + 8\pi\,\frac{\sin^{2}\!\left(\Theta / 2\right)}{1 - \sin^{2}\!\left(\Theta / 2\right)}\,\ln\!\left(\sin\!\left(\Theta / 2\right)\right). \tag{\ref{eq:vectorialAstrCorr}}
\end{align}
The functions \(\Gamma^{\mathsc{X}}_{x\theta} (\Theta)\), \(\Gamma^{\mathsc{X}}_{y\phi} (\Theta)\), \(\Gamma^{\mathsc{Y}}_{y\phi} (\Theta)\), \(\Gamma^{\mathsc{Y}}_{x\theta} (\Theta)\), and \({\Gamma_{x\theta}^{\mathsc{X}, \mathsc{Y}} (\Theta) = \Gamma_{y\phi}^{\mathsc{X}, \mathsc{Y}}(\Theta)}\), are plotted in Fig.~\ref{fig:vectorialExtraPlot}.

\section{The Scalar Longitudinal Astrometric Integrals}
\label{app:longIntegralDer}
\noindent
In this Appendix the evaluation of spatial correlation integrals for the longitudinal scalar \textsc{gw} polarization state, \(L\), will be discussed. Using the method established in the previous Appendices will yield divergent curves. These result are still useful, and reasons for the anomaly are discussed in Section~\ref{sec:longitudinalModes}. A more careful calculation which removes the divergence is presented in Appendix~\ref{app:numericalintegral}. The integration is very similar to those for the \(+\) and \(\times\) states described in Appendix~\ref{app:PlusCrossInt}. The relevant correlation integrals were defined in eqs.~(\ref{eq:defGammaxtheta}) and (\ref{eq:defGammayphi}) as
\begin{subequations}
\begin{align}
\Gamma_{x\theta}^{\mathsc{L}} (\Theta) = \int_{S^{2}}\!\!\!\dd\Omega_{\vect{q}}\;\delta n_{x}^{\mathsc{L}} (\vect{q})\,\delta m_{\theta}^{\mathsc{L}} (\vect{q}) = \int_{0}^{2\pi}\!\!\!\!\dd\phi\int_{0}^{\pi}\!\!\!\dd\theta\,\sin\theta \,\delta n_{x}^{\mathsc{L}} (\theta,\phi)\,\delta m_{\theta}^{\mathsc{L}} (\theta,\phi) \,, \label{eq:appLintONE} \\
\Gamma_{y\phi}^{\mathsc{L}} (\Theta) = \int_{S^{2}}\!\!\!\dd\Omega_{\vect{q}}\;\delta n_{y}^{\mathsc{L}} (\vect{q})\,\delta m_{\phi}^{\mathsc{L}} (\vect{q}) = \int_{0}^{2\pi}\!\!\!\!\dd\phi\int_{0}^{\pi}\!\!\!\dd\theta\,\sin\theta \,\delta n_{y}^{\mathsc{L}} (\theta,\phi)\,\delta m_{\phi}^{\mathsc{L}} (\theta,\phi) \,. \label{eq:appLintTWO}
\end{align}
\end{subequations}
where the vector \(\vect{q}=(\sin\theta\cos\phi,\sin\theta\sin\phi,\cos\theta)\) is the direction on the sky from which the \textsc{gw} originates. The components of the astrometric deflections may be evaluated from the definitions in eqs.~(\ref{eq:dnx}), (\ref{eq:dmtheta}), (\ref{eq:dny}), and (\ref{eq:dmphi}), using the formula for the astrometric deflection in eq.~(\ref{eq:AstroDefFinal}), the expression for the \textsc{gw} basis tensor in eq.~(\ref{eq:tensorLong}), and the expressions for the basis vectors tangent to the sphere in eqs.~(\ref{eq:defux}), (\ref{eq:defutheta}), (\ref{eq:defuy}), and (\ref{eq:defuphi}):

\begin{subequations}
\noindent\begin{minipage}[l]{0.485\linewidth}
\begin{flalign}
&\hspace*{13pt}\delta n_{x}^{\mathsc{L}} = - \frac{1}{\sqrt{2}}\,\frac{\sin\theta \cos\theta \cos\phi}{1 - \cos\theta},&
\end{flalign}
\end{minipage}\hspace*{0.03\linewidth}
\begin{minipage}[l]{0.48\linewidth}
\begin{flalign}
&\delta n_{y}^{\mathsc{L}} = - \frac{1}{\sqrt{2}}\,\frac{\sin\theta \cos\theta \sin\phi}{1 - \cos\theta},&
\end{flalign}
\end{minipage}\\[2pt]
\begin{align}
\begin{split}
&\;\;\delta m_{\theta}^{\mathsc{L}} = \frac{1}{\sqrt{2}\left(1 - \cos\Theta \cos\theta - \sin\Theta \sin\theta \cos\phi\right)} \left(\frac{1}{8}\,\sin\!\left(2\Theta\right) (1 + 3\cos\!\left(2\theta\right)) + \frac{1}{4}\,\cos\!\left(2\Theta\right) \sin\!\left(2\theta\right)\, \cos\phi \right. \\
&\quad\quad\quad\quad\quad\quad\quad\quad\quad\quad\quad\quad\quad\quad\quad\quad\quad\quad\quad\quad\quad\quad\quad\quad\quad\quad\quad\quad\;\;\; \left. + \frac{1}{8}\,\sin\!\left(2\Theta\right) \left(\cos\!\left(2\theta\right) - 1\right) \cos\!\left(2\phi\right)\!\right), \quad\quad\quad\end{split}
\\[5pt]
\begin{split}
&\;\;\delta m_{\phi}^{\mathsc{L}} = \frac{1}{\sqrt{2}\left(1 - \cos\Theta \cos\theta - \sin\Theta \sin\theta \cos\phi\right)} \left(\frac{1}{2}\,\cos\Theta \sin\!\left(2\theta\right) \sin\phi + \frac{1}{4}\,\sin\Theta\,(1 - \cos\!\left(2\theta\right))\,\sin\!\left(2\phi\right)\!\right).
\end{split}
\end{align}
\end{subequations}\\[5pt]
The integrals for \(\Gamma_{y\phi}^{\mathsc{L}} (\Theta)\) and \(\Gamma_{y\phi}^{\mathsc{L}} (\Theta)\) in eqs.~(\ref{eq:appLintONE}) and (\ref{eq:appLintTWO}) may be evaluated in the same way as those for \(\Gamma_{x\theta}^{+,\times} (\Theta)\) and \(\Gamma_{y\phi}^{+,\times} (\Theta)\) in Appendix~\ref{app:PlusCrossInt} to give the results presented in Section~\ref{sec:longitudinalModes} of the main text,

\begin{subequations}
\noindent\begin{minipage}[l]{0.555\linewidth}
\begin{align}
\Gamma_{x \theta}^{\mathsc{L}} (\Theta) = - \frac{10\pi}{3} + \frac{8\pi}{3}\,\sin^{2}\!\left(\Theta / 2\right) - 2\pi\,\frac{\ln\!\left(\sin\!\left(\Theta / 2\right)\right)}{1 - \sin^{2}\!\left(\Theta/2\right)}, \tag{\ref{eq:longCorr1}}
\end{align}
\end{minipage}\hspace*{0.03\linewidth}
\begin{minipage}[l]{0.41\linewidth}
\begin{align}
\Gamma_{y \phi}^{\mathsc{L}} (\Theta) = - \frac{4\pi}{3} - 2\pi\,\frac{\ln\!\left(\sin\!\left(\Theta / 2\right)\right)}{1 - \sin^{2}\!\left(\Theta/2\right)}. \tag{\ref{eq:longCorr2}}
\end{align}
\end{minipage}
\end{subequations}\\[7pt]
As evident from Fig.~\ref{fig:longitudinalPlot}, where these two curves are plotted, they diverge at \(\Theta = 0\). Obviously, this implies that in the context of the scalar longitudinal mode, the method for calculating the overlap reduction function in inapplicable. Below is presented a modified approach which solves the issue of the divergent curves.

\subsection{Numerical Correlation Integrals on the Sphere}
\label{app:numericalintegral}
\noindent
The cause of the divergence in eqs.~(\ref{eq:longCorr1}) and (\ref{eq:longCorr2}) is the use of the use of the distant-source version of the astrometric response in eq.~(\ref{eq:AstroDefFinal}). If the full astrometric response in eq.~(\ref{eq:astrometricsignalFULL}) is used instead the resulting spatial correlation is always finite (although no longer analytically tractable). All of the general discussion of the spatial correlation in Section~\ref{sec:backgroundMapping} proceeds exactly as before, except eq.~(\ref{eq:defDelta}) now becomes the distance dependent expression
\begin{align}
\begin{split}
\tensor{\Delta}{_{i}^{jk}} \left(\vect{n}, \vect{q}, d\right) = &\Bigg(\!\left\{1 + \frac{\rmi (2 - q^{r} n_{r})}{d (1 - q^{\ell} n_{\ell})}\,\Big(1 - \exp\left(-\rmi d (1 - q^{s} n_{s})\right)\!\!\Big)\!\right\} n_{\Imath} \vphantom{\Bigg(\Bigg)^{2}} \\[-5pt]
& \quad\quad\quad\quad \;\, - \left\{1 + \frac{\rmi}{d (1 - q^{\ell} n_{\ell})}\,\Big(1 - \exp\left(-\rmi d (1 - q^{s} n_{s})\right)\!\!\Big)\!\right\} q_{\Imath}\Bigg) \frac{n^{j}n^{k}}{2(1 - q^{\ell} n_{\ell})} \\[5pt]
& \quad\quad\quad\quad\quad\quad\quad\quad - \left\{\frac{1}{2} + \frac{\rmi}{d (1 - q^{\ell} n_{\ell})}\,\Big(1 - \exp\left(-\rmi d (1 - q^{s} n_{s})\right)\!\!\Big)\!\right\} {\delta_{i}}^{j} n^{k}\,,
\end{split}\end{align}
where \(d = \omega \lambda_{\mathsc{S}} \Omega\) is a measure of the distance to the star (in gravitational wavelengths). With this form of \(\tensor{\Delta}{_{i}^{jk}}\), the integral becomes regular for all relevant values of \(\Theta\). The correlation curve is now a 3-parameter function of the angle on the sky \(\Theta\) and the distances to both stars, \(d_{n}\) and \(d_{m}\):
\begin{subequations}
\begin{align}
\Gamma_{x\theta}^{\mathsc{L}} (\Theta, d_{n}, d_{m}) = \int_{S^{2}}\!\!\!\dd\Omega_{\vect{q}}\;\delta n_{x}^{\mathsc{L}} (\vect{q}, d_{n})\,\delta m_{\theta}^{\mathsc{L}}(\vect{q}, d_{m}) = \int_{0}^{2\pi}\!\!\!\!\dd\phi\int_{0}^{\pi}\!\!\!\dd\theta\,\sin\theta \,\delta n_{x}^{\mathsc{L}} (\theta, \phi, d_{n})\,\delta m_{\theta}^{\mathsc{L}} (\theta, \phi, d_{m}) \,, \label{eq:appLint1} \\
\Gamma_{y\phi}^{\mathsc{L}} (\Theta, d_{n}, d_{m}) = \int_{S^{2}}\!\!\!\dd\Omega_{\vect{q}}\;\delta n_{y}^{\mathsc{L}} (\vect{q}, d_{n})\,\delta m_{\phi}^{\mathsc{L}}(\vect{q}, d_{m}) = \int_{0}^{2\pi}\!\!\!\!\dd\phi\int_{0}^{\pi}\!\!\!\dd\theta\,\sin\theta \,\delta n_{y}^{\mathsc{L}} (\theta, \phi, d_{n})\,\delta m_{\phi}^{\mathsc{L}} (\theta, \phi, d_{m}) \,. \label{eq:appLint2}
\end{align}
\end{subequations}
The vector \(\vect{q}=(\sin\theta\cos\phi,\sin\theta\sin\phi,\cos\theta)\) is the direction on the sky from which the \textsc{gw} originates. The components of the astrometric deflections may be evaluated from the definitions in eqs.~(\ref{eq:dnx}), (\ref{eq:dmtheta}), (\ref{eq:dny}), and (\ref{eq:dmphi}), using the formula for the astrometric deflection in eq.~(\ref{eq:astrometricsignalFULL}), the expression for the \textsc{gw} basis tensor in eq.~(\ref{eq:tensorLong}), and the expressions for the basis vectors tangent to the sphere in eqs.~(\ref{eq:defux}), (\ref{eq:defutheta}), (\ref{eq:defuy}), and (\ref{eq:defuphi}). Unfortunately, the integrals in eqs.~(\ref{eq:appLint1}) and (\ref{eq:appLint2}) can no longer be solved using the same method as above, and need to be evaluated numerically. Let \(\mathcal{L}_{x\theta} (\Theta)\) be the result of numerically evaluating \(\Gamma_{x\theta}^{\mathsc{L}} (\Theta)\) and \(\mathcal{L}_{y\phi} (\Theta)\) be the result of numerically evaluating \(\Gamma_{y\phi}^{\mathsc{L}} (\Theta)\). In Fig.~\ref{fig:longitudinalPlot} two curves \(\mathcal{L}_{x\theta} (\Theta)\) and \(\mathcal{L}_{y\phi} (\Theta)\) are shown as examples of these integrals for \((d_{n}, d_{m}) = (100, 200)\). It becomes evident that the numerical curves tend to the two analytic functions (\ref{eq:longCorr1}) and (\ref{eq:longCorr2}) for large \(\Theta\).

\subsection{Correlation integrals for stars in the same direction on the sky}
\label{app:corrSurfaces}
\noindent
Integrals (\ref{eq:appLint1}) and (\ref{eq:appLint2}) can be solved analytically for the case \(\Theta = 0\) (and also for \(\Theta = \pi\)) to give a 2-parameter family of functions which quantifies the correlation between the astrometric deflections of start with the same position on the sky but located at different distances,
\begin{align}\label{eq:longCorrSurfDef}
\Sigma^{\mathsc{L}} (d_{\vect{n}}, d_{\vect{m}}) = \left.\int_{0}^{2\pi}\!\!\!\!\dd\phi\int_{0}^{\pi}\!\!\!\dd\theta\,\sin\theta \,\delta n_{x}^{\mathsc{L}} (d_{\vect{n}}, \theta,\phi)\,\delta m_{\theta}^{\mathsc{L}} (d_{\vect{m}}, \theta,\phi) \right|_{\Theta = 0} = \int_{0}^{2\pi}\!\!\!\!\dd\phi\int_{0}^{\pi}\!\!\!\dd\theta\,\sin\theta \,\delta n_{x}^{\mathsc{L}} (d_{\vect{n}}, \theta,\phi)\,\delta n_{x}^{\mathsc{L}} (d_{\vect{m}}, \theta,\phi).
\end{align}
The derivation involves a simple azimuthal integral which yields another straightforward polar integral. The components of the astrometric deflections (at \(\Theta = 0\)) may be evaluated from the definitions in eqs.~(\ref{eq:dnx}), (\ref{eq:dmtheta}), (\ref{eq:dny}), and (\ref{eq:dmphi}), using the formula for the astrometric deflection in eq.~(\ref{eq:astrometricsignalFULL}), the expression for the \textsc{gw} basis tensor in eq.~(\ref{eq:tensorLong}), and the expressions for the basis vectors tangent to the sphere in eqs.~(\ref{eq:defux}), (\ref{eq:defutheta}), (\ref{eq:defuy}), and (\ref{eq:defuphi}):
\begin{align}
\begin{split}
&\delta n_{x}^{\mathsc{L}} (d, \theta, \phi) = \delta n_{y}^{\mathsc{L}} (d, \theta, \phi) = - \frac{1}{\sqrt{2}} \left(\!\frac{\cos^{3}\!\left(\theta/2\right)}{\sin\!\left(\theta/2\right)} - \sin\!\left(\theta/2\right)\cos\!\left(\theta/2\right)\!\right)\cos\phi \\
&\quad\quad\quad\quad\quad\quad\quad\quad\quad\quad\quad\quad + \frac{1}{2 \sqrt{2}\,d}\left(3 - 2 \cos^{2}\!\left(\theta/2\right)\right)\!\left(\!\frac{\cos^{3}\!\left(\theta/2\right)}{\sin^{3}\!\left(\theta/2\right)} - \frac{\cos\!\left(\theta/2\right)}{\sin\!\left(\theta/2\right)}\!\right)\sin\!\left(2d\sin^{2}\!\left(\theta/2\right)\right)\cos\phi\,.
\end{split}
\end{align}
The integral for \(\Sigma^{\mathsc{L}} (d_{\vect{n}}, d_{\vect{m}})\) in eq.~(\ref{eq:longCorrSurfDef}) may be evaluated using a computer algebra package to give
\begin{align}\label{eq:longSurf}
&\Sigma^{\mathsc{L}} (d_{\vect{n}}, d_{\vect{m}}) = \pi\,\gamma + \pi \ln(2) - \frac{29\pi}{30} + \frac{33\pi}{d_{\vect{n}}^{4}} + \frac{33\pi}{d_{\vect{m}}^{4}} + \frac{11\pi}{2\,d_{\vect{n}}^{2}} + \frac{11\pi}{2\,d_{\vect{m}}^{2}} - \frac{90\pi}{\left(d_{\vect{n}}^{2} - d_{\vect{m}}^{2}\right)^{2}} + \pi\ln\!\left(d_{\vect{n}} d_{\vect{m}}\right) \nonumber\\
&\quad\quad - \pi\,\frac{\left(d_{\vect{n}} + d_{\vect{m}} - 2\right)\left(d_{\vect{n}} + d_{\vect{m}} + 2\right)}{4\,d_{\vect{n}} d_{\vect{m}}}\,\ln\!\left(d_{\vect{n}} + d_{\vect{m}}\right) + \pi\,\frac{\left(d_{\vect{n}} - d_{\vect{m}} - 2\right)\left(d_{\vect{n}} - d_{\vect{m}} + 2\right)}{4\,d_{\vect{n}} d_{\vect{m}}}\,\ln\!\left|d_{\vect{n}} - d_{\vect{m}}\right| \nonumber\\
&\quad\quad - \pi\,\Ci\!\left(2 d_{\vect{n}}\right) - \pi\,\Ci\!\left(2 d_{\vect{m}}\right) - \frac{9\pi}{2\,d_{\vect{n}}}\,\Si\!\left(2 d_{\vect{n}}\right) -  \frac{9\pi}{2\,d_{\vect{m}}}\,\Si\!\left(2 d_{\vect{m}}\right) \nonumber\\
&\quad\quad + \pi\,\frac{\left(d_{\vect{n}} + d_{\vect{m}} - 2\right)\left(d_{\vect{n}} + d_{\vect{m}} + 2\right)}{4\,d_{\vect{n}} d_{\vect{m}}}\,\Ci\!\left(2\!\left(d_{\vect{n}} + d_{\vect{m}}\right)\right) - \pi\,\frac{\left(d_{\vect{n}} - d_{\vect{m}} - 2\right)\left(d_{\vect{n}} + d_{\vect{m}} + 2\right)}{4\,d_{\vect{n}} d_{\vect{m}}}\,\Ci\!\left(2\!\left|d_{\vect{n}} - d_{\vect{m}}\right|\right) \nonumber\\
&\quad\quad + \frac{15 \pi \left(d_{\vect{n}} + d_{\vect{m}}\right)}{4\,d_{\vect{n}} d_{\vect{m}}}\,\Si\!\left(2\!\left(d_{\vect{n}} + d_{\vect{m}}\right)\right) - \frac{15 \pi \left(d_{\vect{n}} - d_{\vect{m}}\right)}{4\,d_{\vect{n}} d_{\vect{m}}}\,\Si\!\left(2\!\left|d_{\vect{n}} - d_{\vect{m}}\right|\right)\!\big) \\
&\quad\quad - \pi\,\frac{5d_{\vect{n}}^{2} - 54}{2\,d_{\vect{n}}^{4}}\,\cos\!\left(2d_{\vect{n}}\right) - \pi\,\frac{5d_{\vect{m}}^{2} - 54}{2\,d_{\vect{m}}^{4}}\,\cos\!\left(2d_{\vect{m}}\right) + \pi\,\frac{d_{\vect{n}}^{4} + 11 d_{\vect{n}}^{2} - 60}{2\,d_{\vect{n}}^{5}}\,\sin\!\left(2d_{\vect{n}}\right) + \pi\,\frac{d_{\vect{m}}^{4} + 11 d_{\vect{m}}^{2} - 60}{2\,d_{\vect{m}}^{5}}\,\sin\!\left(2d_{\vect{m}}\right) \nonumber\\
&\quad\quad - \frac{10\pi}{\left(d_{\vect{n}}^{2} - d_{\vect{m}}^{2}\right)^{2}}\,\cos\!\left(2d_{\vect{n}}\right)\cos\!\left(2d_{\vect{m}}\right) - \pi\,\frac{\left(d_{\vect{n}}^{2} - d_{\vect{m}}^{2}\right)^{3} - 12 \left(d_{\vect{n}}^{2} - d_{\vect{m}}^{2}\right)^{2} + 100 \left(d_{\vect{n}}^{2} + 3 d_{\vect{m}}^{2}\right)}{4\left(d_{\vect{n}}^{2} - d_{\vect{m}}^{2}\right)^{3}d_{\vect{m}}}\,\cos\!\left(2d_{\vect{n}}\right)\sin\!\left(2d_{\vect{m}}\right) \nonumber\\
&\quad\quad + \pi\,\frac{\left(d_{\vect{n}}^{2} - d_{\vect{m}}^{2}\right)^{3} + 12 \left(d_{\vect{n}}^{2} - d_{\vect{m}}^{2}\right)^{2} - 100 \left(d_{\vect{n}}^{2} + 3 d_{\vect{m}}^{2}\right)}{4\left(d_{\vect{n}}^{2} - d_{\vect{m}}^{2}\right)^{3}d_{\vect{n}}}\,\sin\!\left(2d_{\vect{n}}\right)\cos\!\left(2d_{\vect{m}}\right) \nonumber\\
&\quad\quad + \frac{\pi}{8\,d_{\vect{n}}d_{\vect{m}}}\left(\!\frac{20}{\left(d_{\vect{n}} + d_{\vect{m}}\right)^{2}} + \frac{20}{\left(d_{\vect{n}} - d_{\vect{m}}\right)^{2}} + 31\!\right)\,\sin\!\left(2d_{\vect{n}}\right)\sin\!\left(2d_{\vect{m}}\right). \nonumber
\end{align}
Here, \(\gamma = 0.57722\) is the E\"{u}ler-Mascheroni constant, and \(\Ci(\bullet)\) and \(\Si(\bullet)\) are the cosine and sine integrals, respectively.

\end{appendices}
\end{document}